\documentclass[traditabstract]{aa}

\usepackage{txfonts}
\usepackage{graphicx}
\usepackage{pdflscape}
 \usepackage{fixltx2e}
\usepackage{natbib}
\bibpunct{(}{)}{;}{a}{}{,} 

\def\NH2{N$_{\mathrm{H}_{2}}$}
\def\mum{$\mu$m}
\def\grad{^{\circ}}
\def\sun{$_{\odot}$}
\def\Her{\textit{Herschel}}
\def\Hyp{\textit{Hyper}}
\def\Cut{\textit{Cutex}}
\def\Get{\textit{getsources}} 
\def\herschel{\textit{Herschel}} 

\title{\textit{Hyper}: Hybrid Photometry and Extraction Routine}

\author{A. Traficante\inst{\ref{inst:1}} \thanks{e-mail:alessio.traficante@manchester.ac.uk} \and G. A. Fuller\inst{\ref{inst:1},\ref{inst:2}} \and J. E. Pineda\inst{\ref{inst:2},\ref{inst:3}} \and S. Pezzuto\inst{\ref{inst:4}}}

\institute{Jodrell Bank Centre for Astrophysics, School of Physics and Astronomy, University of Manchester, Oxford Road, Manchester M13 9PL, UK \label{inst:1} \and 
UK ARC Node, Jodrell Bank Centre for Astrophysics, School of Physics and Astronomy, University of Manchester, Manchester, M13 9PL, UK \label{inst:2} \and 
European Southern Observatory (ESO), Garching, Germany \label{inst:3}
\and IAPS - INAF, via Fosso del Cavaliere, 100, I-00133 Roma, Italy \label{inst:4} }

\begin{document}

\date{}

\label{firstpage}

\abstract{We present a new Hybrid Photometry and Extraction Routine: \Hyp. It is
  designed to do compact source photometry allowing for varying spatial resolution and  sensitivity in multi-wavelength surveys. \Hyp\ combines multi-Gaussian fitting with aperture photometry to provide reliable photometry in regions
  with variable backgrounds and in crowded fields. The background is evaluated
  and removed locally for each source using polynomial fits of various
  orders. Source de-blending is done through simultaneous multi-Gaussian
  fitting of the main source and its companion(s), followed by the
  subtraction of the companion(s). \Hyp\ allows also simultaneous multi-wavelength photometry by setting a fixed aperture size independent of the map resolution and evaluating the source flux within the
  \textit{same} region of the sky at multiple wavelengths at the same time. This new code has been initially
  designed for precise aperture photometry in complex fields such as the
  Galactic plane observed in the far infrared (FIR) by the \herschel\ infrared survey of the Galactic plane (Hi-GAL). \Hyp\ has
  been tested on both simulated and real \herschel\ fields to quantify the quality of the
  source identification and photometry. The code is highly modular and fully
  parameterisable, therefore it can be easily adapted to different
  experiments. Comparison of the \Hyp\ photometry with the catalogued sources in the Bolocam Galactic Plane survey (BGPS), the 1.1 mm survey of the Galactic Plane carried out with the Caltech Sub-millimeter observatory, demonstrates the versatility of \Hyp\ on different datasets. It is fast and light in its memory usage, and it is freely available to the scientific community.}


\maketitle

\section{Introduction}
Increasingly high resolution and sensitive surveys at a range of wavelengths are
revealing the complex structure of the sky emission with a high level of
detail.  As a consequence, the source extraction and photometry task is an
increasingly challenging problem.  There are at least three main issues that
this task has to carefully manage: the sources are often observed on top of a complex, highly variable
background; the sources  are not necessarily isolated
but are often blended into multiple, closely spaced
objects; and the instruments are often designed to gather multi-wavelength
emission from the sky, which is essential for understanding the
underlying astrophysics, but the spatial resolution and sensitivity of
the observations are then wavelength dependent.

For example, the \herschel\ space observatory \citep{Pilbratt10} has observed
the far-infrared/sub-mm sky at various wavelengths in the range
$70\leq\lambda\leq500$ \mum\ with unprecedented sensitivity and resolution,
requiring photometry routines tailored for its capabilities.  So far, several
different approaches have been produced to estimate the photometry in \herschel\
data, either adapted from previous existing algorithms or newly developed
specifically for \herschel.

The standard \herschel\ data processing pipeline, HIPE \citep[\herschel\
Interactive Processing Environment,][]{HIPE} allows the user to estimate the
photometry by choosing between SUSSEXtractor \citep{Savage07} and
DAOPHOT \citep{daophot}. SUSSEXtractor has been
optimised in particular for SPIRE \citep{Griffin10}, one of the two photometry instruments on-board
\herschel. It convolves the image with a kernel
derived from the point response function and the convolved image is used to
identify source peaks and to estimate source fluxes. DAOPHOT is based on a
combination of \verb+find+ and \verb+aper+ IDL{\footnote{Interactive Data Language}} tasks. The sources are
identified in the image convolved with a DAOPHOT convolution kernel, and the
source flux is estimated with a standard aperture photometry assuming a uniform
background.

Alternative algorithms such as \textit{starfinder} \citep{Diolaiti00}, which
adopts the instrumental point spread function (PSF) as a template for the source identification, has
been used in extragalactic \herschel\ surveys such as the PACS Evolutionary
Program \citep[PEP,][]{Lutz11}. On the other hand, some widely-used source identification algorithms
such as \textit{clumpfind} \citep{Williams94} do not work well in regions with
extended emission, since the method ignores the presence of
background emission \citep[e.g.,][]{Pineda11}.

In addition, at least three approaches based on different techniques have been
developed specifically for these new \herschel\ data.  The MADX algorithm
\citep{Rigby11} has been designed to identify point-like sources in the \herschel\
extragalactic key-project survey ATLAS. The \Cut\ algorithm \citep{Molinari11} has also been developed
for the photometry of point-like and compact sources in crowded fields, whereas the \Get\ algorithm  \citep{Men'shchikov12} has been designed to extract sources ranging from point-like to very extended.

\Cut\ identifies compact sources in the second derivative
image of the sky. The second derivative is very sensitive to strong signal
variations such as point sources and not significantly affected by smooth
signal variations such as those due to diffuse structures. This allows 
source identification in fields with strong background contamination. The flux
is then estimated by fitting each source with a 2d Gaussian model. The major
limitation of this approach is with multi-wavelength analyses, as the Gaussian
fit is constrained by the beam size. In the \herschel\ data the beam size varies
by a factor of 7 (from $\simeq5$\arcsec at 70 \mum\ up to $\simeq35$\arcsec at 500 \mum),
therefore the Gaussian fits encompass larger regions at longer
wavelengths. To account for this bias, the flux of a source at longer wavelengths needs to be
rescaled assuming a reference beam at a fixed wavelength. However, the
rescaling procedure requires some assumed \textit{a-priori} knowledge about the
source properties \citep[e.g.][]{Nguyen-Luong11,Giannini12}.

On the other hand, \Get\ addresses the multi-wavelength
problem based on a fine decomposition of the original images over a wide range
of spatial scales and across all wavelengths. This approach appears to work in
both test-cases and real data analysis, however it is time-consuming and may
require considerable storage space \citep{Men'shchikov12} which could both be
significant limitations in producing source catalogues across large regions.

An alternative approach is classical aperture photometry
\citep{DaCosta92}.  Aperture photometry can easily be adapted to
multi-wavelength analysis, circumventing wavelength dependent resolution by
using a fixed aperture at all wavelengths. This removes the need for any
wavelength-resolution dependent correction factor when comparing source fluxes
(unlike, for example, \Cut) or super-resolution of sources (unlike
\Get).  Aperture photometry does not require detailed source
modelling and the computation time is in general short. However, there are
known issues in using classical aperture photometry in the presence of strong
background contamination, as well as in crowded fields where sources are
blended. 

Here we present a new source extraction and photometry algorithm: \Hyp\
(Hybrid photometry and extraction routine) with the goal of providing an
aperture photometry-based alternative method for the identification and
multi-wavelength photometry of point-like and compact (elliptical) sources in
regions (including crowded regions) where there is significant and variable
background emission. \Hyp\ uses an aperture matched to the source size
and shape at a selected wavelength, usually that with the lowest resolution,
to measure the flux of sources, providing a measurement of the emission from a
common volume of material at all wavelengths.

%
%

\begin{itemize}

\item[1.] 2d Gaussian fitting is used to determine the size of the elliptical
  aperture region used for each source.  The
  2d Gaussian fitting takes into account the source elongation however it is used
  only to estimate the region over which to integrate the source flux. This
  approach minimises the flux contamination from the background or the
  companion sources with respect to circular aperture photometry.

\item[2.] The background is estimated locally, evaluating different background models across regions of different sizes, modelling the background with different polynomial orders in each region. The code automatically chooses the background based on the lowest r.m.s. of the residual maps (evaluated for each estimated background).  

\item[3.] In the complicated and frequent case of crowded regions with
  blended sources the flux estimation is done using an hybrid approach between
  aperture photometry and multi-Gaussian fitting. In the case of blended sources \Hyp\
  performs a simultaneous multi-Gaussian fit of the sources and separates the
  reference source from the (modelled) companions.

\item[4.] \Hyp\ is designed for multiple
  wavelengths photometry. It associates sources detected independently in different bands and allows the user to choose a fixed wavelength at which to initially identify sources and a different wavelength (typically, but not necessarily,
  the longest wavelength) at which to fit sources with a 2d Gaussian. The 2d Gaussian fit defines the aperture used to integrate the flux arising from the \textit{same} volume of gas and dust at all wavelengths, bypassing the issue of wavelength dependent spatial resolutions.

\item[5.] Finally, \Hyp\ is very fast and it allows the user to identify
  hundreds of sources in wide fields in few minutes, making the code well
  suited to producing complete, multi-wavelengths source catalogue even in
  very complex and crowded fields.

\end{itemize}

The algorithm has been initially designed to produce a catalogue of proto- and pre-stellar clumps
\citep{Traficante14_cat} embedded in the infrared dark clouds
(IRDCs) observed with the \herschel\ survey of the Galactic plane, Hi-GAL
\citep{Molinari10_PASP}. This survey has mapped the whole Galactic plane in
the FIR in 5 bands, using both PACS \citep{Poglitsch10} and SPIRE
\citep{Griffin10} photometry instruments in parallel mode to observe the sky at 70, 160
\mum\ (PACS), 250, 350, 500 \mum\ (SPIRE). However, \Hyp\ is highly adaptable
and it can be used to produce source catalogues and photometry from different
surveys or instruments.

\Hyp\ is divided into two main blocks: source identification and source
photometry. The source identification strategy is described in Sect.
\ref{sec:source_identification} while the aperture photometry is described in
detail in Sect. \ref{sec:source_aperture_photometry}. In Sect.
\ref{sec:simulated_data} and \ref{sec:test_realdata} we describe the
photometry results on both simulated and real data, respectively. Finally, in
Sect. \ref{sec:conclusions} we present our conclusions.

\section{Source identification}\label{sec:source_identification}

The source identification, if done by eye, is a relatively simple exercise which allows experienced people to recognize source peaks even in very confused regions. On the other hand, setting up an automatic algorithm is a complicated task since the sources can be blended together, affected by noise and distributed on top of very variable background in the most complicated cases. In order to recognize sources in every environment, \Hyp\ starts with the assumption that if we decompose each map into its spatial frequencies, the higher spatial frequencies describe the point-like and compact sources while all the background and environment variations are described by the lower frequencies. Following this hypothesis, \Hyp\ identifies sources in a modified high-pass filtered map. The high-pass filtered map is obtained by convolving the map with a squared kernel designed to emphasise the brightness of the central pixel relative to neighbouring pixels. It is designed to have a positive peak value in the centre, surrounded by negative values. The filtering suppresses background variations
and emphasises the source peaks as well as small-scale noise and small-scale background variations, which, however, do not affect the source extraction as we show in Sect. \ref{sec:simulated_data}. The size of the kernel used to filter the
map is a user-defined
parameter. It has to be set considering the beam size, the pixel size
and the structures that the user wants to preserve. If the sky is properly
Nyquist-sampled and the pixel size is (as usual) one third of the beam
size, then a kernel of 5-7 pixels filters out most of the diffuse emission and
it is able to isolate point-like, PSF-shaped sources, while a larger kernel is more
suitable to find both PSF-shaped and slightly extended objects. The default
value is set to 9, a value which also preserves the structure of the elongated
sources. The convolution of the map with the chosen kernel is calculated in pixel space and it generates a ringing due to the nature of the
kernel itself. The main ring is visible as negative pixels across bright
sources in the filtered map, while the secondary rings give a
negligible contribution to the map and do not affect the source identification step. Since these negative pixels cannot be real sources, they are masked before proceeding with source identification. \Hyp\ sets the negative values to zero and we called the map ``clean filtered" map. Note that since the filtered map is used only to identify the peak position of the sources, this filtered map can be modified to improve the identification without altering the flux estimation.


In the clean filtered map only PSF-shaped or the slightly elongated sources (plus noise and small-scale background variations) survive, set on top of a smoothed, or clean, residual background. This simplifies the peak identification on this map which is now easier and less dependent on the original complexity of the map. The peaks in the clean filtered map are identified by applying in sequence the \verb+find+ and \verb+gcntrd+ IDL routines. \verb+find+ searches
for perturbations in the maps by means of Gaussian fitting, the most
suitable procedure to identify compact source peaks. The \verb+gcntrd+ routine
computes the source centroid starting with the \verb+find+ identification and
it is used after the \verb+find+ routine to better constrain the source
centroids. The \verb+find+ routine requires a threshold, $\sigma_{\mathrm{t}}$, a user-defined parameter expressed as a multiple of the clean filtered map r.m.s. We underline that the clean filtered map r.m.s. is unrelated to the true r.m.s. of the sky map. Instead, it depends on the distribution of sources in each field since in the clean filtered map the non-zero pixels contain only the source peaks and the background noise residuals. Therefore, $\sigma_{\mathrm{t}}$ determines the intensity of perturbation which is considered a source, and is the number of recovered sources in each field, as well as the likelihood of detecting false sources. As often occurs in approaches which require the selection of a threshold to analyse data which are potentially very different, there are no recipes to fix \textit{a
  priori} the threshold, since it is influenced by the crowding of each specific field. For example, this approach is similar to that used in \Cut, since the source
identification is done in the second derivative images and the threshold is
fixed in the derivative domain \citep{Molinari11}. Also, it is similar to $\sigma$-clipping algorithms which require the definition, \textit{a-priori}, of an unknown level to exclude outliers in the analysis. Users are therefore encouraged to
test different values and choose \textit{a posteriori} the most suitable value
for their purposes. The algorithm evaluates the peak flux of the faintest identified source. This value is reported in the output file, providing the user feedback on the ``depth" of the source extraction for the chosen threshold.

\begin{figure*}
\centering
\includegraphics[width=15cm]{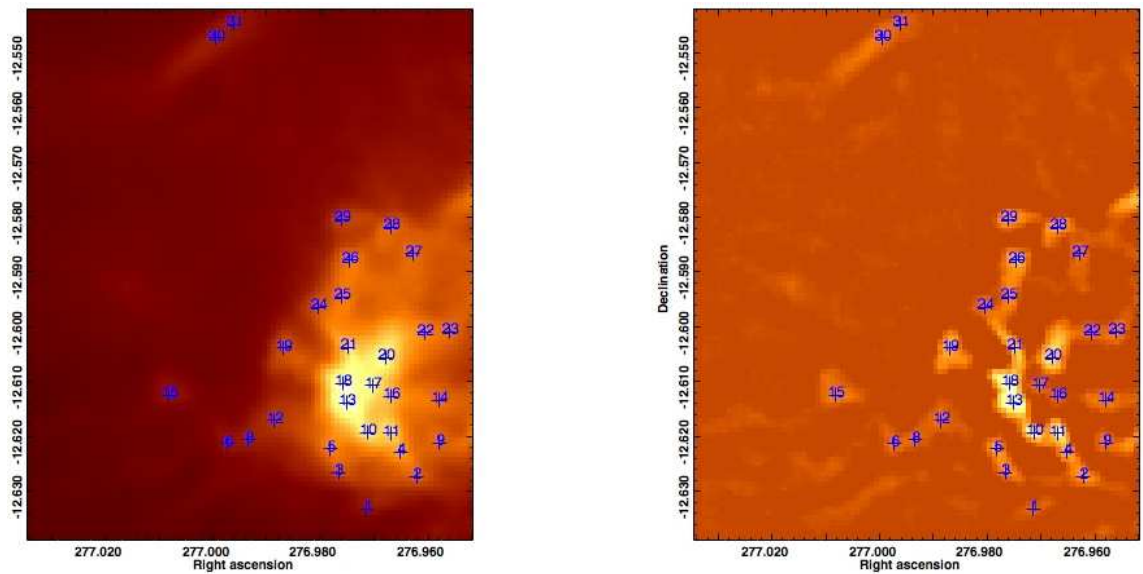}
\caption{Hi-GAL 70 \mum\ counterpart of the IRDC
  SDC19.073-0.602 (left panel). The blue crosses are the sources
  identified in the clean high-pass filtered image (right panel). The
  high-pass filtering emphasises the peaks while suppressing the
  diffuse emission.}
\label{fig:HGL19.073-0.602}
\end{figure*}

In Sect. \ref{sec:simulated_data} we discuss the effect of varying the threshold values for each \herschel\ wavelength. As we show with simulations in Sect. \ref{sec:simulated_data}, the code correctly identifies the sources in the fields and the number of false positives
identified can be very low, depending on the chosen threshold. An example of source
identification in the 70 \mum\ Hi-GAL counterpart of the IRDC SDC19.073-0.602
\citep{Peretto09} is shown in Fig. \ref{fig:HGL19.073-0.602}. The high-pass
filter simply suppresses the diffuse emission while the point sources survived
the filtering step, making them easily identifiable by the \verb+find+ and
\verb+gcntrd+ IDL routines. In this example we used a threshold of
$\sigma_{\mathrm{t}}=5.5$, which is suitable for this dataset \citep{Traficante14_cat}.


\section{Source aperture photometry}\label{sec:source_aperture_photometry}

Once identified, the sources are modelled in the original (unfiltered) image with a 2d Gaussian profile.
When analyzing a single wavelength this Gaussian fit is used to define
both the source position and size as well as the elliptical region
across which \Hyp\ integrates the source flux. The process for
analysing observations at multiple wavelenghts is described in
Sect.~\ref{sec:multiwavelength}.  This approach is a significant
improvement with respect to a standard circular aperture photometry
with two main advantages: it minimises the integration of the residual emission which does not belong to the source in
particular for the most elongated sources, for which a circular shape
is a poor approximation; it also minimises the flux contamination
arising from nearby sources.



The 2d Gaussian fitting is done using the \texttt{mpfit2dpeak} IDL routine
\citep{Markwardt09}. For each source \Hyp\ estimates seven parameters which
can vary within specific ranges. The results are the best fit
values of these parameters, for which an initial guess is needed. These
parameters are broken down into four groups:

\begin{enumerate}
\item \textit{background}: The code cuts out a small region around
  each source and the background level is initially considered
  constant within the extracted region. The initial guess is fixed by
  taking the mean value of the region, evaluated with a sigma-clipping
  procedure to minimise the contribution of possible companion
  sources.
\item \textit{source centroids}: The centroid of each source is fixed
  to the value estimated in the source identification step. It can
  vary within a radius that can be set in the parameter file.
\item \textit{source peak flux}: The source peak flux is initially
  taken as the difference between the peak value and the background value
  and it can vary without bounds.

\item \textit{2d Gaussian FWHMs and the position
    angle, $\phi$}: By default the FWHMs can vary from a minimum of
  $\Delta\theta_{\lambda}$ up to $2.0\cdot\Delta\theta_{\lambda}$, where
  $\Delta\theta_{\lambda}$ is the beam FWHM of the observations. The range can be changed in a parameter file. The position angle can vary without limits.
  
\end{enumerate}
 
The elliptical aperture used for the photometry is defined to have semi-minor and semi-major axes (the aperture radii) equal to the FWHMs obtained from the 2d Gaussian fit.

In two distinct cases \Hyp\ forces the semi-axes to be equal to the average of the minimum and maximum FWHM values chosen in the parameter file: if the semi-axes ratio is greater than 2.5, which can be the result of Gaussian fits forced by filaments where many sources are observed; if the \texttt{mpfit2dpeak} routine does not converge. In
the output files the status of the fit is reported and it can assume three
different values: 0 if the fit converged; $-$1 if the result of the fit was
too elongated; $-$2 if the routine did not converge. 

Before evaluating the source fluxes the background is
estimated and removed and nearby sources are de-blended as described
in Sect. \ref{sec:background} and \ref{sec:deblending}
respectively.

\subsection{Background estimation}\label{sec:background}
\Hyp\  follows an iterative procedure to estimate and remove the local background. The background is evaluated in a rectangular region across each source and it is modelled with a function described by a 2d-polynomial. The size of the region can vary from a minimum of twice up to four times the major FWHMs of each source. The background is evaluated using different sized regions and different polynomial orders (from zero up to the fourth) with the \texttt{mpfit2dfun} routine \citep{Markwardt09}. The source is masked in the centre to avoid source contamination in the background estimation. All the different backgrounds are subtracted from the original map and the residual map with the lowest \textit{r.m.s.} is selected to determine which is the best polynomial fit (and the region size) to model the background. This approach automatically accounts for both very smooth and highly variable local backgrounds. A keyword in the output file (``pol.'') identifies the best-order polynomial fit. 



\begin{figure*}
\centering
\includegraphics[width=5.5cm]{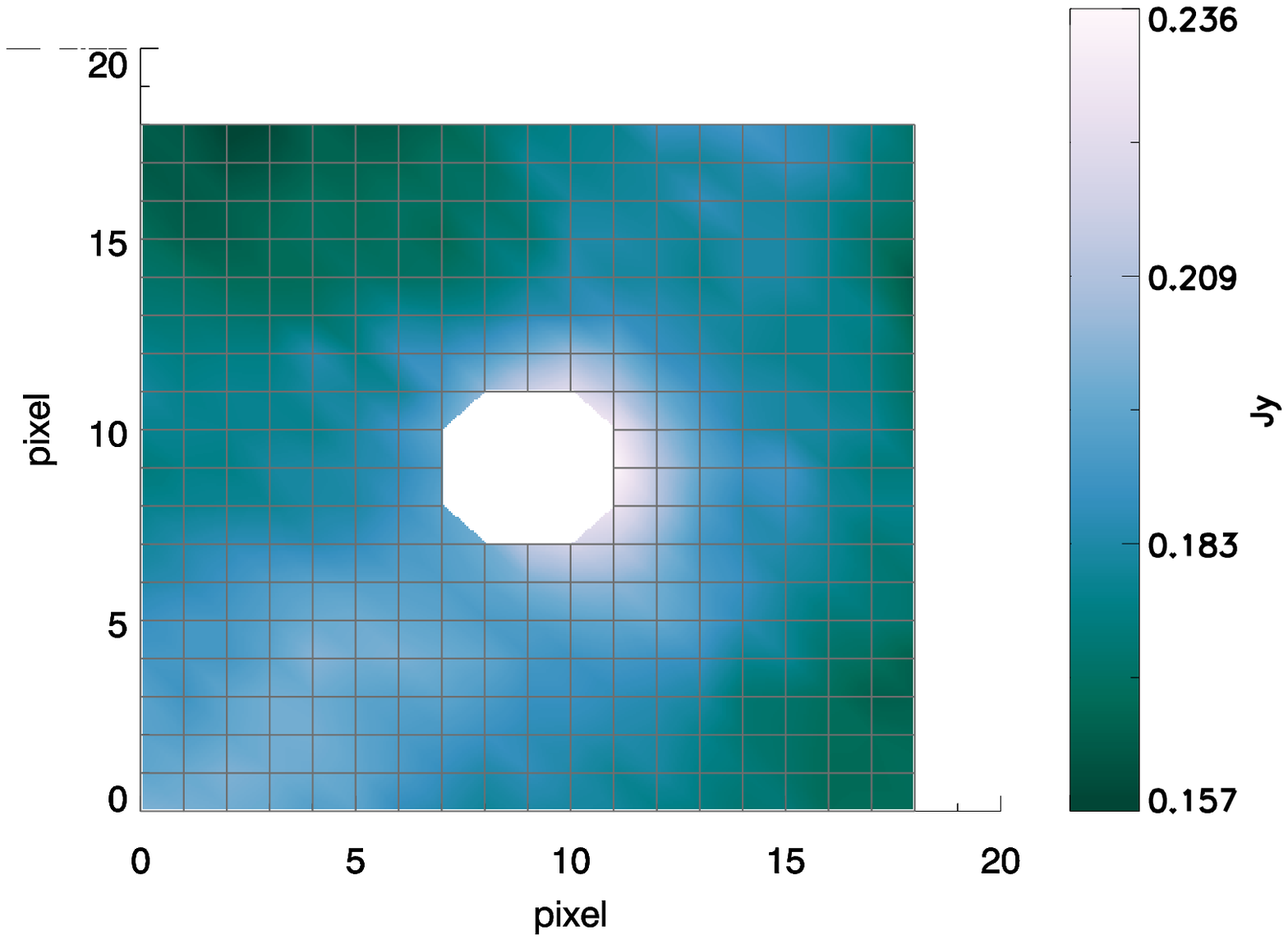}
\includegraphics[width=5.5cm]{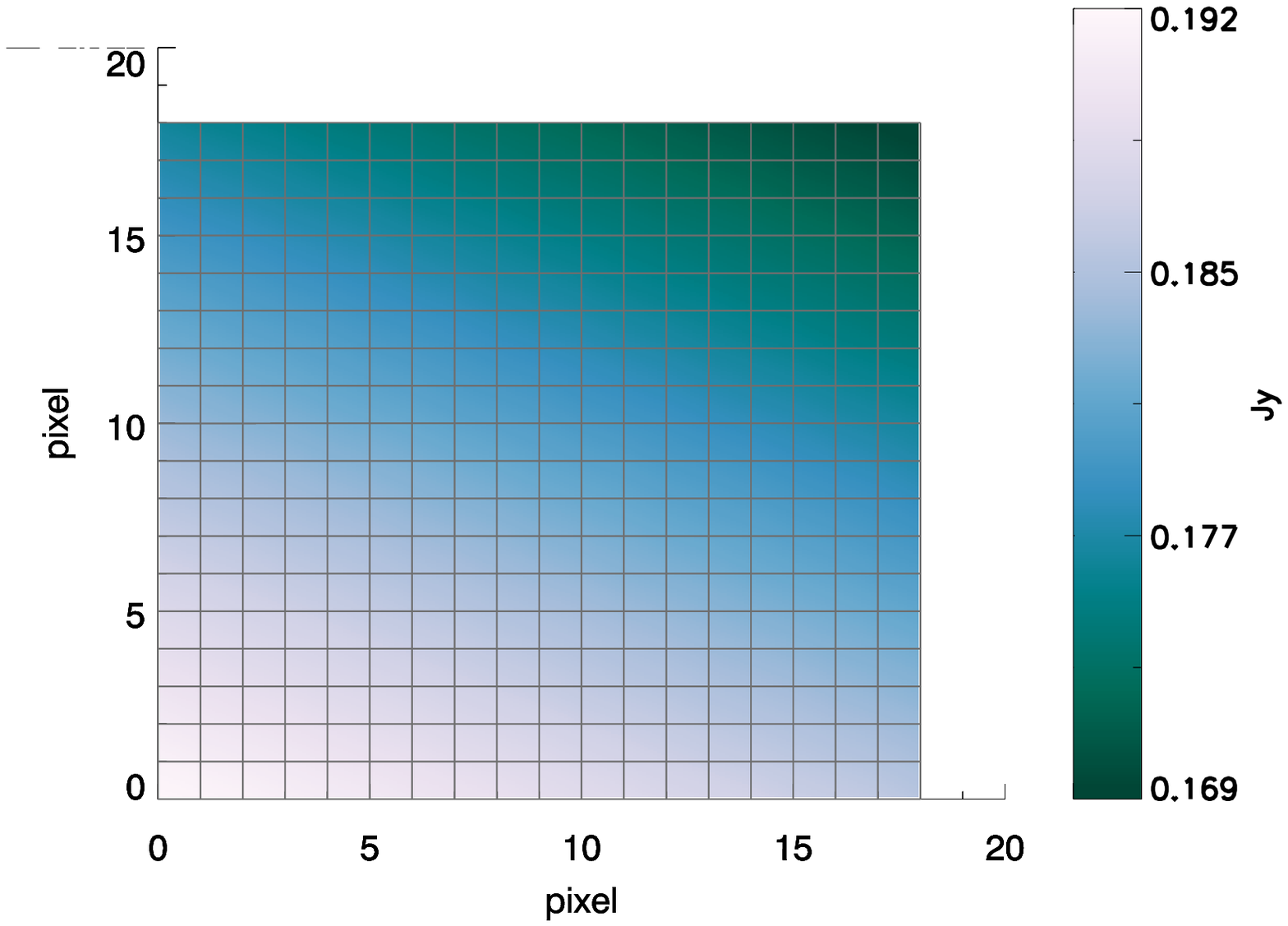}
\includegraphics[width=5.5cm]{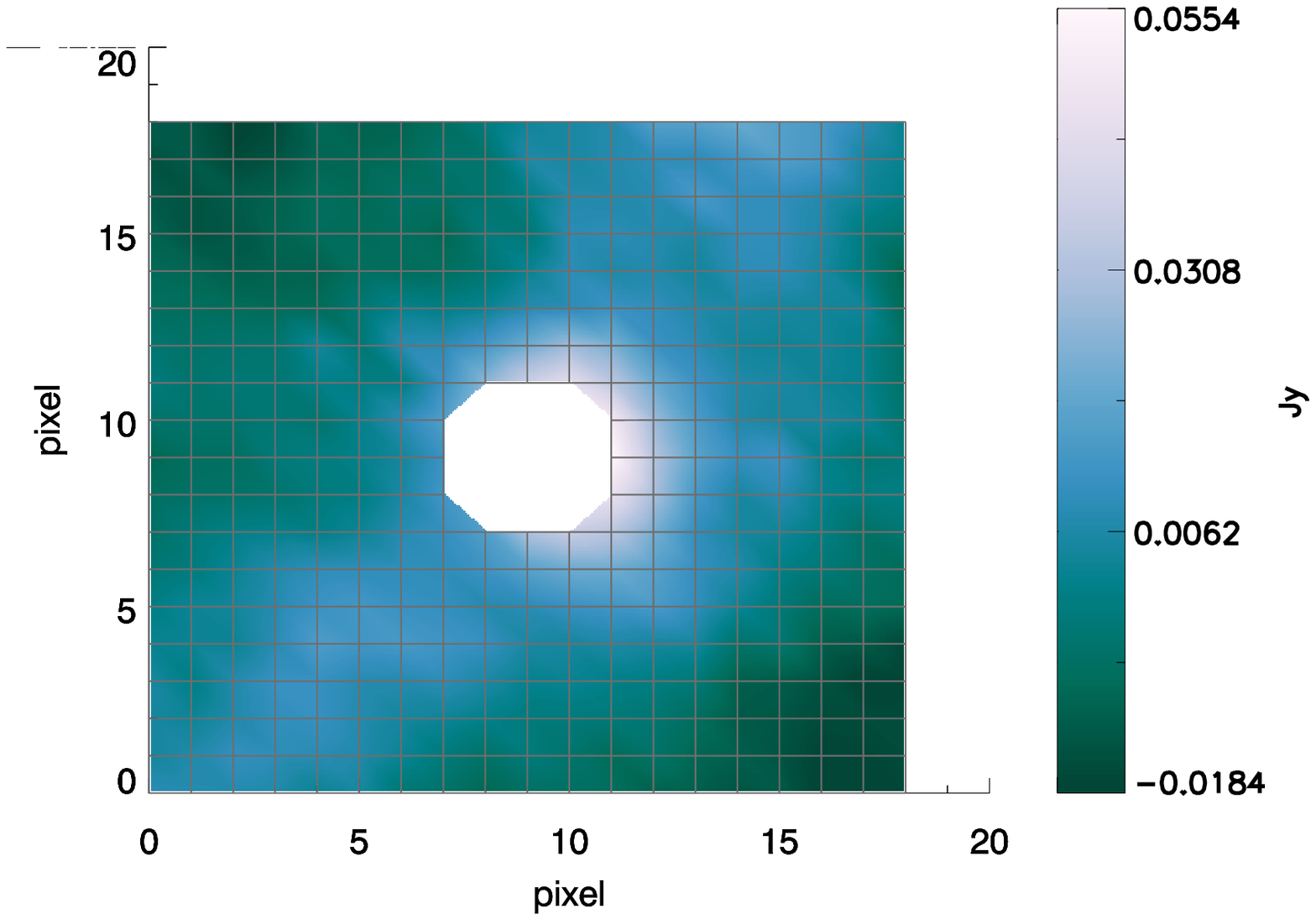}\\
\includegraphics[width=5.5cm]{map_l055_red_map.eps}
\includegraphics[width=5.5cm]{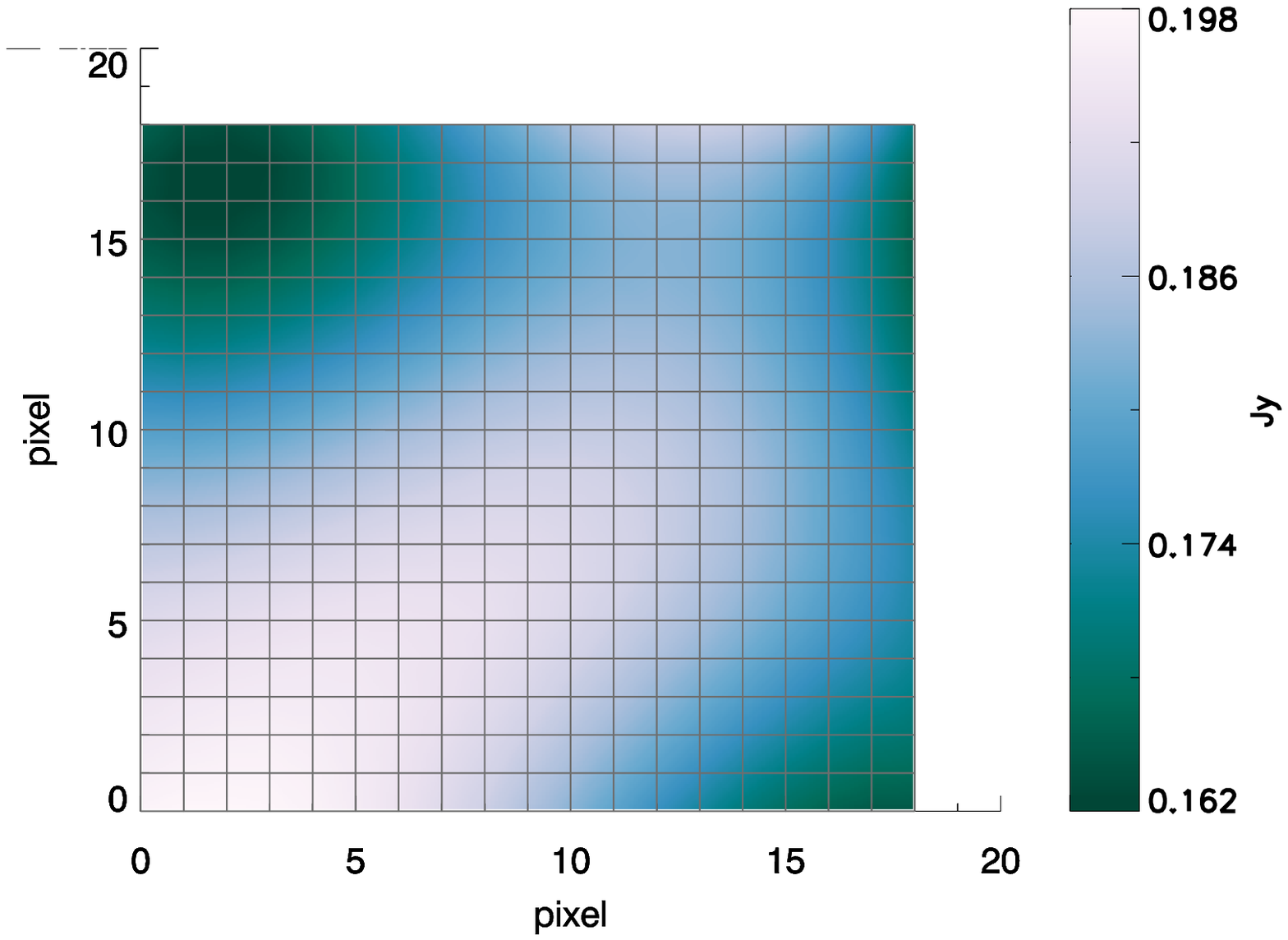}
\includegraphics[width=5.5cm]{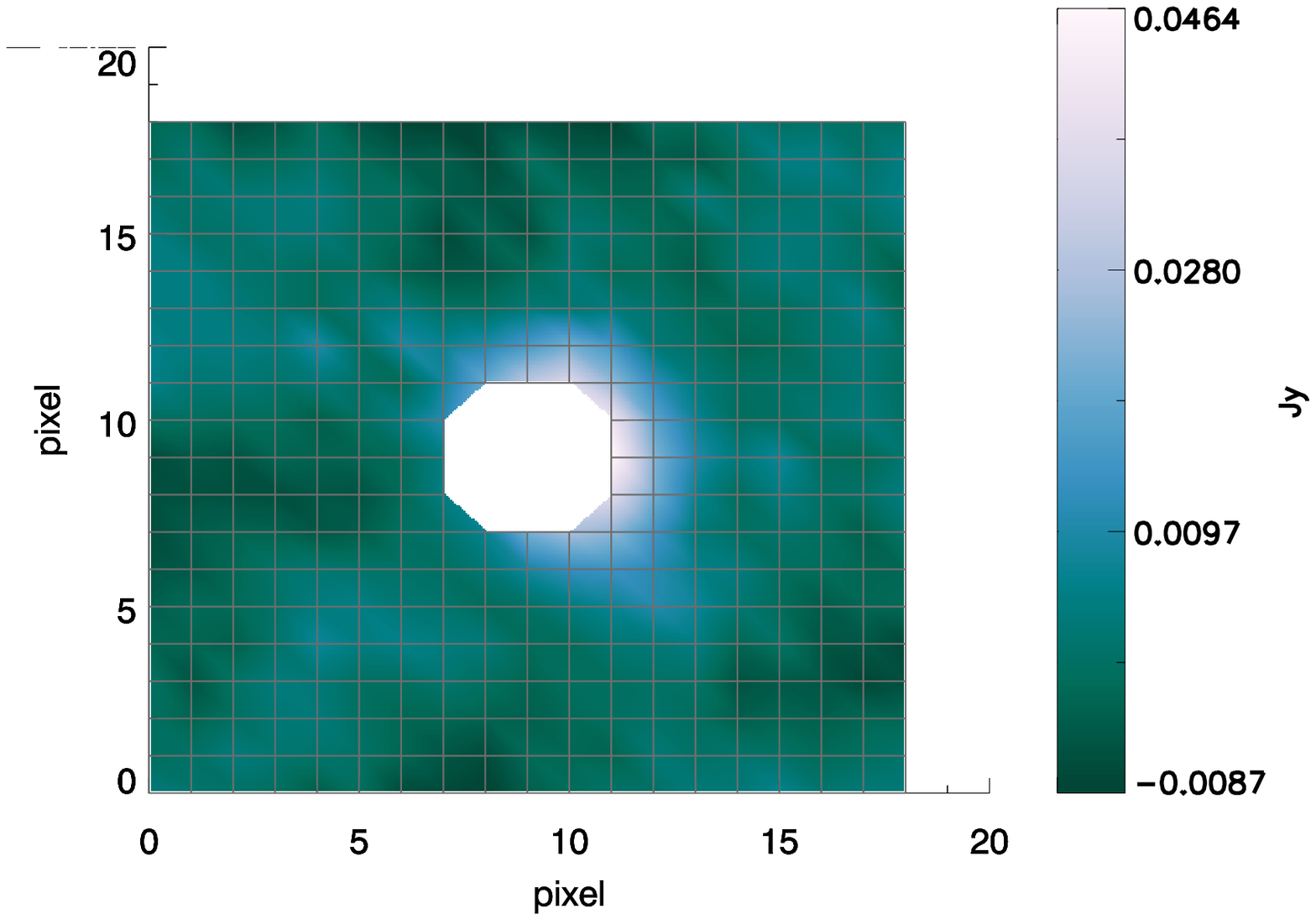}

\caption{\textit{Upper and lower left}: Simulated map of a 2d Gaussian source added on top of
  a variable background, extracted from a region centred on longitude
  $l=55\grad$ in the Galactic plane from the Hi-GAL survey at 160 \mum. The
  source itself is masked in the centre to improve the polynomial
  fit to the background. \textit{Center}: Background estimate adopting a first-order (plane)
  approximation (upper center) and a fourth-order approximation (lower center). \textit{Right}: Residual map assuming a first-order polynomial fit for the background estimation (upper right) or a fourth-order (lower right) polynomial fit. The high-order polynomial takes into account the complexity of the background.}
\label{fig:surface_map}
\end{figure*}

To show an example where a high-order polynomial improves the background estimation, we simulated a 2d Gaussian source that we added
on top of a highly variable, real background extracted from a Hi-GAL region
centred on longitude $l$=55$\grad$ at 160 \mum. The source has a peak flux of
$\simeq53$ mJy and its integrated flux is $0.840$ Jy. The peak flux is $\simeq
4.5\times\sigma$ where $\sigma=12$ mJy/pix is the standard deviation of the
map evaluated after a sigma-clipping procedure. Fig. \ref{fig:surface_map} shows the simulated
map, the first-order background and the best-order polynomial background
(fourth-order) fits and the residuals respectively. Table \ref{tab:polynomial_flux} shows the
fluxes evaluated using different background fits up to the fourth-order
fit. Due to the high variation in the background, a simple first-order
approximation gives rise to a severe overestimate of the flux and \Hyp\
measured a total flux of $1.452$ Jy, $\simeq73\%$ higher than the true
value. The discrepancy is significantly reduced using higher orders. With a
fourth-order polynomial the code reduces the flux difference to $\simeq19\%$,
with the flux measured equal to $1.003$ Jy. Also the r.m.s. of the background
residual after subtraction, evaluated in the same rectangular region used to
fit the background, is lower for the higher degree polynomial fits (see Table
\ref{tab:polynomial_flux}).

\begin{table}
\caption{\Hyp\ flux estimation for the injected test source (Fig. \ref{fig:surface_map}) using different polynomial fits to model the background.}
\begin{center}
\begin{tabular}{c|c|c|c}
\hline
\hline
Polynomial & \Hyp\ flux & Difference & r.m.s.\\ 
 order & (Jy) & (\%) & (mJy)\\   
\hline
1 & 1.451 & 73 & 11.2\\
2 & 1.075 & 28 & 7.6\\
3 & 1.062 & 26 & 7.3\\
4 & 1.003 & 19 & 6.8\\
\hline
\end{tabular}
\end{center}
\tablefoot{The reference flux of the injected source is 0.840 Jy. The fourth-order polynomial improves the flux estimation up to $\simeq55\%$ with respect to the first-order polynomial. Also the residual background r.m.s. around the source is lower in the fourth-order polynomial approximation of the background.}
\label{tab:polynomial_flux}
\end{table}

\subsection{Companion sources removal}\label{sec:deblending}

The blending of overlapping sources is a major problem for aperture
photometry and it can lead to a severe source flux
overestimation if the companion sources are not carefully identified
and disentangled. 

The parameter which determines if a reference source is assigned
one (or more) companions is the distance between its centroid and the
centroids of other sources in the field. All sources with a distance
less than a fixed value are flagged as companion sources by \Hyp. The distance
is a user specified parameter but by default is automatically set to twice
the chosen maximum aperture radius (see Sect.
\ref{sec:source_aperture_photometry}). This is the maximum distance at
which two sources can have their 2d Gaussian fits (and therefore their
integration areas) partially overlapping.

In the case of a source with companions \Hyp\ uses a hybrid combination of
multi-Gaussian source fitting and aperture photometry. It simultaneously fits
2d Gaussians to all the sources plus a first-order polynomial
background fit as a first guess. \Hyp\ does not use a higher order in this step
since the confusion due to the companion emission could affect the
background fitting.

The 2d Gaussian parameters of the companions are used to subtract the corresponding Gaussians
 from the initial image. The final image contains the reference source plus the background and the
residuals from the subtraction of each companion. The 2d Gaussian fitting
of the source is then used to identify the region to integrate the
flux as is in the single source case (Sect.
\ref{sec:aperture_photometry}), after the standard background subtraction described in Sect. \ref{sec:background}.

\begin{figure*}
\centering
\includegraphics[width=18cm]{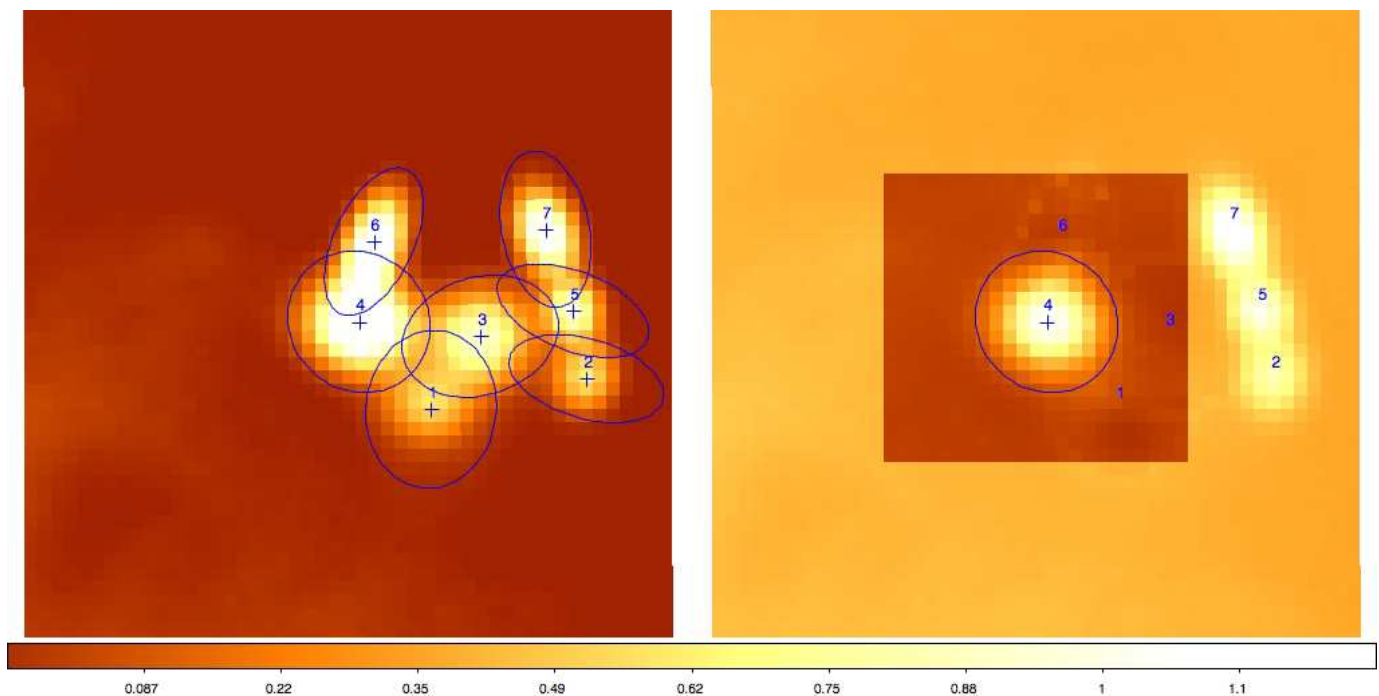}
\caption{Left: Simulation of seven blended sources with the corresponding \Hyp\ profiles. Each ellipse represents the aperture area of the sources. The semi-axes are equal to the FWHMs obtained from the 2d Gaussian fit. Right: Map of source 4 after companions removal done with the simultaneous multi-Gaussian fit and local background subtraction as described in the text. Only three sources contribute to the flux at the position of source 3 (sources 1, 3 and 6) and have been de-blended. The rectangular region in the right panel is the local region where the background for the source 3 is evaluated after the de-blending as described in Sect. \ref{sec:background}. The residuals are low and they do not lead to severe flux errors for the reference source (Table \ref{tab:sources_7_deblended})}.
\label{fig:multigauss_7sources}
\end{figure*}
\begin{figure}
\centering
\includegraphics[width=8cm]{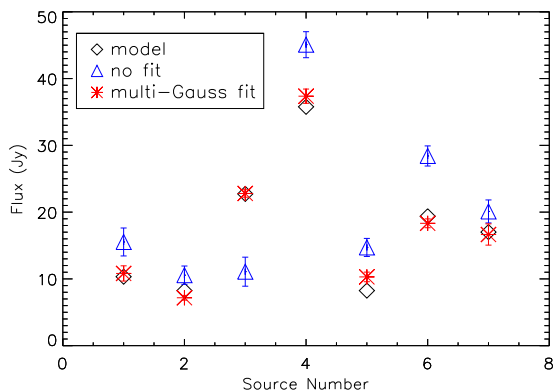}
\caption{Results of de-blending the seven partially overlapping sources shown in Fig. \ref{fig:multigauss_7sources}.
  The flux of the 2d Gaussian source model for each source is shown as
  a black diamond. The source flux evaluated without any companions
  removal is shown as a blue triangle, and after companion removal is shown as a red asterisk.}
\label{fig:multigauss_7sources_plot}
\end{figure}

To test the algorithm we have simulated a toy model in which we have injected
7 partially overlapping sources on top of a real background extracted from a
patch of Hi-GAL map observed with PACS at 160 \mum. \Hyp\ recognises all the seven sources with a position accuracy of $1.14\pm0.72\arcsec$, therefore better than one third the pixel size (4.5$\arcsec$ at 160 \mum). Fig.
\ref{fig:multigauss_7sources} shows the map of the seven sources and the map
of the residuals after the companion and background subtraction for the
brightest source. In this example, the flux of the source
model is 35.772 Jy. The integrated flux estimated by \Hyp\ without subtracting any companions
is 45.076 Jy, a difference of $\simeq26\%$. After
the companions removal (Fig. \ref{fig:multigauss_7sources_plot}) the \Hyp\ flux is 37.366 Jy, a difference of
$\simeq4\%$. The model fluxes compared with the fluxes recovered with and without
companions de-blending for all the seven sources are shown in Table
\ref{tab:sources_7_deblended}. The mean difference between the source
model fluxes and the Hyper fluxes is $\simeq43\%$ without companions
removal. The difference drops substantially after the de-blending, with a mean
value of $\sim7.8\%$ and in 5 out of 7 cases the flux recovered by \Hyp\ is within
5\% of the flux of the model.

\begin{table}
\caption{Comparison between the integrated flux for the seven blended sources in Fig. \ref{fig:multigauss_7sources}.}
\begin{center}
\begin{tabular}{c|c|c|c}
\hline
\hline
Source & Reference flux & No removal & Removal\\ 
& (Jy) & (Jy) & (Jy)\\   
\hline
1 & 10.332 & 15.532 & 10.818 \\
2 & 8.235 & 10.557 & 7.175 \\
3 & 22.729 & 11.081 & 22.796\\
4 & 35.772 & 45.076 & 37.366 \\
5 & 8.249 & 14.715 & 10.298\\
6 & 19.380 & 28.413 & 18.327 \\
7 & 17.008 & 20.105 & 16.651\\

\hline
\end{tabular}
\end{center}
\tablefoot{Col. 2: reference flux of the seven sources; Col. 3 \& 4: flux evaluated without and with de-blending the companions for each source respectively.}
\label{tab:sources_7_deblended}
\end{table}

\subsection{Source photometry}\label{sec:aperture_photometry}

Once the source background has been removed and the companions have been
subtracted, the flux is evaluated by integrating all the flux within the
elliptical aperture defined as described at the beginning of this Section.

The precision of the flux integral depends on the precision in evaluating the percentage of each pixel that belongs to the integration region, in particular for the pixels which are only partially included within the elliptical aperture. \Hyp\ follows an oversampling procedure to evaluate the flux integral. Each pixel is expanded in a fixed number of sub-pixels using the \verb+frebin+ IDL routine, which preserves the total pixel flux. The \verb+frebin+ routine requires the number of bins per side of each pixel, which will determine the final number of sub-pixels. All the sub-pixels that are within the ellipse defining the aperture are counted in the evaluation of the flux integral. 

The precision in the integration also depends on the number of bins and the number of points describing the ellipse. The higher the number of bins, the higher the number of points describing the ellipse required to include all the sub-pixels within the integration area. 

\begin{figure}
\centering
\includegraphics[width=8cm]{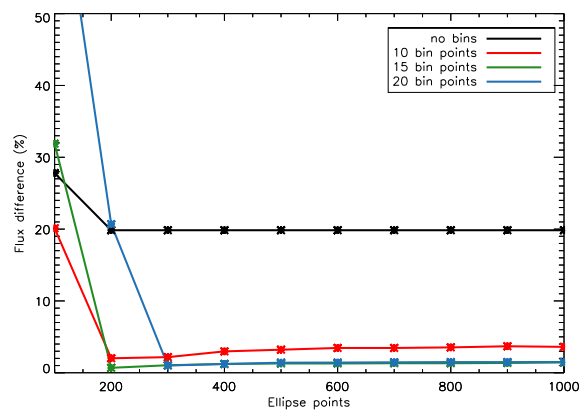}
\caption{Flux difference in percentage with varying the number of points
  describing the ellipse for different bins used to oversample each
  pixel with respect to the reference model flux. The differences are $\leq1.5\%$ using more than
  200 point to describe each ellipse and with at least a bin value per side of 15. Using no bins the percentage is remarkably high and independent from the number of points describing the ellipse (using at least 200 points), since it implies to integrate the entire pixel values for all the pixels in which the ellipse points is described.}
\label{fig:ellipse_points}
\end{figure}

We tested different configurations, varying the number of bins as well as the
points describing the aperture ellipse. In order to determine the quality of
the integration routine we produced a simple test map with constant pixel values of 10
(arbitrary units). We integrate in a circular region with a radius equal to 2
pixels, therefore the area of the circle is equal to 125.664. In Fig.
\ref{fig:ellipse_points} we report the percentage difference of the integral
evaluated by \Hyp\ for varying the number of ellipse points and bins.

Without oversampling (no bins) the pixels can either fall completely within
the area or fall completely outside, therefore the code does not converge to
the correct value irrespective of the number of points describing the
ellipse. At high numbers of bins and low numbers of points describing the
ellipse, too many sub-pixels are associated with the region and the area
results severely overestimated. However, increasing the number of points
significantly improves the estimation and the difference from the true area
and the estimated area is $\simeq1.5\%$ using 1000 points for the ellipse and
with bin value per side of 20 (e.g. oversampling the map of a factor 400). The number of bins as well as the points which describe the ellipse
will determine the accuracy of the flux estimation. Increasing these values
will enhance the flux estimation, but at the expenses of memory request and
execution time. The default settings, used in the tests described in Sect. \ref{sec:simulated_data} and \ref{sec:test_realdata}, are 20 bins and 1000 points.

\subsection{Multiple wavelength source photometry}\label{sec:multiwavelength}

A major problem with multi-wavelength
analysis is the different spatial resolution of the maps at the
different wavelengths. \Hyp\ (and the aperture photometry approach in
general) can overcome this limitation by setting a \textit{fixed}
aperture radius over which to integrate the source flux regardless of the
wavelength. We note that the algorithm leaves the user free to choose the minimum and maximum apertures as described above, however the aperture region should never be smaller than the beam size of the lowest resolution map. 

This approach assures that the emission arising from the same volume of gas and dust is included in the flux integration at all wavelengths. The background is estimated at each  wavelength using a square region with a length of a side equal to at least as the aperture, as described in Sect. \ref{sec:background}. The flux measured for a source will include any perturbation in high resolution maps with a spatial frequency higher than the order of the best polynomial. This is appropriate as the emission from such perturbations contributes to the emission measured within the beam of the low resolution observations. In Sect. \ref{sec:test_multiwave} we test the \Hyp\ multi-wavelength approach by injecting sources  in a complex field at each \Her\ wavelengths and measuring the source flux in the region estimated at the longest \Her\ wavelength, the 500 \mum\ band. In Sect. \ref{sec:test_multiwave} we also demonstrate that the method works well at all wavelengths and the largest uncertainties arise at 70 \mum, where is strongest the contribution from background structures not subtracted within the 500 \mum\ aperture.

If the user wants to obtain a multi-wavelength photometry with different
apertures at each wavelength, \Hyp\ can be easily run separately for each
wavelength and the results combined a-posteriori.

To process maps at various wavelengths \Hyp\ can read a list of maps from its
input file and four parameters can be set to control the multiple wavelength
analysis. These parameters are:

\begin{enumerate}

\item\textit{Multiple wavelengths}: This parameter defines all the wavelengths at which the user wants to evaluate the flux within the same aperture region.
 
\item\textit{Reference wavelength}: This parameter defines the wavelength at
  which the sources are initially identified.

\item\textit{Fitting wavelength}: The wavelength at which the source
  size is determined and used to define the elliptical photometry
  aperture. It can be different from the reference
  wavelength. 

\item\textit{Common source distance}: This radius sets the maximum
  distance at which sources observed at two different wavelengths are
  considered counterparts of the same object. By default it is equal to half the FWHM of the highest resolution map among the various selected wavelengths.

\end{enumerate}

When running \Hyp\ on multi-wavelength data the source identification
and 2-d Gaussian fitting is carried out independently at each
wavelength. As well as providing the source positions, this procedure allows
the de-blending of the sources at each wavelength. The common source distance is then used to determine whether a source detected at the
reference wavelength has a counterpart at the other wavelengths. The
position of a source at the \textit{reference wavelength} and of a source at another wavelength must be within the common source distance
for them to be associated. The elliptical photometry aperture is then
defined by the 2d Gaussian fit to the source at the fitting wavelength. The integrated flux is then
evaluated within this \textit{same} aperture at each wavelength. Therefore, \Hyp\ evaluates the source flux only for the sources with counterparts at all the multiple wavelengths set in the parameter file. However, \Hyp\ also produces a catalog and a region file with the source peak positions of all the sources identified independently at each wavelength. The user can identify a posteriori sources which are detected at only some wavelengths and were therefore not included in the final output.

If there are multiple sources within the aperture in a high resolution
map, the counterpart is assumed to be the source closest to the
reference source. The fluxes of all the resolved sources contribute to
the total flux if they are within the elliptical aperture. In the
output file the \textit{cluster} parameter indicates the number of the
resolved sources observed at each wavelengths which fall within the
aperture (see Sect. \ref{sec:test_realdata}).

\section{Simulated data}\label{sec:simulated_data}

To test the quality of the extraction and photometry in presence of a highly variable,
realistic background we produced several maps starting from real data
extracted from the Hi-GAL survey. The Galactic plane seen in the FIR shows a
strong background and it is crowded with sources
\citep[e.g.,][]{Molinari10_PASP}, so it is a suitable test of the code in
extreme conditions. 

We tested separately the source extraction routine, in order to explore the identification of false positives as function of the threshold  $\sigma_{\mathrm{t}}$ (Sect. \ref{sec:source_identification}), and the full \Hyp\ capabilities on different series of simulated maps.

\subsection{Testing the source extraction routine}
We first generated a series of map to test the effect of varying the threshold $\sigma_{\mathrm{t}}$ (Sect. \ref{sec:source_identification}) which determines the
percentage of sources recovered, as well as of the number of false
identifications. In order to show how the threshold $\sigma_{\mathrm{t}}$ influence the goodness of the source extraction, we extracted a squared
region of the Galactic plane with smooth background and few point sources. The region has been extracted from the PACS 160
\mum\ map centred on $(l,b)=(20.6\grad,0.7\grad)$ and with
a size of 15$\arcmin$ per side. It contains a few identified point sources, depending on the value of the threshold $\sigma_{\mathrm{t}}$. We produced 5 runs with different thresholds $\sigma_{\mathrm{t}}=[5.5,6.5,7.5,9.0,10.0]$. The number of sources identified in the field are $[12,3,3,1,1]$ respectively.

We generated 80 random Gaussian sources modelled as 2d Gaussian with random FWHMs along their two axes and position angles. For this test we aimed to avoid confusion due to overlapping sources, therefore we generated sources with a minimum centroid to centroid separation of 50\arcsec. The FWHM was allowed to vary in the range
$\Delta\theta_{\lambda}\leq\mathrm{FWHM}\leq1.5\cdot\Delta\theta_{\lambda}$. In order to improve the statistics, we generated 20 different templates, for a total of 1600 sources, divided in 2 groups of 800 sources.

\begin{figure}
\centering
\includegraphics[width=8cm]{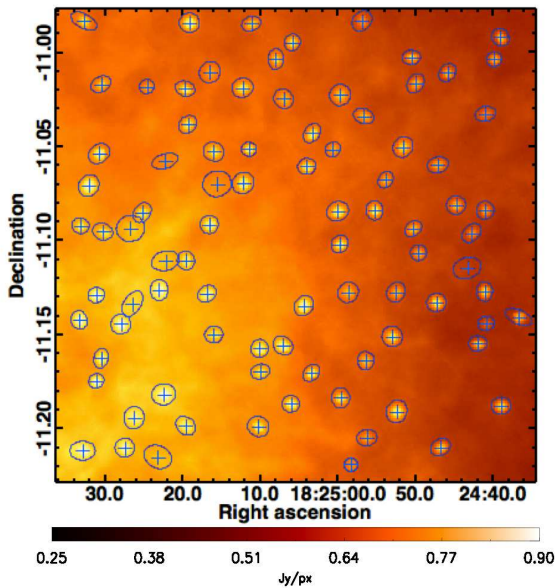}
\caption{One realisation of the TS1 run with 80 injected sources on top
  of true background, extracted from the Hi-GAL 160 $\mu$m map centred
  at on $(l,b)=(20.6\grad,0.7\grad)$. The blue ellipses are the source
  identified by \Hyp\ using a threshold $\sigma_\mathrm{t}$ of 6.5. The ellipses describe the aperture regions of the sources, evaluated as discussed in Section \ref{sec:source_aperture_photometry}.}
\label{fig:T1_run}
\end{figure}

\begin{figure}
\centering
\includegraphics[width=8cm]{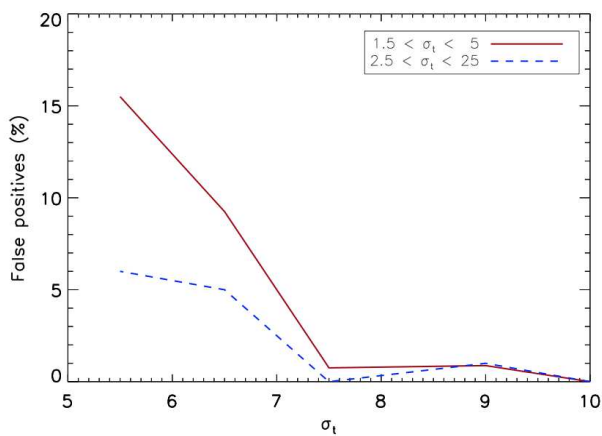}
\caption{Percentage of false positives identified for various thresholds $\sigma_{\mathrm{t}}$ for two simulated fields, TS1 with faint
  sources (red line) and TS2 with bright sources (blue dashed line). The fraction of false positives decreases with increasing threshold
  values.}
\label{fig:source_false_positives}
\end{figure}

In the first group (hereafter TS1), the source peak flux, $F_{p}$, varies from $1.5\sigma_{test}$ up to $5\sigma_{test}$ and in the second (TS2) 2.5$\sigma_{test}\leq F_{p}\leq25\sigma_{test}$, with  $\sigma_{test}$ being the standard deviation of the flux in the map. TS1 has been carried out to test the ability of the code to identify faint objects. An example of TS1 realisation is in Fig. \ref{fig:T1_run}. We ran the code with different thresholds for both groups and the results are shown in Fig.
\ref{fig:source_false_positives}. The false
positives are due to perturbations in the background, therefore at low $\sigma_{\mathrm{t}}$ values the
percentage of false positives is larger in TS1 than in TS2. At high
$\sigma_{\mathrm{t}}$ values the number of false positives strongly
decreases with less than 2\% of false positives identified in both groups. This trend however can vary from observation to observation
and there is no universal $\sigma_{\mathrm{t}}$ value (Sect. \ref{sec:source_identification}); it has to be optimised
for each different source extraction analysis. In both TS1 and TS2
the percentage of true sources recovered is 100\%. 

Note that the threshold depends on both the intensity of the sources in the field and the wavelength. The next series of tests, aimed to demonstrate the quality of the overall \Hyp\ capabilities, have been performed at all wavelengths and the thresholds have been chosen to minimise the false positives and maximise the true source recovery. The chosen threshold can be taken as a reference for each \herschel\ wavelength, however the users are encouraged to test different values to optimise the extraction for their own data.

\subsection{The photometry accuracy of Hyper}\label{sec:test_photometry}
The full \Hyp\ pipeline, from the source extraction up to the source photometry, has been tested and validated on complex, extended fields, with sources injected on top of real background extracted from \Her\ observations. We have run two different series of tests: in the first we injected sources randomly across wide regions with variable backgrounds and in the second the sources has been injected on top of specific, highly background contaminated regions where real compact sources are more likely to be found, namely filaments.

In the first series of test we extracted a large patch for each wavelength of the Hi-GAL data in order to include different variable backgrounds. The patch belongs to a region included in the Galactic range $35\grad\leq l\leq41\grad$ depending on the wavelength with an extent of $\simeq 0.5\grad\times0.5\grad$. 

\begin{figure*}
\centering
\includegraphics[width=9cm]{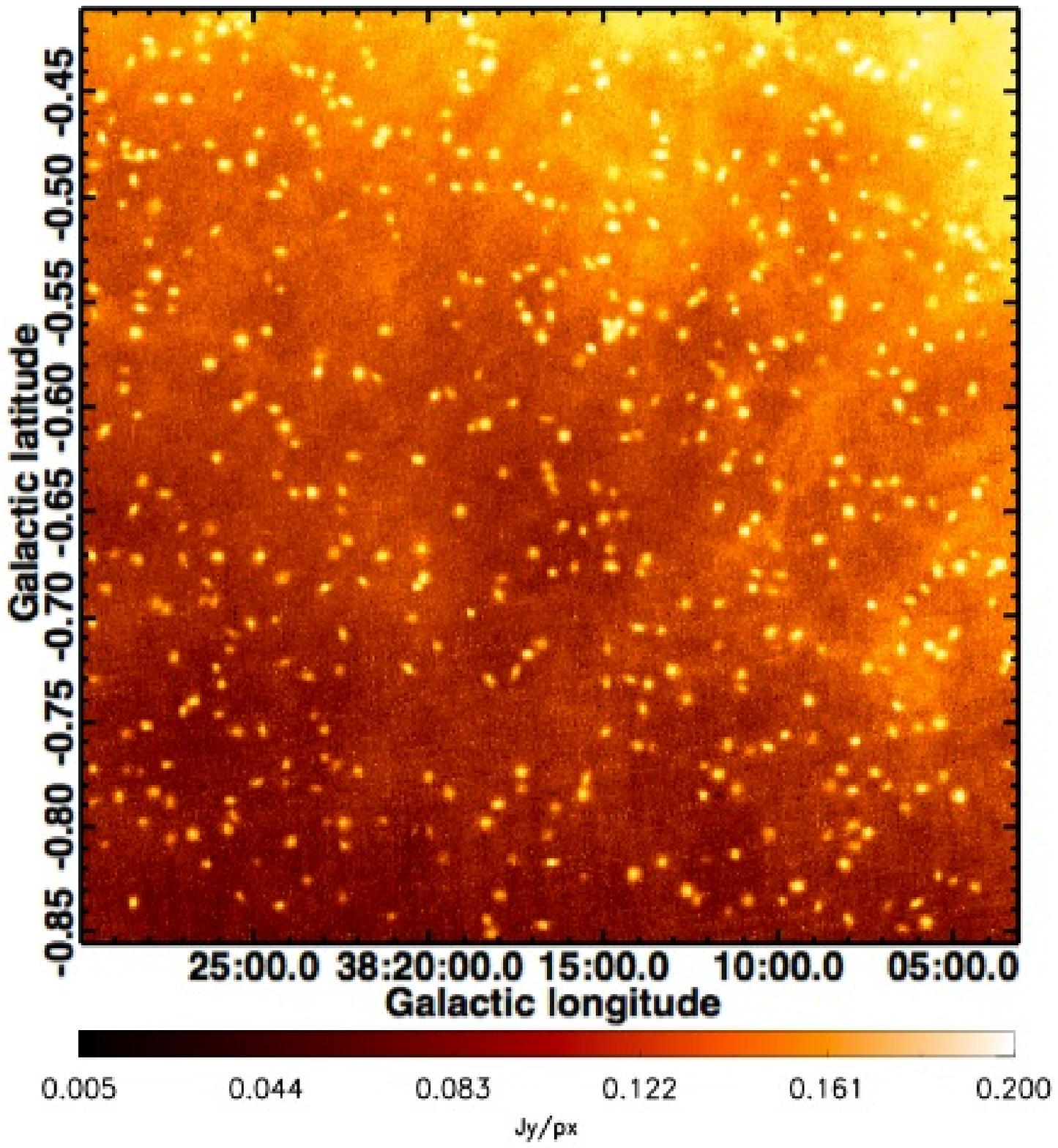}
\includegraphics[width=9.1cm]{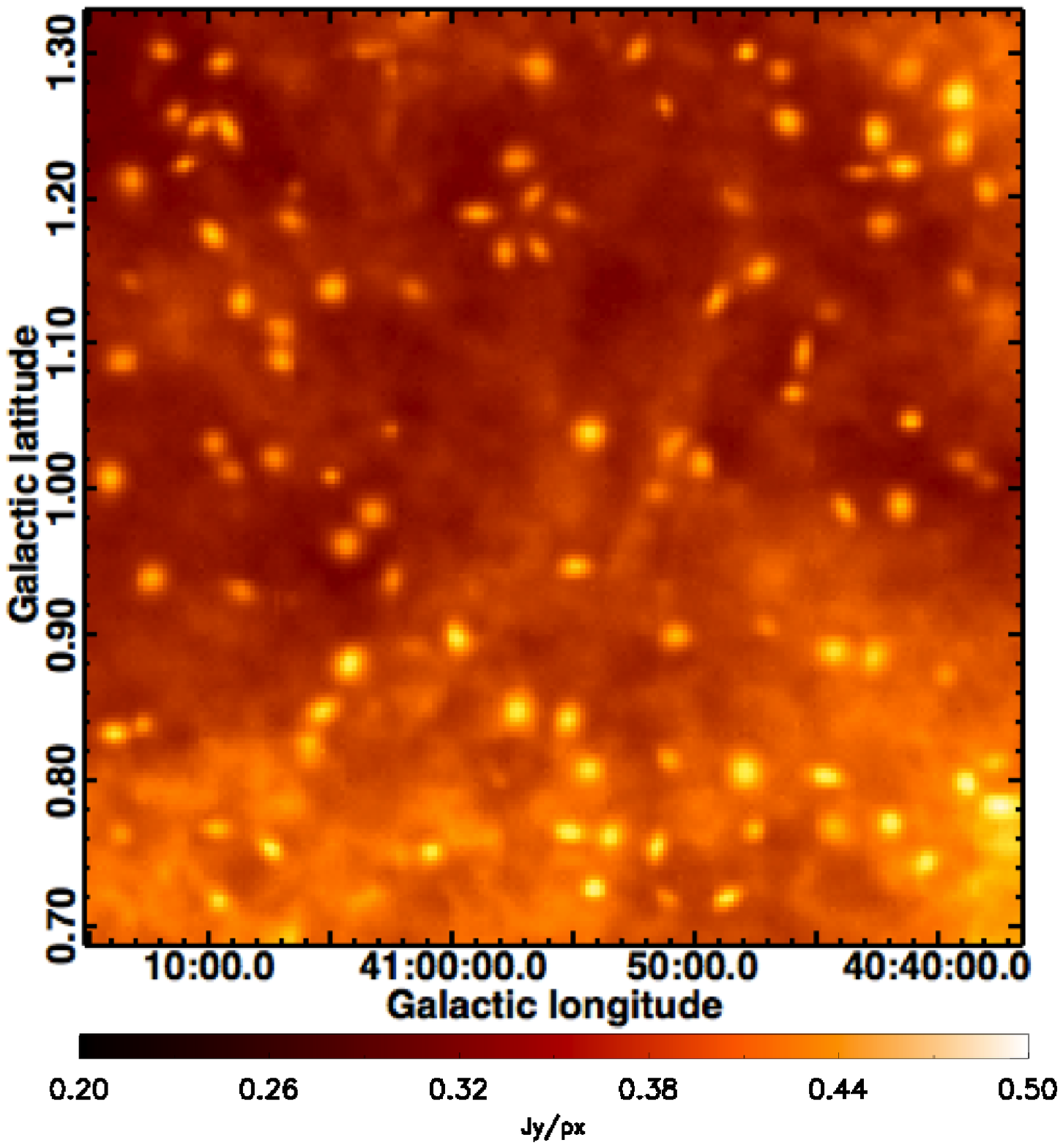}
\caption{One realisation of T1 run for a field extracted from PACS 70 \mum\ map (left panel) and SPIRE 500 \mum\ map (right panel). There are 500 sources and 100 sources injected per realisation in the 70 \mum\ and 500 \mum\ maps respectively.}
\label{fig:70_500_sources_map}
\end{figure*}

For each wavelength (except 350 and 500 \mum), we generated 6 templates of 500 sources each modelled as 2d Gaussian as in the TS1 and TS2 tests. Due to the different pixel size of the longer Hi-GAL wavelengths, we injected fewer sources (100) in the 350 and 500 \mum\ maps and produced 20 template realisations. In total, we generated 13000 model sources (Table \ref{tab:tot_sources_simulation}). In order to include more complex situations, in each realisation the source FWHMs could vary in the range $\Delta\theta_{\lambda}\leq\mathrm{FWHM}\leq2.0\cdot\Delta\theta_{\lambda}$. The sources have minimum centroid to centroid separation to avoid excessive overlapping of $2.0\cdot\Delta\theta_{\lambda}$. As in the case of TS1 and TS2, we ran two different series of tests. In the first run (hereafter T1), the source peak flux, $F_{p}$, varies from 1.5$\sigma_{test}$ up to 5$\sigma_{test}$ and in the second (T2) $5.5\sigma_{test}\leq F_{p}\leq25\sigma_{test}$, with  $\sigma_{test}$ being the standard deviation of the flux in each map. Examples of T1 realisations for PACS 70 \mum\ and SPIRE 500 \mum\ are shown in Fig. \ref{fig:70_500_sources_map}.



\begin{table*}
\caption{Total number of sources injected at various wavelengths, for both T1 and T2 test cases.}
\begin{center}
\begin{tabular}{c|c|c|c|c|c|c|c}
\hline
\hline
Wavelength & sources per real. & Tot. realisation & Tot. sources & sources per real. & Tot. realisation & Tot. sources & Total sources\\ 
(\mum) & T1 & T1 & T1 & T2 & T2 & T2 & T1+T2 \\   
\hline
70 & 500 & 3 & 1500 & 500 & 3 & 1500 & 3000 \\
160 & 500 & 3 & 1500 & 500 & 3 & 1500 & 3000 \\
250 & 500 & 3 & 1500 & 500 & 3 & 1500 & 3000 \\
350 & 100 & 10 & 1000 & 100 & 10 & 1000 & 2000 \\
500 & 100 & 10 & 1000 & 100 & 10 & 1000 & 2000 \\
\hline
\end{tabular}
\end{center}
\tablefoot{(col. 1): reference wavelength; (col. 2-3-4): number of sources injected in each realisation, the number of realisations and the total number of injected sources for run T1; (col. 5-6-7): same of col. 2-3-4 but for run T2; (col. 8): total number of sources injected at each wavelength combining T1 and T2 runs.}
\label{tab:tot_sources_simulation}
\end{table*}

We fixed different thresholds for each run, chosen after few trials (less than 5) and visual inspections of the extraction results. The thresholds have been chosen to recover as many sources as possible without extracting false positives. The number of sources recovered at each wavelength, with the corresponding thresholds, are in Table \ref{tab:tot_sources_extraction_sigma15_5} for both the test cases. For T1, the most compelling test, the percentage of recovered sources is 100$\%$ in all the SPIRE bands and only 1 out of 1500 sources is not recovered in the PACS 160 map.

The only exception is the 70 \mum\ map with 76 out of 1500 sources not recovered. A lower threshold would recovered all the sources, but at the cost of contamination with false sources. The chosen thresholds however circumvent this effect in both T1 and T2. For T2, the percentage of recovered sources is always 100$\%$ with the only exception of SPIRE 250 \mum, in which \Hyp\ misses 2 out of 1500 sources.


\begin{table}
\caption{Number and percentage of recovered sources at each wavelength for T1 and T2 runs using different thresholds.}
\begin{center}
\begin{tabular}{c|c|c|c|c}
\hline
\hline
Wavelength & Test run & $\sigma_{\mathrm{t}}$ & Recovered & Percentage\\ 
(\mum) & & & sources & ($\%$) \\   
\hline
70 & T1 & 8.5 & 1436 & 95.7\\
160 & T1 &  8.5 & 1500 & 100\\
250 & T1 &  8.0 & 1500 & 100\\
350 & T1 &  4.0 & 996 & 99.6\\
500 & T1 &  4.0 & 998 & 99.8\\
\\

70 & T2 &  10.5 & 1500 & 100\\
160 & T2 & 7.5 & 1496 & 99.7\\
250 & T2 & 8.0 & 1500 & 100\\
350 & T2 & 3.0 & 1000 & 100\\
500 & T2 & 3.0 & 1000 & 100\\
\hline
\end{tabular}
\end{center}
\tablefoot{(col. 1): reference wavelength; (col. 2): reference test run; (col. 3): thresholds fixed at each wavelength; (col. 4 \& 5): number and percentage of recovered sources  respectively for each wavelength and test run. With these thresholds \Hyp\ finds no false positives, and for T1 the percentage of recovered sources is very close to 100\% in all bands except the 70 \mum\ map, where the recovered sources are 95.7\% (col. 4 \& 5). For T2 the percentage of recovered sources is 100\% at all wavelengths except at 160 \mum, where \Hyp\ misses only 4 sources out of 1500. This is due to limiting the false positives generated by the noise in the maps as described in the text.}
\label{tab:tot_sources_extraction_sigma15_5}
\end{table}


The flux of each source, estimated as described in Sect. \ref{sec:aperture_photometry}, is compared with the flux of the injected sources. The flux of the models is evaluated within the same area that we expect to recover with \Hyp, therefore the flux integral is evaluated within a one FWHM radius. This area corresponds to $93.75\%$ of the total area of a 2d Gaussian equal to $2*A*\pi\sigma_{a}\sigma_{b}$, with $A$ being the intensity of the source peak and $\sigma_{a}$ and $\sigma_{b}$ being the standard deviation of the Gaussian along the two semi-axes $a$ and $b$ respectively.

The flux differences between the value recovered by \Hyp\ and each source model for all wavelengths, as function of the peak flux expressed as multiples of the map standard deviation (peak SDEV), are shown in Fig.
\ref{fig:source_sdev_15_5} and \ref{fig:source_sdev_55_25} for T1 and T2
respectively.

We also compared the source sizes evaluated from the model and measured by \Hyp.  For the model, the source size is defined as the geometrical mean of the FWHM in each direction used to build the 2d Gaussian of each source. For
\Hyp, it is defined as the geometrical mean of the FWHMs estimated by the \texttt{mpfit2dpeak} routine, which is equivalent to the aperture radius. The comparison plots are shown in Fig. \ref{fig:source_radius_15_5} and
\ref{fig:source_radius_55_25} for runs T1 and T2 respectively. 

In Table \ref{tab:radius_rms} there is the r.m.s. of the size distribution as obtained from \Hyp\ compared with the ``true" size of the source models for both T1 and T2 runs. There are some sources for which \Hyp\ cannot properly estimate the source shape and forces, internally in the \texttt{mpfit2dpeak} routine, the 2 FWHMs to be equal to the minimum or maximum aperture radius. These points are seen as the points distributed along fixed radii corresponding to the minimum and maximum aperture radius chosen at each wavelength. If the fit does not converge at all (the status of the fit is $-1$ or $-2$), \Hyp\ forces the FWHMs to be equal to the geometrical mean of the minimum and the maximum aperture chosen in the parameter file. These points are distributed along fixed radii in the middle of the plots.


\begin{table}
\caption{r.m.s. of the size distribution for both T1 and T2 runs at each wavelength. The size distributions are in good agreement with the model for both T1 and T2 runs.}
\begin{center}
\begin{tabular}{c|c|c}
\hline
\hline
Wavelength & r.m.s. T1 & r.m.s. T2 \\ 
(\mum) & (\%) & (\%) \\   
\hline
70 & 11.26 & 8.44 \\
160 & 11.19 & 8.24 \\
250 & 9.49 & 7.72 \\
350 & 14.76 & 10.49 \\
500 & 13.37 & 9.40 \\
\hline
\end{tabular}
\end{center}
\label{tab:radius_rms}
\end{table}

\subsection{Testing Hyper in a complex environment: sources along filaments}\label{sec:polaris}
\begin{figure*}
\centering
\includegraphics[width=18cm]{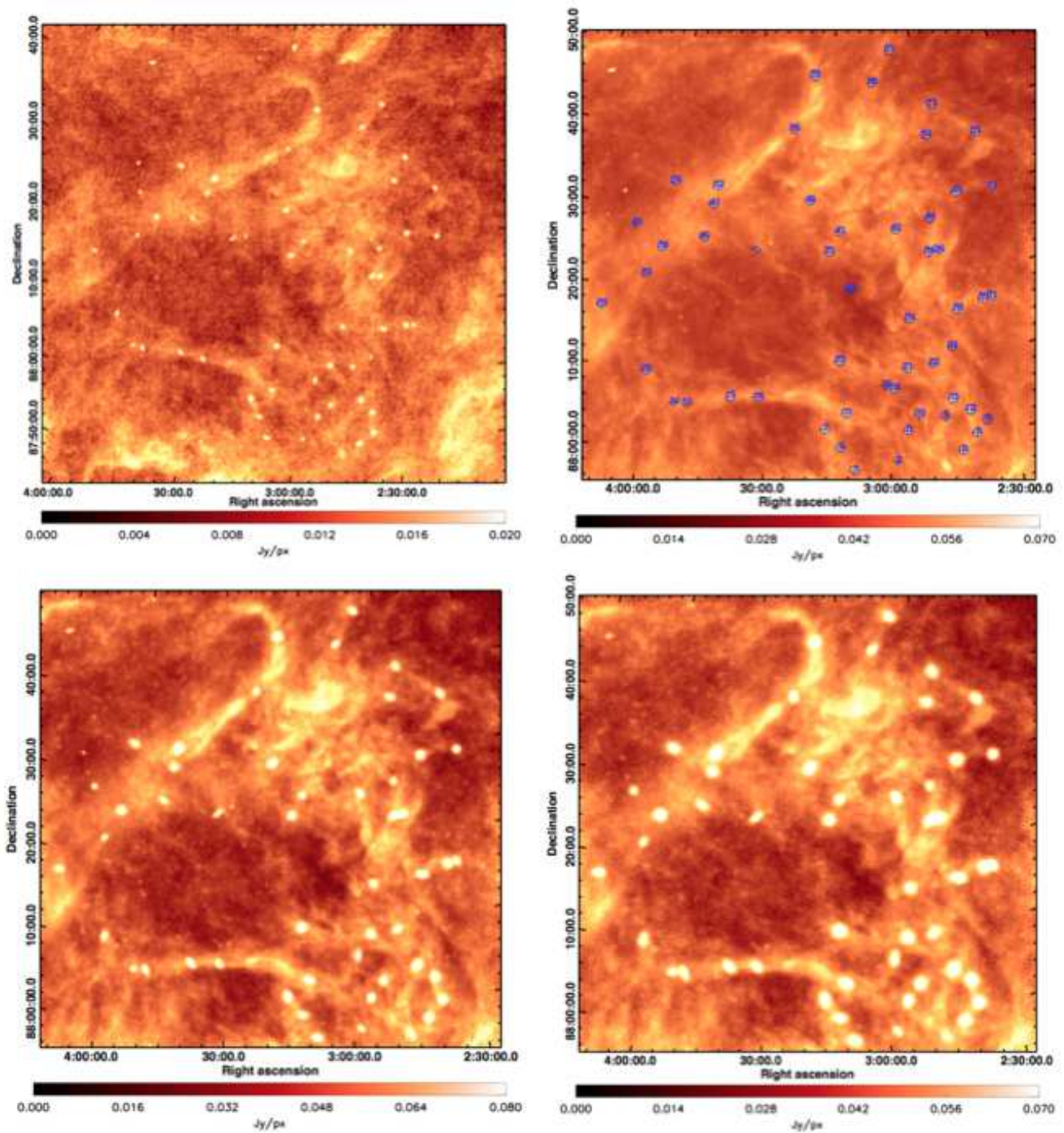}
\caption{Section of the Polaris Flare region with filamentary structures as seen by \Her\ at 160 (\textit{top left}), 250 (\textit{top right}), 350 (\textit{bottom left}) and 500 (\textit{bottom right}) \mum\ \citep[from HGBS,][]{Andre10}. Fifty random 2d Gaussian sources have been injected at each wavelength and positioned on top of the filaments, as discussed in Section \ref{sec:polaris}. The 250 \mum\ map also shows the 50 sources positions and 2d Gaussian fit as obtained with \Hyp.}
\label{fig:Polaris_sources}
\end{figure*}

We further validated \Hyp\ across regions with strong local background variations where real compact objects are likely to be found. \Her\ reveals that star formation mostly occurs along filamentary structures \citep[e.g.][]{Molinari10_PASP,Andre10,Arzoumanian11}, therefore  for this test we selected real filaments observed in the Polaris Flare region. Polaris has been observed as part of the \Her\ Gould Belt Survey program \citep[HGBS,][]{Andre10} and its reduced maps were downloaded from the HGBS Archives\footnote{http://gouldbelt-herschel.cea.fr/archives}. This region is rich in cirrus emission and filamentary structures \citep[e.g.][]{Men'shchikov10,Ward-Thompson10} but the absence of star formation activity makes it suitable for source injection. We isolated a $\simeq1\grad$ wide sub-region at 160, 250, 350 and 500 \mum. Since star formation has not yet begun, the region does not emit at $\lambda=70$ \mum. We injected 50 2d Gaussians at each wavelength with random axes, orientations and fluxes along the filamentary structures and avoiding real sources. The 2d Gaussians have been generated at each wavelength using the same procedure described in Sect. \ref{sec:test_photometry}. The fluxes vary in the range $6\sigma_{Pol}\leq F_{p}\leq 10\sigma_{Pol}$, with $\sigma_{Pol}$ being the \textit{r.m.s.} of the map at each wavelength. The radii of the sources are in the range $\Delta\theta_{\lambda}\leq\mathrm{FWHM}\leq2.0\cdot\Delta\theta_{\lambda}$, with $\Delta\theta_{\lambda}$ being the beam of each observation. Figure \ref{fig:Polaris_sources} shows the region and the injected sources at the four \Her\ wavelengths.

We performed the test by injecting weak sources with the brightest source from $\simeq1.5$ to $\simeq10$ times fainter than the densest starless cores observed within the Polaris field, depending on the wavelength \citep{Ward-Thompson10}. The minimum and maximum source fluxes, the average flux difference between the \Hyp\ and the model fluxes and the 1-sigma dispersion of this difference at each wavelength are given in Table \ref{tab:hyper_polaris}. Although the background structures generated by the filaments are difficult to model and the injected sources are relatively faint, the agreement between the recovered and the modelled fluxes is $\simeq20\%$ on average, with a 1-sigma dispersion of $\simeq15\%$ on average.

\begin{table}
\caption{Average flux difference between the \Hyp\ fluxes and model fluxes for the 50 sources injected in the Polaris field as showed in Figure \ref{fig:Polaris_sources} at 160, 250, 250 and 500 \mum. Also shown are the integrated fluxes of the faintest and the brightest source injected at each wavelength.}
\begin{center}
\begin{tabular}{c|c|c|c|c}
\hline
\hline
Band & Min source flux & Max source flux &  Flux diff. & \textit{1-sigma}\\ 
(\mum) & (Jy) & (Jy) & (\%) & (\%)\\   
\hline
160 & 0.13 & 0.42 & 16.3 & 13.9\\
250 & 0.40 & 1.25 & 25.1 & 17.2\\
350 & 0.58 & 1.80 & 18.2 & 12.4 \\
500 & 0.49 & 1.51 & 20.1 & 16.2 \\
\hline
\end{tabular}
\end{center}
\label{tab:hyper_polaris}
\end{table}


\subsection{Testing \Hyp\ multi-wavelength analysis}\label{sec:test_multiwave}
In order to test the \Hyp\ multi-wavelength approach described in Sect. \ref{sec:multiwavelength} we selected a complex region from Hi-GAL, namely a $2\grad\times2\grad$ region centred on $(l,b)=(50\grad,0\grad)$ in the Galactic Plane and we injected across the central region 200 identical 2d Gaussian sources at all Hi-GAL wavelengths: 70, 160, 250, 350 and 500 \mum. We modelled the source spectral energy distribution (SED) at wavelengths $160\leq\lambda\leq500$ \mum\ assuming a single-temperature greybody model with a spectral index of $\beta=2.0$ \citep{Elia10,Giannini12}. We assumed a source with a temperature of 15 K, mass of 300 M\sun\ in a radius of 0.5 pc and located at 4 kpc, typical values of compact protostars observed in the Galactic Plane \citep[e.g.][]{Traficante14_cat}. The corresponding fluxes are $F=(31.50,29.84,17.37,7.28)$ Jy at $\lambda=(160,250,350,500)$ \mum\ respectively. At 70 \mum\ the emission from young sources is usually observed in excess compared to a single-temperature greybody model \citep[e.g.][]{Motte10,Giannini12}. Therefore we adopted a 70 \mum\ flux of 3 Jy for this source (in comparison to the 0.9 Jy from the 15 K greybody model). We fixed the centroids of the sources equal at all wavelengths, but we let the FWHMs and the position angles at each wavelength vary. Each source realisation has been convolved with the appropriate Herschel beam before being injected into the corresponding map.

\begin{figure}
\centering
\includegraphics[width=6cm]{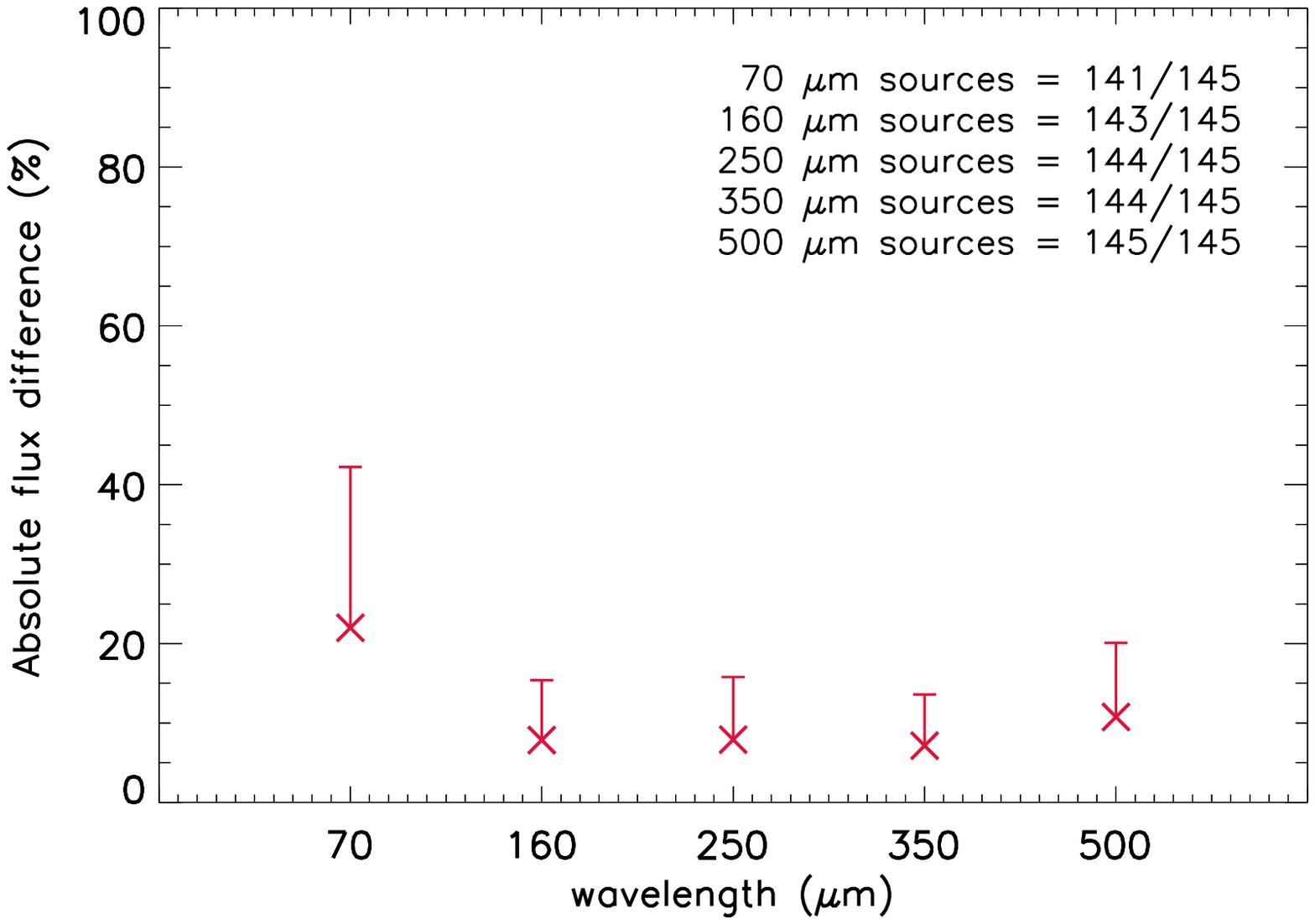}
\caption{Average absolute flux differences between the flux models and the flux recovered by \Hyp\ within the 500 \mum\ fit aperture region at each wavelength. The accordance is on average  $\simeq10\%$ at all wavelengths except at 70 \mum, where within the 500 \mum\ aperture the flux estimation includes residual background contribution not removed with the background removal as discussed in Section \ref{sec:test_multiwave}.}
\label{fig:flux_distribution_multiwavelength}
\end{figure}

\begin{figure*}
\centering
\includegraphics[width=8cm]{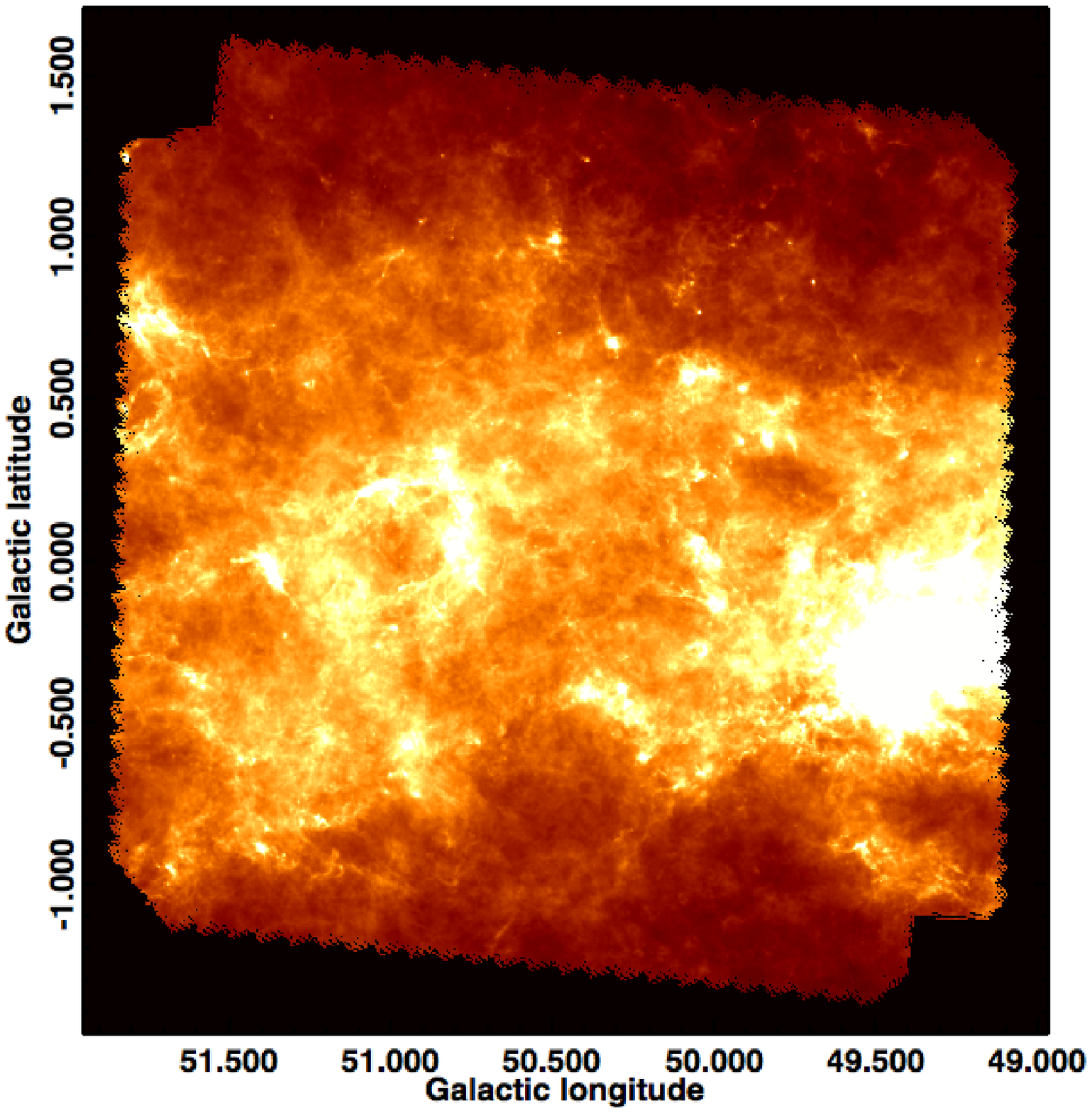} \qquad
\includegraphics[width=8cm]{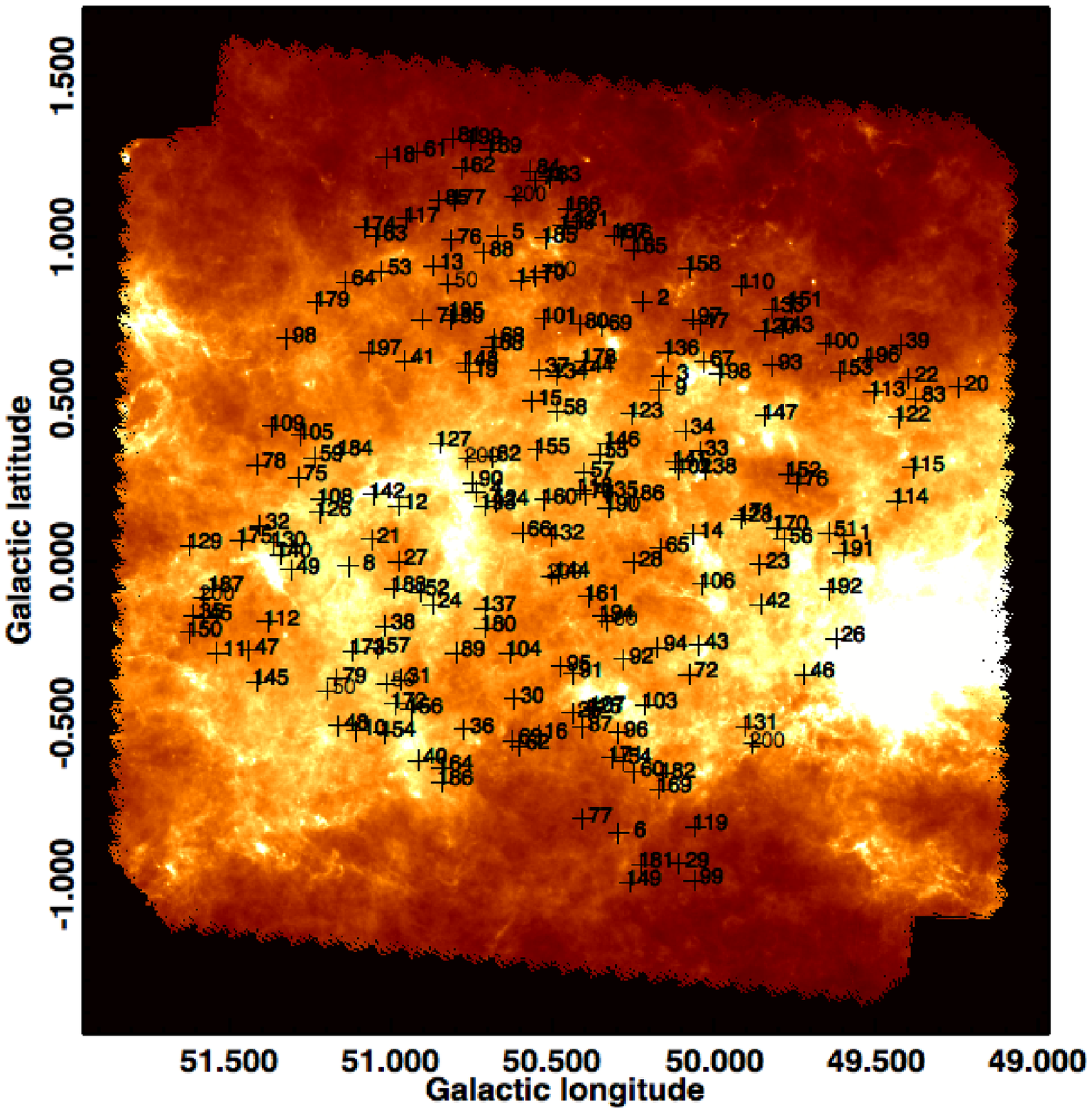}\\
\quad \includegraphics[width=16cm]{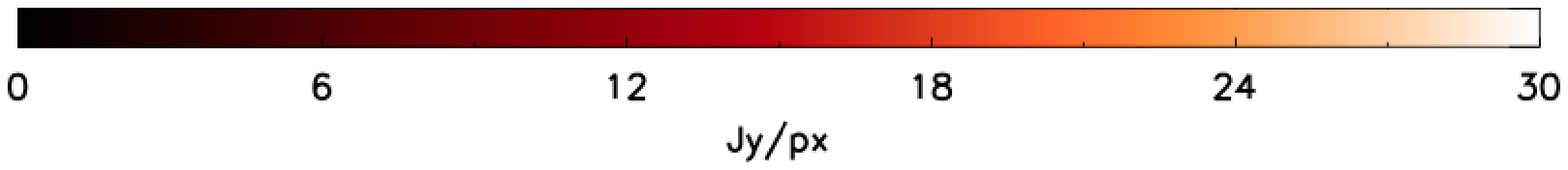}\\

\caption{\textit{Left}: Hi-GAL map of the $2\grad\times2\grad$ region chosen centred on $(l,b)=(50\grad,0\grad)$ observed at $\lambda=250$ \mum\ chosen to test the \Hyp\ multi-wavelength approach described in Sect. \ref{sec:test_multiwave}. \textit{Right}: same region with overlapped the 200 injected sources as described in Sect. \ref{sec:test_multiwave}.}
\label{fig:l050_global}
\end{figure*}

\begin{figure*}
\centering
\includegraphics[width=3.5cm]{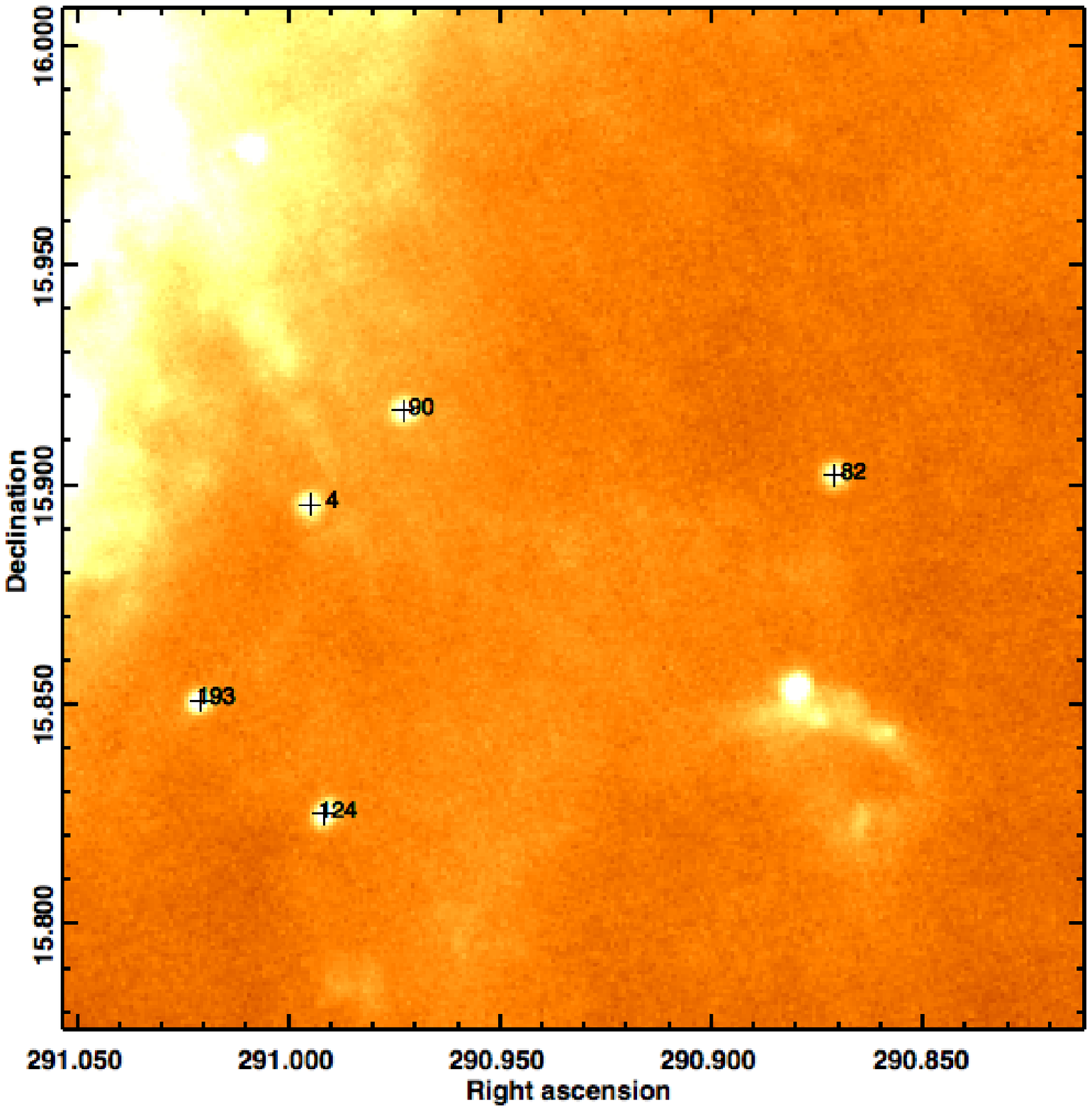}
\includegraphics[width=3.5cm]{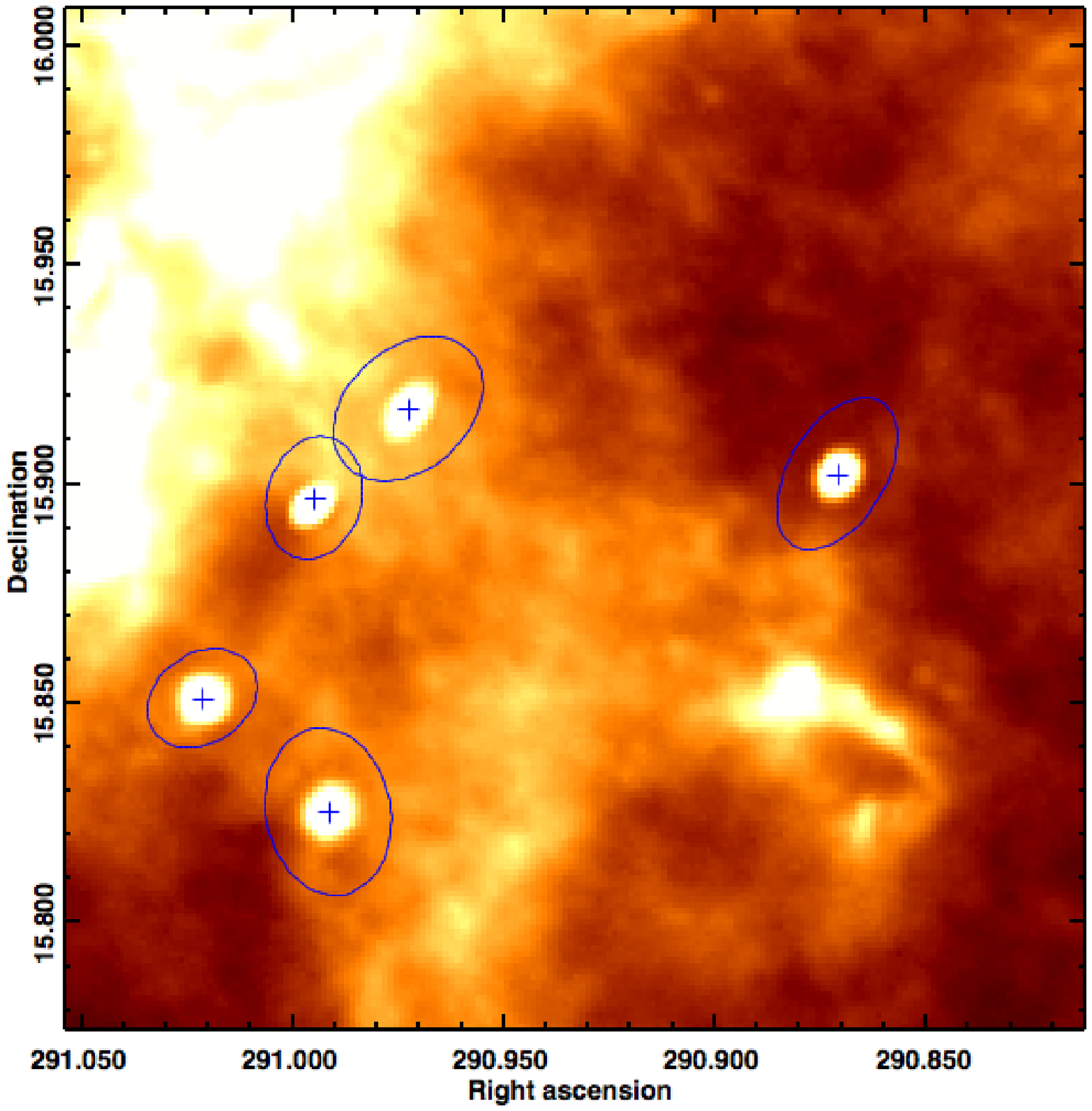}
\includegraphics[width=3.5cm]{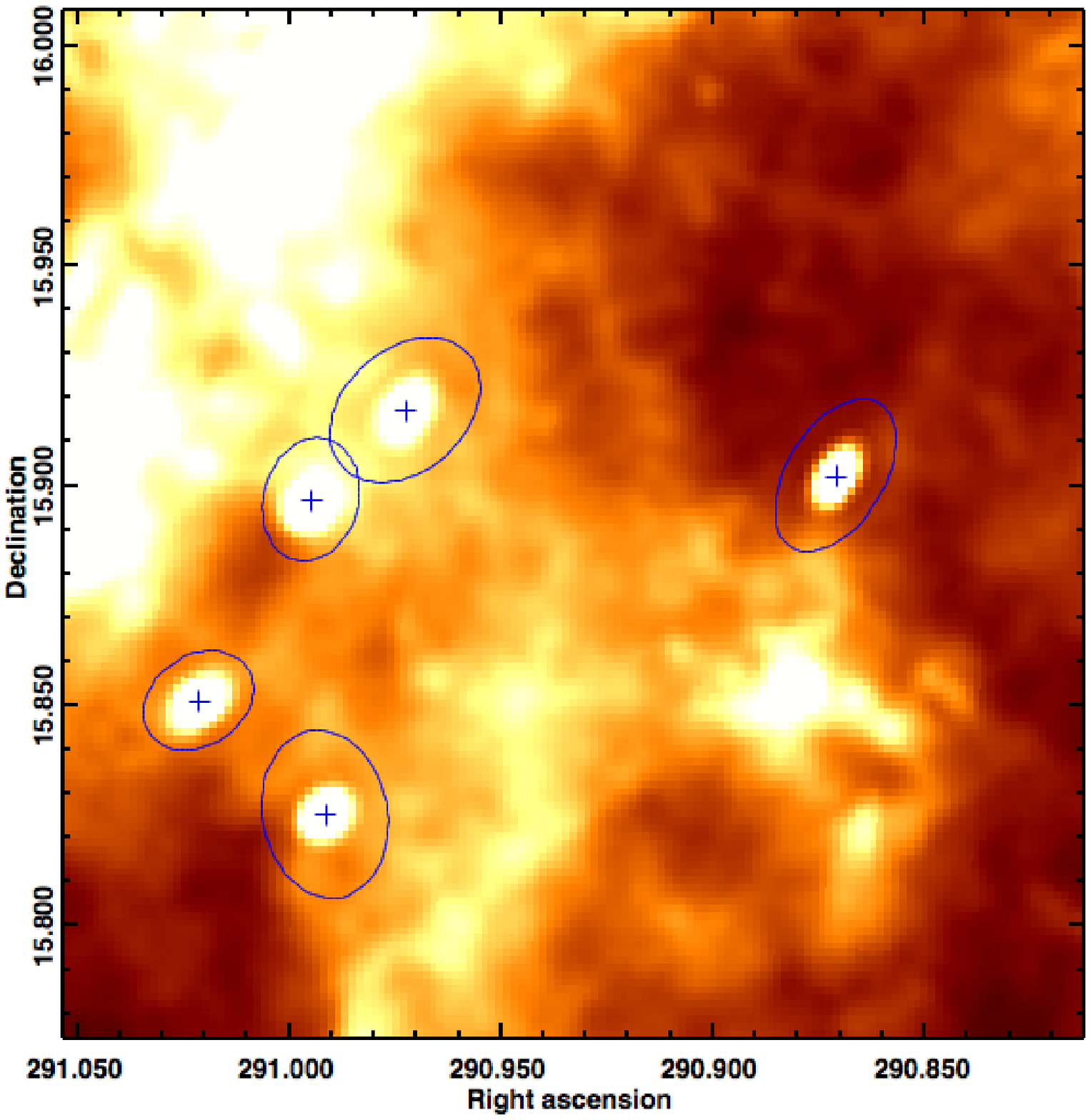}
\includegraphics[width=3.5cm]{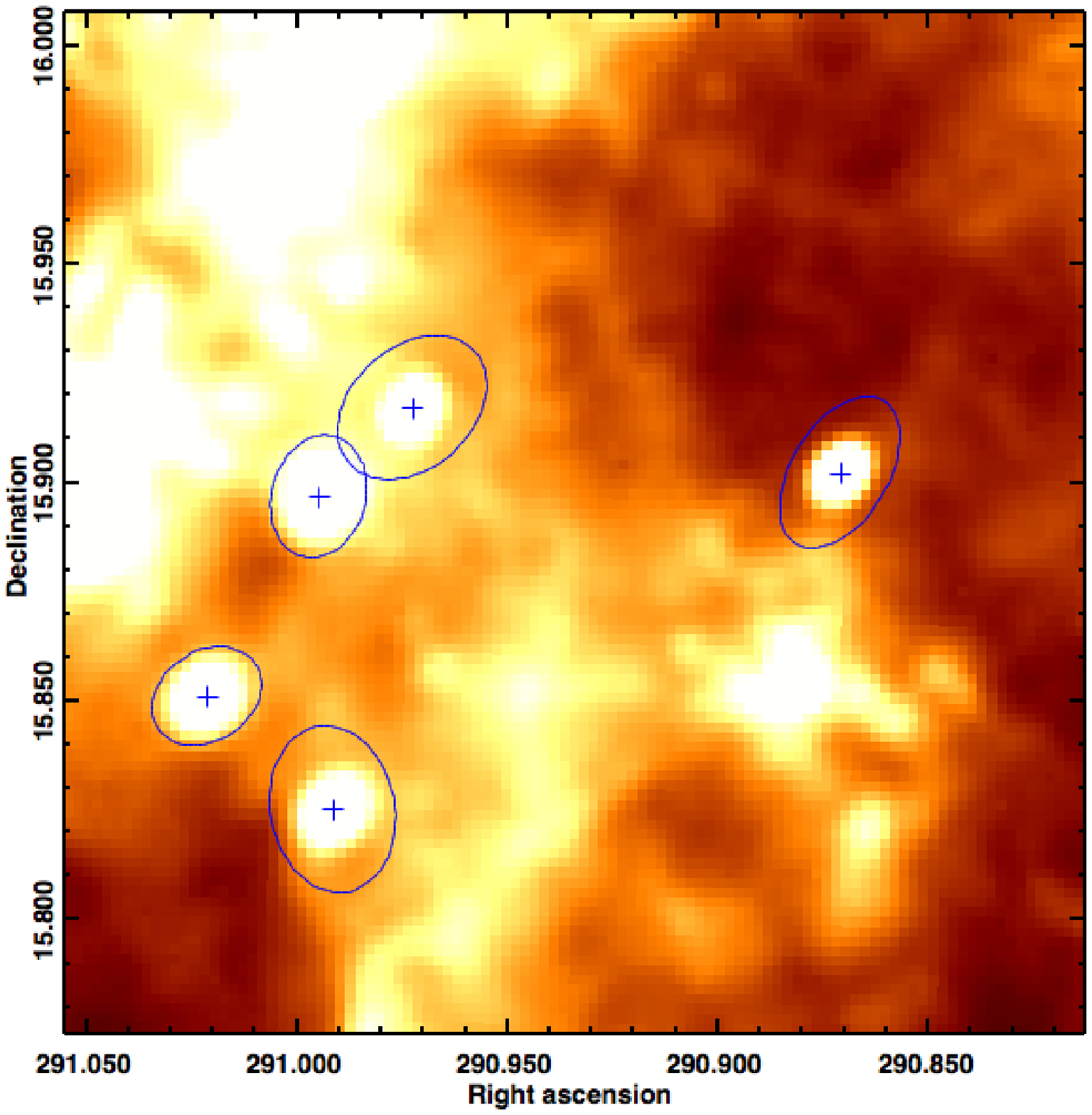}
\includegraphics[width=3.5cm]{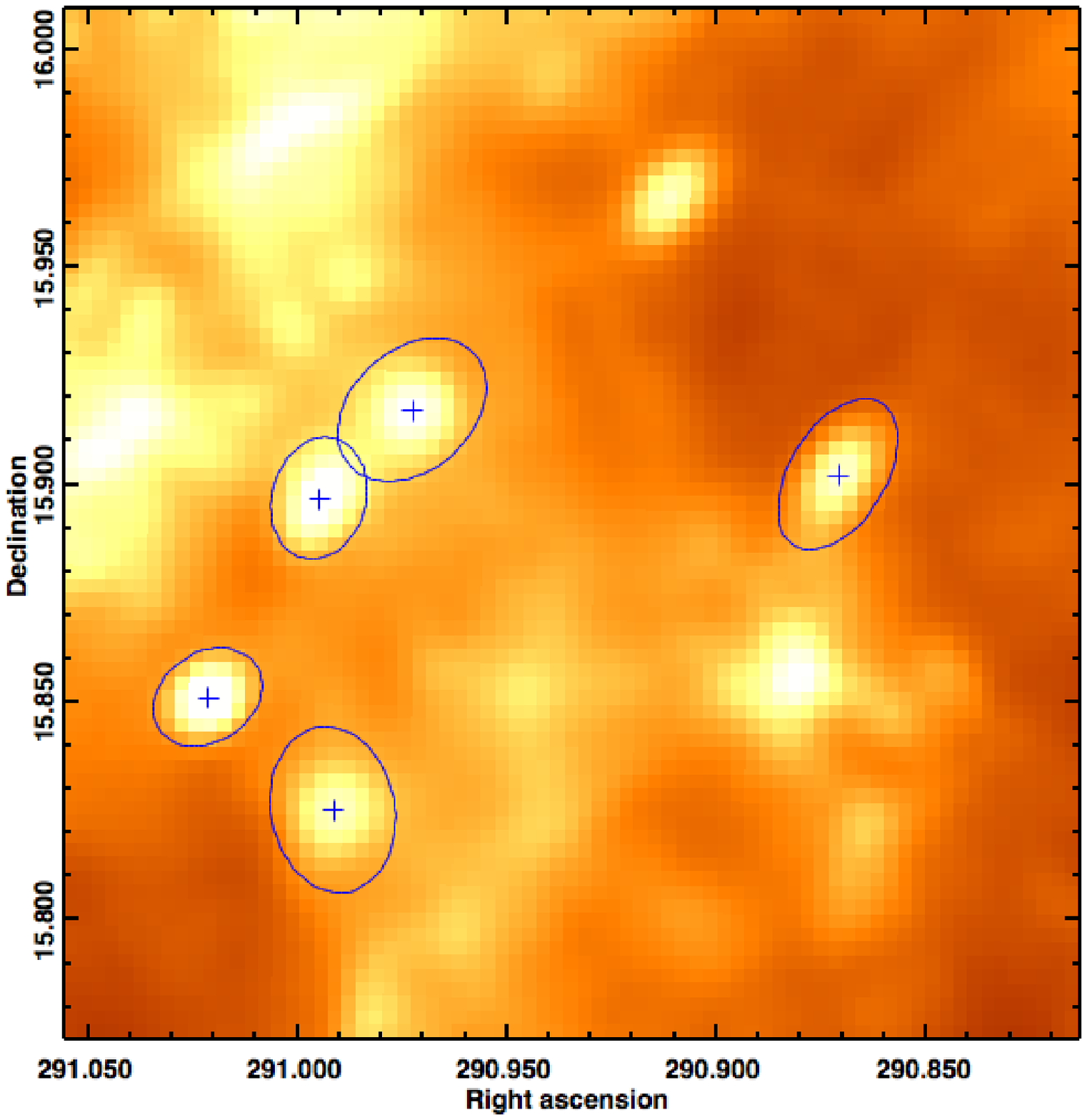}

\caption{Zoom of a region with 5 injected sources extracted from the map showed in Fig. \ref{fig:l050_global}. \textit{From left to right}: the same region observed at 70, 160, 250, 350 and 500 \mum. The black cross in the 70 \mum\ image identify the 5 source centroids. The blue ellipses in the other maps represent the \Hyp\ source fit done in the 500 \mum\ map which determines the aperture region at all wavelengths as described in Section \ref{sec:test_multiwave}.}
\label{fig:l050_zoom}
\end{figure*}

The aperture region was estimated at 500 \mum\ and the flux is evaluated at all wavelengths within the same aperture. \Hyp\ identified 194/200 sources using a threshold $\sigma_{t}=3.0$. After a visual inspection of the 6 missing sources, 5 were randomly injected at the borders of the map and only 1 has been missed by the algorithm. An image of the complexity of the region in which we injected the sources at 250 \mum\ and a zoom of a region including 5 injected sources at all wavelengths are in Fig. \ref{fig:l050_global} and \ref{fig:l050_zoom} respectively. The run at all five wavelengths simultaneously on this region requires $\simeq15$ minutes on a 2.2 GHz Intel Core i7 with 8GB of RAM.

We have compared the \Hyp\ fluxes with the injected fluxes for a subset of sources, namely all the sources which do not show any identified clustered companions. Clustered sources are in fact real sources identified within the integration region at shorter wavelengths and possibly overlapped with the injected ones at the longer wavelengths (see Sect. \ref{sec:multiwavelength}). We obtain 145 sources with no clustered companions. In Appendix C the flux difference distribution histograms at each wavelength are shown. We further excluded from the comparison the sources with an absolute flux difference greater than $5\cdot\sigma_{d}$ the mean of the flux difference, with $\sigma_{d}$ being the dispersion of the distribution. These few sources (4 at 70 \mum, 2 at 160 \mum\ and 1 at 250 and 350 \mum\ respectively) are close to bright, diffuse emission which contaminates the aperture region and was not fully removed by the background removal. We ended up with [141,143,144,145,145] sources at $\lambda=(70,160,250,350,500)$ \mum\ respectively. The mean flux differences between the injected models and the \Hyp\ sources are shown in Fig. \ref{fig:flux_distribution_multiwavelength}. Even if the background emission in the region is extremely complex and variable across the map, the absolute flux differences are $[22, 8, 8, 7, 11] \%$  with a $\sigma_{d}$ of $[20, 7, 8, 6, 9]\%$ at $\lambda=(70,160,250,350,500)$ \mum\ respectively. 
The difference is slightly higher on average at 70 \mum, likely due to high-spatial frequency variations of the background within the integration region that are not accounted in the background estimation.


If the user wants to measure the flux of each single resolved source in the high resolution images,and so more accurately remove the background on a smaller region in particular for the 70 \mum\ counterparts, \Hyp\ can be easily run separately at each different wavelength.


\section{Real data applications}\label{sec:test_realdata}

\begin{figure}
\centering
\includegraphics[width=8cm]{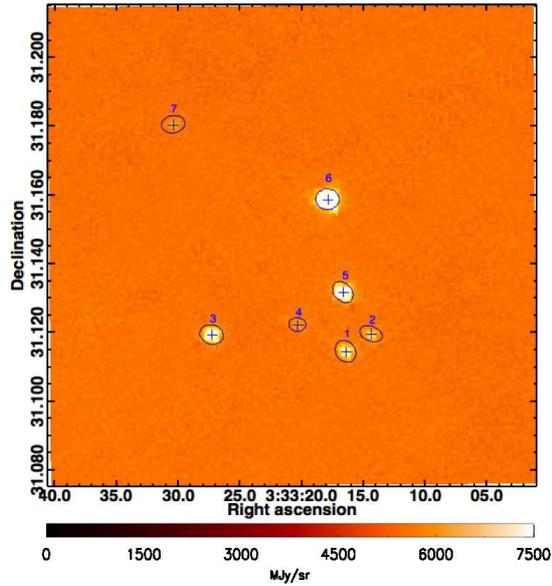}\\ 
\caption{A small area of the Perseus B1 region observed with PACS 70 \mum\ as part of the HGBS \citep{Andre10}. The background is flat all across the region and the seven sources (the blue ellipses) are easily identified by the code.}
\label{fig:Perseus_B1}
\end{figure}

\begin{figure}
\centering
\includegraphics[width=8cm]{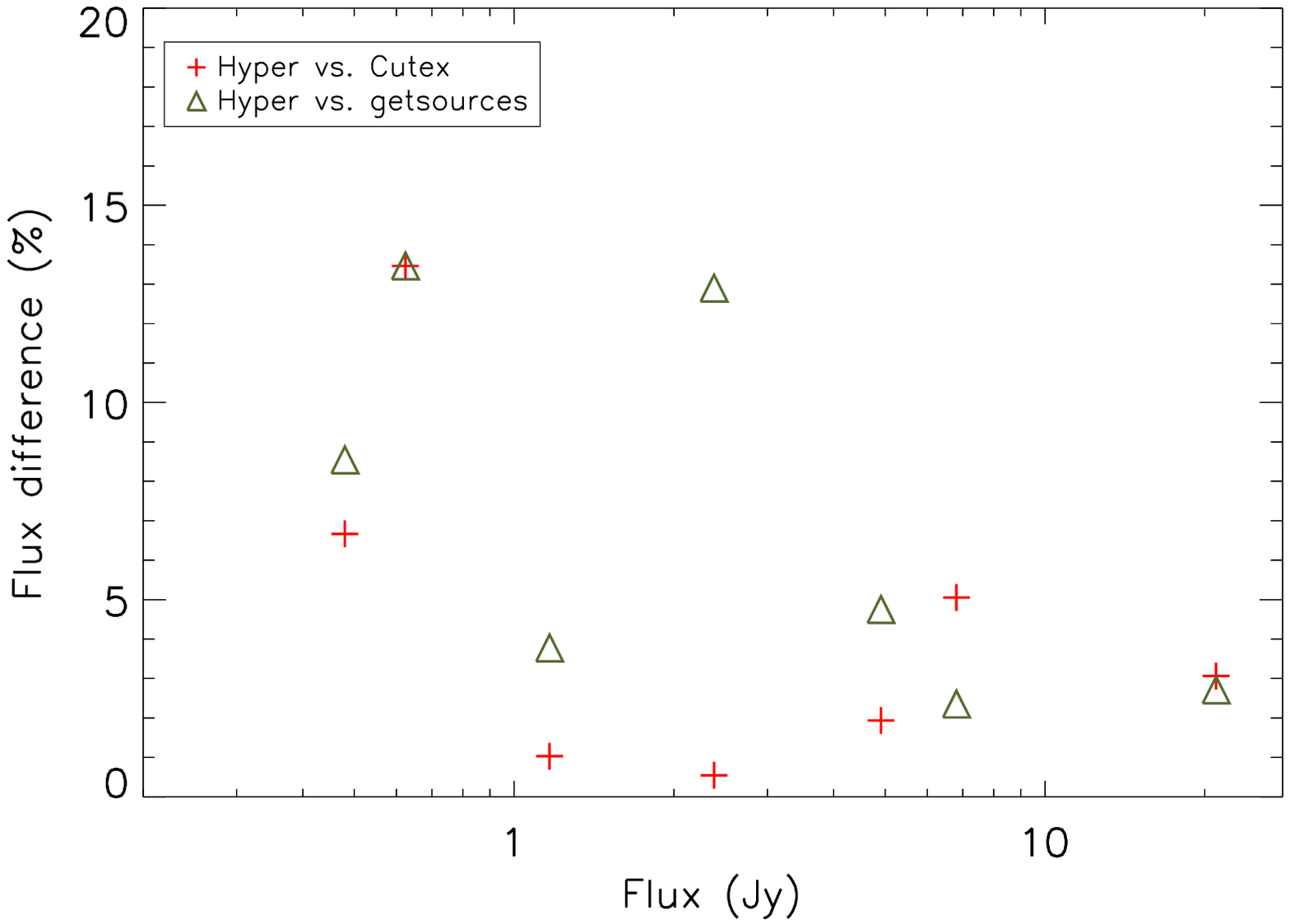}\\
\caption{Flux comparison between \Hyp\ and \Cut\ (red crosses) and \Get\ (green triangles)for the seven sources identified in the Perseus 70 \mum\ field. The \Hyp\ flux are in very good agreement with the \Cut\ and the \Get\ fluxes, with a discrepancy on average of 7\% and 13\% with \Cut\ and \Get\ fluxes respectively.}
\label{fig_cutex_getsources_hyper}
\end{figure}

Testing photometry algorithms on real dataset is not an easy task since the correct value of the source flux is not known nor easily predictable, in particular in complex fields.It is not surprising that two different methods give different fluxes. The main reasons are the differences in the removal of the background and the de-blending of the sources. A useful illustration has been shown in \citet[][Table 1]{Pezzuto12}, in which the fluxes of the two first hydrostatic core candidates \citep[FHSC, e.g.][]{Commercon12} identified by \herschel\ in the Perseus region can differ up to $\simeq50\%$ if evaluated with \Cut\ or \Get. In that work the authors have chosen \Cut\ to estimate the flux and combined the \Cut\ and \Get\ photometry to estimate the flux uncertainties \citep{Pezzuto12}. Recently, Sadavoy et al. (in preparation) found for several sources in Perseus star forming region differences in fluxes greater than a factor 2 when comparing the Spitzer c2d catalog \citep{Evans03} with the \citet{Gutermuth09} catalog, also based on Spitzer data.


Nonetheless, testing \Hyp\ on real data is crucial to validate the algorithm and, in order to compare \Hyp\ with some other existing codes, we selected two different test-cases: in the first we selected a region of few relatively isolated sources for which we expect that the photometry should be least sensitive to the algorithm used (Sect. \ref{sec:test_perseus}). In the second we selected a catalogue of $\simeq1000$ compact sources across the Galactic Plane, with integrated fluxes and background properties varying by several order of magnitudes (Sect. \ref{sec:test_irdc}). The comparison of  \Hyp\ with other algorithms in such a complex case gives a good estimation of the intrinsic uncertainties in the flux estimation due to the different approaches.

Finally, we compared the \Hyp\ photometry with the published catalogue of source fluxes extracted from a survey of the Galactic Plane made with a ground-based telescope, the 1.1 mm Bolocam Galactic Plane survey \citep[BGPS,][]{Aguirre10} carried out with the Caltech Sub-millimeter observatory (Sect. \ref{sec:BGPS}). This test shows the versatility of the aperture photometry approach applied to different surveys and instrument specifications.

\subsection{Test on real data: isolated sources in B1 Perseus field}\label{sec:test_perseus}
As region with relatively isolated sources we selected the B1 Perseus star forming region observed with \herschel\
at 70 $\mu$m as part of the HGBS program \citep[][]{Andre10}. 

The Perseus star-forming region is active in forming stars as demonstrated,
e.g., by the large number of FHSC candidates reported in literature, e.g. L1451-mm
\citep{Pineda11}. In the B1 part of the Perseus region
\citep[see][]{Sadavoy12}, \citet{Pezzuto12} have recently found other two FHSC
candidates, B1-bN and B1-bS, thanks to \herschel\ observations of the region. We
have extracted from B1 a small square region where the sources are point-like,
isolated and with a very faint background easily allowing a direct comparison
of different photometry algorithms and philosophies. Importantly for this
region the photometry is also available for both the \Cut\ and \Get\
algorithms.


The image of the region observed with PACS at 70 \mum\ is shown in Fig.
\ref{fig:Perseus_B1}. Seven sources can be seen by eye and all of them have
been identified by \Hyp\ using $\sigma_{\mathrm{t}}=6.5$. The same sources have also been
identified with \Cut\ and \Get. The background is very clean and
different values of the \Hyp\ threshold do not give rise to other, possibly false
detections. The \Hyp\ output file for these sources with all the parameters
estimated by \Hyp\ is shown in Table \ref{tab:hyper_output_file}. 

Table \ref{tab:hyper_cutex_flux} compares the flux and size measurements from
\Hyp, \Cut\ and \Get. For the \Hyp\ fluxes in this table we have been applied
the aperture corrections, which are source-size and wavelength dependent. The correction
curve is publicly
available\footnote{$http://herschel.esac.esa.int/twiki/pub/Public/PacsCalibrationWeb\\
  /pacs\_bolo\_fluxcal\_report\_v1.pdf$}. The \Cut\ fluxes have also been
corrected for a size-dependent correction factor. The photometry for the
different algorithms is compared in Fig. \ref{fig_cutex_getsources_hyper}.
The source FWHM has been evaluated as the geometrical mean of the FWHMs
obtained from the 2-d fits for both \Hyp\ and \Cut\ algorithms.  The agreement
between the three codes is very good, with a mean difference in flux of
$<2\%$ and $\simeq7\%$ between \Hyp\ and \Cut\ and \Get\
respectively. The source FWHMs differ by $\simeq6\%$ and $\simeq7\%$ with \Cut\ and \Get\ respectively.

\begin{landscape}
\begin{table}
\caption{\Hyp\ output parameters.}
\centering
\begin{tabular}{c|c|c|c|c|c|c|c|c|c|c|c|c|c|c|c|c|c|c|c|c}
\hline
\hline
map & source & band & peak & peak${\_}$jy & flux & err${\_}$flux & sky${\_}$nob. & sky & pol. & FWHM & FWHM  & PA & status & glon & glat &  ra & dec & deb. & dist${\_}$ref & clust\\
& & (\mum) & (MJy / sr) & (Jy) & (Jy) & (Jy) & (Jy) & (Jy) &  & ($\arcsec$) & ($\arcsec$) & ($^{\circ}$) & & ($^{\circ}$) & ($^{\circ}$) & ($^{\circ}$) & ($^{\circ}$) & &  ($\arcsec$) &\\
\hline

B1-bN    &   1  &    70  &   559.4 & 0.135 &  2.380   &  0.082   & 0.013  &  0.013  &     4   &   10.02   &   12.26    &  215.65   &  0 &  -- -- & -- -- &   -- -- & -- --  &    0  &    0.00  &     1\\
B1-bN   &    2  &    70  &  344.3 & 0.083 &  1.166  &   0.103  &  0.019  &  0.019   &    3  &    7.65   &   12.26  &    249.12  &  0 &   -- -- & -- --  &      -- -- & -- -- &   0   &   0.00   &    1\\
B1-bN   &    3  &    70   &   1261.4 & 0.304 & 4.909   &  0.143  &  0.023  &  0.023   &    1   &   10.12   &   12.14 &     250.60 &  0 &    -- -- & -- --  &      -- -- & -- -- &    0  &    0.00  &     1\\
B1-bN   &    4   &   70   & 115.1 & 0.028 &  0.480  &   0.019  & 0.004 &  0.004  &     1  &    7.15  &    8.86   &   262.30  &  0 &   -- -- &  -- --  &     -- -- & -- -- &     0   &   0.00   &    1\\
B1-bN    &   5   &   70  & 1481.6 & 0.357  &  6.808  &    0.210   &  0.036  &  0.036   &     3  &    9.09   &   12.26  &    222.67 &   0 &   -- -- &  -- -- &      -- -- & -- -- &    0   &   0.00   &    1\\
B1-bN   &    6   &   70   &  5114.1 & 1.231 &  20.999  &    0.636  &   0.100   &  0.099   &    3   &   10.87   &   11.93  &    268.83   &  0 &  -- -- &     -- --&    -- -- & -- -- &    0   &   0.00  &     1\\
B1-bN    &   7    &  70   &  155.5 & 0.037 & 0.624  &   0.022   & 0.004  & 0.004  &     4  &    9.08  &   12.26  &    100.58  &  0 &   -- -- &  -- --  &     -- -- & -- -- &     0  &    0.00  &     1\\

\hline
\end{tabular}
\tablefoot{(col. 1): map name; (col. 2): \Hyp\ source
  number; (col. 3): reference wavelength of the source; (col. 4 \& 5): source peak flux expressed in MJy/sr and Jy respectively; (col. 6 \& 7): source
  flux and source flux error. The source flux is corrected for the aperture size but not for
  colour corrections. The flux error is estimated from the sky
  r.m.s. multiplied by the square root of the number of pixels in the area
  over which the source flux is integrated; (col. 8 \& 9): r.m.s. of the sky
  evaluated in the rectangular region used to model the background before and
  after the background subtraction respectively; (col. 10): polynomial order
  used to model the background; (col. 11 \& 12): FWHM minor and FWHM major of the
  source. They correspond to the aperture radii used to evaluate the
  flux; (col. 13): Source Position Angle; (col. 14): goodness of the
  2d Gaussian fit. Status can be equal to [0,-1,-2] as described in Sect.
  \ref{sec:source_aperture_photometry}; (col. 15 $-$ 18): source centroids
  Galactic and Equatorial coordinates. Issues related to the proprietary
  nature of these data require that the coordinates can not be presented
  here. However the positions are in complete agreement with \Cut\ and \Get\ positions and will be soon released by the HGBS consortium; (col. 19): number of sources identified as companions and de-blended; (col. 20): distance from the source centroids in the wavelength used to identify the source and the source counterparts at the other wavelengths; (col. 21): number of source counterparts at each wavelength. It is usually equal to 1 but it can be higher if the source is resolved in more than one counterparts in the high resolution maps.}
\label{tab:hyper_output_file}
\end{table}
\end{landscape}

\begin{table*}
\caption{Fluxes and FWHM of the seven sources in part of the Perseus B1 region measured with \Hyp, \Cut\ and \Get\ respectively. The \Hyp\ fluxes are corrected for the aperture corrections. \Cut\ fluxes are rescaled for a factor determined by the source sizes. \Cut\ and \Get\ fluxes are extracted from the final table of sources in the Perseus star-forming region, under preparation by the HGBS consortium.}
\begin{center}
\begin{tabular}{c|c|c|c|c|c|c}
\hline
\hline
Source & \Hyp\ flux & \Cut\ flux & \Get\ flux & \Hyp\ FWHM & \Cut\ FWHM & \Get\ FWHM \\ 
number & (Jy) & (Jy) & (Jy) & ($\arcsec$) & ($\arcsec$)& ($\arcsec$)\\   
\hline
1 & 2.380 & 2.367 & 2.073 & 11.08 & 10.02 & 10.6 \\
2 & 1.166 & 1.154 & 1.122 & 9.68 & 9.54 & 11.0\\
3 & 4.909 & 5.004 & 4.676 & 11.14 & 9.39 & 10.6\\
4 & 0.480 & 0.448 & 0.439 & 7.96 & 9.55 &11.3\\
5 & 6.808 & 7.152 & 6.648 & 10.56 & 9.26 &10.2\\
6 & 20.999 & 21.642 & 20.430 & 11.55 & 9.14 &10.6\\
7 & 0.624 & 0.540 & 0.540 &10.55 & 9.36 & 11.6\\
\hline
\end{tabular}
\end{center}
\label{tab:hyper_cutex_flux}
\end{table*}

\subsection{Test on complex real data: protostar clumps in IRDCs}\label{sec:test_irdc}
\begin{figure}
\centering
\includegraphics[width=8cm]{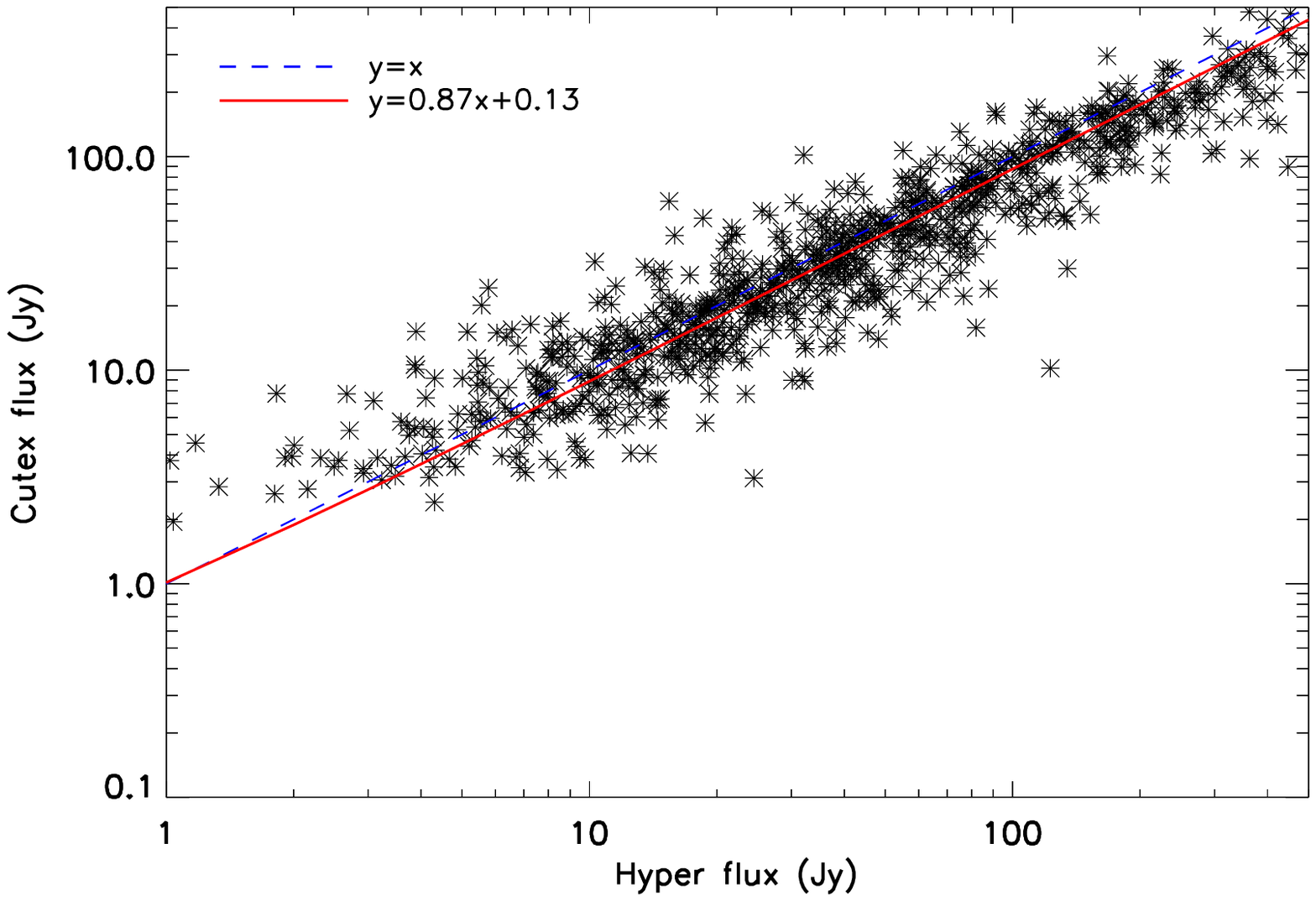}\\
\caption{Comparison between \Hyp\ and \Cut\ 250 \mum\ integrated fluxes for $\simeq1000$ compact sources extracted from the \citet{Traficante14_cat} survey of protostars associated with IRDCs in the Galactic range $15\grad\leq l \leq55\grad$. \textit{Blue dashed line:} y=x straight line. \textit{Red line}: linear fit of the distribution. The coefficient of the fit is $m=0.87$, and the intercept is $0.13$ Jy which indicate no systematics in the algorithms but a slight overestimate of the \Hyp\ fluxes compared with the \Cut\ fluxes. The scatter is mostly due to the different background subtraction and de-blending approaches.}
\label{fig:hyp_vs_cut_irdc}
\end{figure}

As test case of a strongly variable background and various source fluxes, we selected the catalogue of protostellar clumps associated with IRDCs in the Galactic region $15\grad\leq l\leq55\grad$, $\vert b\vert\leq1\grad$ \citep{Traficante14_cat}. These sources have been observed as part of the Hi-GAL survey \citep{Molinari10_PASP} for which the first generation of the compact sources catalogue produced with \Cut\ is being published (Molinari et al. 2014, in preparation), therefore allowing a direct comparison between these two algorithms in very complex and realistic fields. The \Hyp\ catalogue contains $\simeq1000$ clumps identified as \Hyp\ compact sources at the reference wavelength of 160 \mum\ and with counterparts at 70, 250 and 350 \mum. The fitting wavelength is the 250 \mum. Since \Cut\ evaluates the flux independently at each wavelength \citep{Molinari11}, the two approaches are directly comparable only at the wavelength fixed as the fitting wavelength in the \Hyp\ catalogue, therefore at 250 \mum. In Fig. \ref{fig:hyp_vs_cut_irdc} we show the comparison between \Hyp\ and \Cut\ fluxes for the 960 sources in common between the two 250 \mum\ compact sources catalogues. Despite the complexity of the analysis the distributions are in a good agreement, although with a larger dispersion than for the simple case of the Perseus field (Sect. \ref{sec:test_perseus}). The linear fit of the distribution has an intercept of $0.13$ Jy which indicates very few systematics in the algorithms. The coefficient is $m=0.87$, which indicates a slight overestimate of the \Hyp\ fluxes compared to the \Cut\ fluxes on average. The scatter is likely due to the different approaches of the algorithms to estimate the background and to de-blend the sources in crowded fields. However the r.m.s. of the distribution is $\simeq50\%$, in line with the differences observed by \citep{Pezzuto12} between \Cut\ and \Get\ in Perseus.



\subsection{Test on the Bolocam Galactic Plane survey data}\label{sec:BGPS}
\begin{figure}
\centering
\includegraphics[width=8cm]{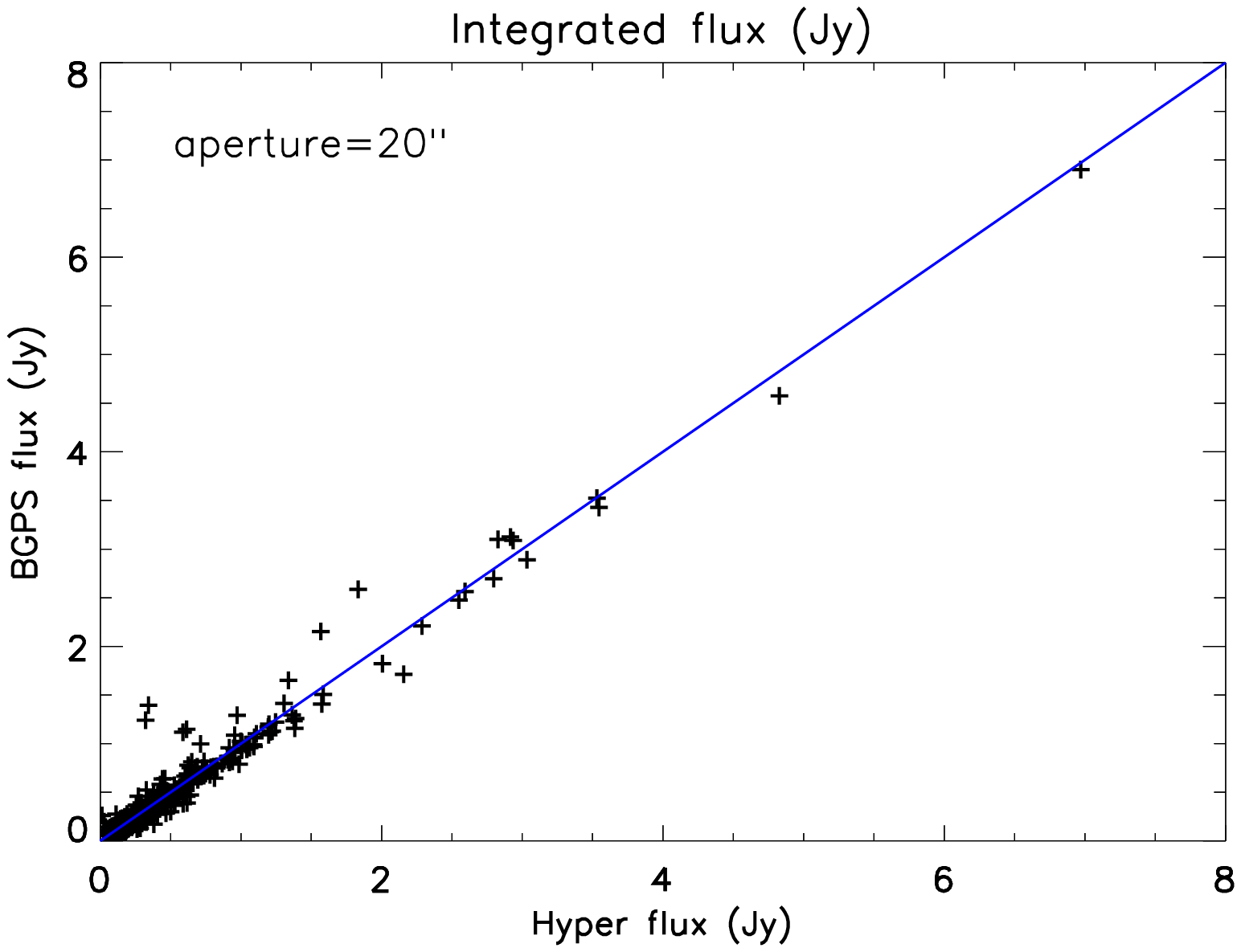}
\includegraphics[width=8cm]{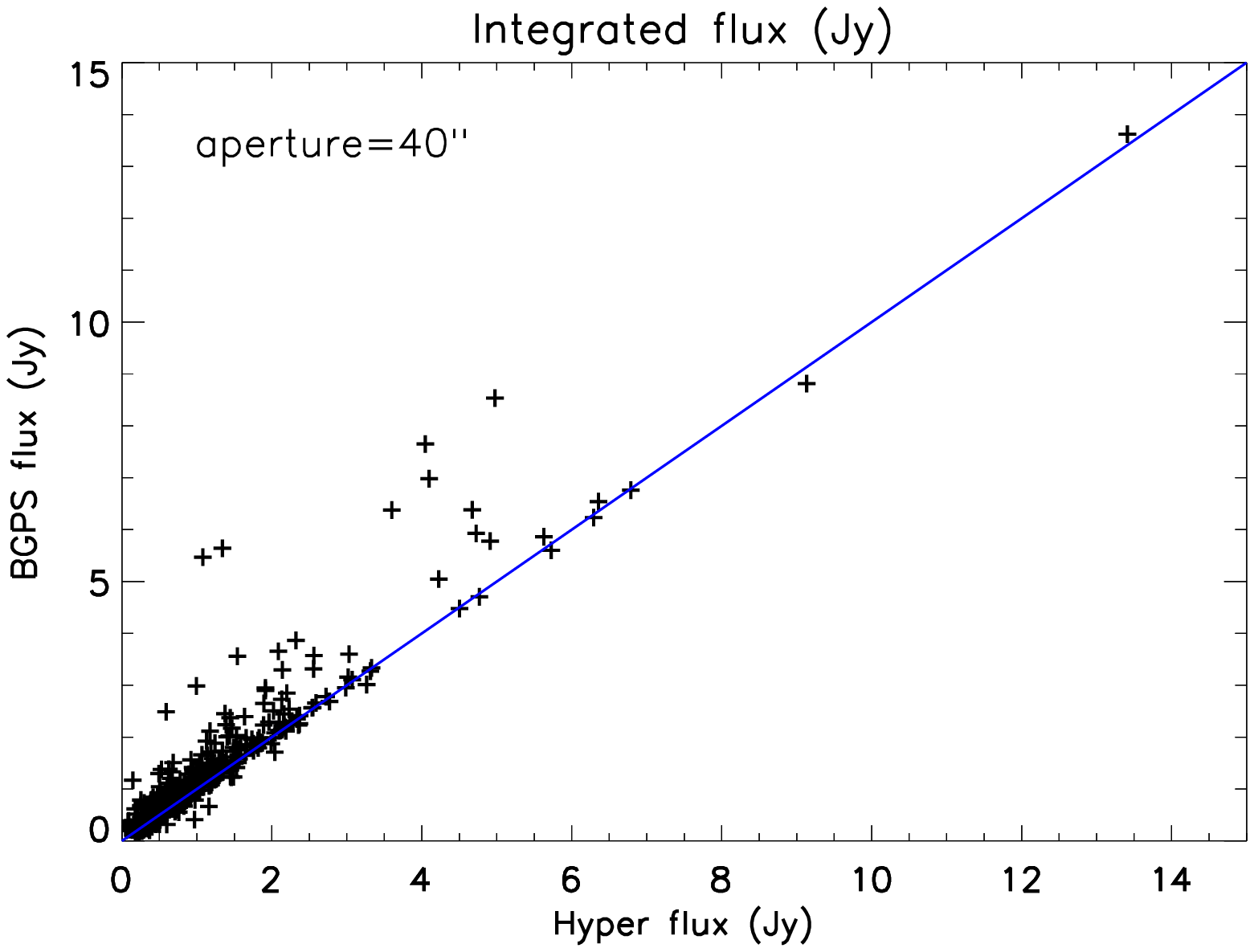}
\includegraphics[width=8cm]{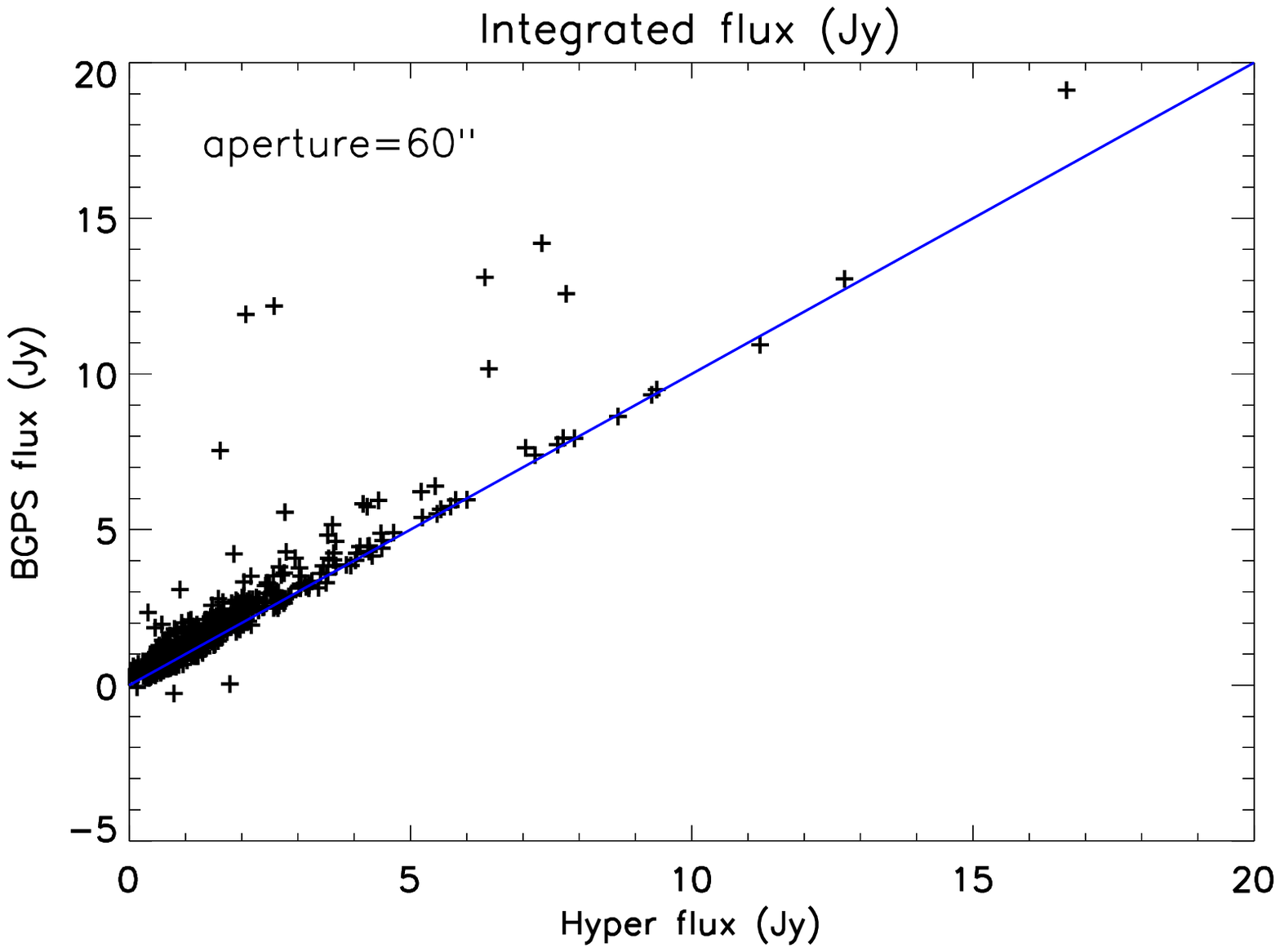}\\
\caption{Flux comparison between the official BGPS flux catalogue and the \Hyp\ flux estimation. The comparison includes 796 sources extracted from the public available BGPS v2 catalogue in the Galactic range $20\grad\leq l\leq28\grad$. The fluxes in the BGPS v2 catalogue are estimated with Bolocat within three different aperture radii: 20\arcsec, 40\arcsec\ and 60\arcsec \citep[for details of the BGPS v2 catalogue generation see][]{Ginsburg13}. \Hyp\ fluxes have been evaluated for all the sources fixing the three different apertures. The agreement is very good for both faint and bright sources at each aperture, with less than 1\% of the source with a flux difference $\geq50\%$.}
\label{fig:BGPS}
\end{figure}

To show the versatility of \Hyp\ we have also tested the algorithm on real data obtained with a completely different instrument. For this purpose we have extracted a sub-region of the BGPS \citep[][]{Aguirre10}. The survey covers approximately the Galactic range $-10.5\grad\leq l\leq90.5\grad$, $\vert b\vert\leq0.5\grad$ and the effective beam FWHM is 33\arcsec \citep{Aguirre10}. The public catalogue of compact source coordinates and flux densities has been produced with a specifically designed algorithm, Bolocat \citep{Rosolowsky10}. We used the BGPS v2 catalogue \citep{Ginsburg13} which contains $\simeq8000$ sources in total with flux densities estimated within three different aperture radii: 20\arcsec, 40\arcsec and 60\arcsec. We have compared the BGPS source catalogue fluxes with the \Hyp\ fluxes measured on the public available BGPS map in a $\simeq8\grad$ wide region in the range $20\grad\leq l\leq28\grad$, $\vert b\vert\leq0.5\grad$. The BGPS v2 catalog in this region contains 796 sources. We evaluated the \Hyp\ fluxes for all the BGPS sources at three fixed aperture radii, 20\arcsec, 40\arcsec and 60\arcsec, in order to allow a direct comparison with the BGPS fluxes. The flux comparison using these three aperture radii are shown in Fig. \ref{fig:BGPS}. The agreement is very good at all the three different apertures, with only few outliers at each aperture radius. Less than 1\% of the sources have a flux difference greater than 50\% at each aperture (67, 34 and 54 using aperture radii of 20\arcsec, 40\arcsec\ and 60\arcsec\ respectively). Visual inspections show that these sources are either very weak or in crowded regions and/or on top of very variable backgrounds. The mean and \textit{r.m.s.} of the flux differences at each aperture radius excluding these outliers are in Table \ref{tab:bgps}. The average difference between the BGPS and \Hyp\ fluxes is $\simeq9\%, \simeq -8\% $ and $\simeq-10\%$ at 20\arcsec, 40\arcsec\ and 60\arcsec\ respectively, with a  \textit{r.m.s.} of $\simeq16.5\%$ for all the apertures, likely due to the different strategies adopted to estimate the background and the source de-blending.

\begin{table}
\caption{Mean flux difference and \textit{r.m.s.} of the flux difference distribution between the flux of 796 BGPS sources extracted from the BGPS v2 catalogue and the flux estimated with \Hyp\ using three different aperture radii. See Sect. \ref{sec:BGPS} for details.}
\begin{center}
\begin{tabular}{c|c|c}
\hline
\hline
Aperture & Mean flux & \textit{r.m.s.} flux  \\ 
radius ($\arcsec$) & diff $(\%)$ & diff $(\%)$ \\   
\hline
20	& 8.8		& 16.9 \\
40	& -8.5	& 16.3 \\
60	& -10.3	& 16.7 \\ 
\hline
\end{tabular}
\end{center}
\label{tab:bgps}
\end{table}

\section{Conclusions}\label{sec:conclusions}

We have developed a new source extraction and photometry algorithm, \Hyp, an
enhanced application of aperture photometry specifically designed for multi-wavelengths photometry on crowded fields in complex
background. The extraction is done in a high-pass filtered map which amplifies
the compact sources while suppressing the diffuse emission, allowing source
identification in regions with highly variable background. The source
photometry is done over an elliptical aperture with a size and shape estimated
from a 2d Gaussian fit using the \texttt{mpfit2dpeak} IDL routine. The 2d Gaussian
fitting allows us to identify the region over which integrate the flux for
both point sources and slightly extended sources, minimising the flux
contamination from the region surrounding the sources. The background is
modelled with different polynomial orders and in squared regions of different sizes. The model which minimises the \textit{r.m.s.} of the residual map is taken as the background estimate and subtracted from the data. Blended sources are
fitted simultaneously with a multiple 2-d Gaussian models and the fit for
companions is subtracted from the original data before evaluating the flux of
the reference source. This de-blending systematically improves the flux
estimates of the sources in crowded fields. The algorithm is designed to allow multi-wavelength flux estimation by fixing the aperture radius at a reference wavelength and integrating simultaneously at all the selected wavelengths across the same volume of gas and dust.

\Hyp\ has been tested on
simulated fields in which model sources have been injected on real observed backgrounds. These simulations show that \Hyp\ can typically recover the model source flux with a high degree of accuracy both in the case of random injected sources and in specific tests with sources injected across filamentary structures. The multi-wavelength approach has been tested at the \Her\ wavelengths demonstrating high degree of accuracy at each wavelength, with a slight flux overestimation in the extreme case of the flux at 70 \mum\ estimated within the aperture fixed at 500 \mum.

\Hyp\ photometry has been tested on real fields showing good agreement with other algorithms and the estimation of the uncertainties both in a very simple case (few isolated sources in the B1 Perseus star-forming regions) and in very complex fields (hundred of sources on top of very variable backgrounds and in crowded regions). Also, the versatility of \Hyp\ has been demonstrated by showing the very good agreement between the \Hyp\ and the publicly available fluxes of $\simeq800$ sources extracted from the BGPS, the 1.1 mm survey of the 
Galactic Plane carried out with the Caltech Sub-millimeter Observatory.


\Hyp\ is also very fast. To measure the fluxes of $\simeq$1600 sources
extracted from Hi-GAL counterparts of IRDCs in the $15^{\circ}\leq
l\leq55^{\circ}$, $\vert b\vert\leq1^{\circ}$ region of the Galactic
plane at four wavelengths simultaneously (70, 160, 250 and 350 \mum) with
the default settings it requires $\simeq30$ minutes on a 2.2 GHz Intel
Core i7 machine, and uses less than 200 Mb of RAM \citep{Traficante14_cat}

 \Hyp\ is an IDL code initially developed
to extract compact sources from \herschel\ surveys, in particular for
the \herschel\ Galactic plane survey, Hi-GAL. However, it is highly
modular and highly parameterisable and allows the user to adapt it to
the specifications of different surveys and observations. 

The code is freely available and it can be downloaded from
the page http://www.irdarkclouds.org.

\section*{Acknowledgements}
This research has made use of data from the \Her\ Gould Belt survey (HGBS) project (http://gouldbelt-herschel.cea.fr). The HGBS is a \Her\ Key Programme jointly carried out by SPIRE Specialist Astronomy Group 3 (SAG 3), scientists of several institutes in the PACS Consortium (CEA Saclay, INAF-IFSI Rome and INAF-Arcetri, KU Leuven, MPIA Heidelberg), and scientists of the Herschel Science Center (HSC). The authors want to thank the HGBS consortium also for using the B1 Perseus map at PACS 70 \mum\ and to publish the \Cut\ and \Get\ fluxes in Table \ref{tab:hyper_cutex_flux}. The authors want to thank Alexander Men'shchikov for having run \Get\ on a part of the B1 Perseus region. JEP has received funding from the European Community's Seventh Framework Programme (/FP7/2007-2013/) under grant agreement No 229517. AT wants to thank Maria del Mar Rubio Diez for the help during the test phase of the \Hyp\ algorithm. AT is supported by STFC consolidated grant to JBCA.


\bibliographystyle{aa} 
\bibliography{Hyper_def_Astroph.bbl}

\newpage

\begin{appendix}
\section{Flux difference between source model and Hyper measurements for T1 and T2}\label{app:fluxes}

\begin{figure*}[!ht]
\centering
\includegraphics[width=8cm]{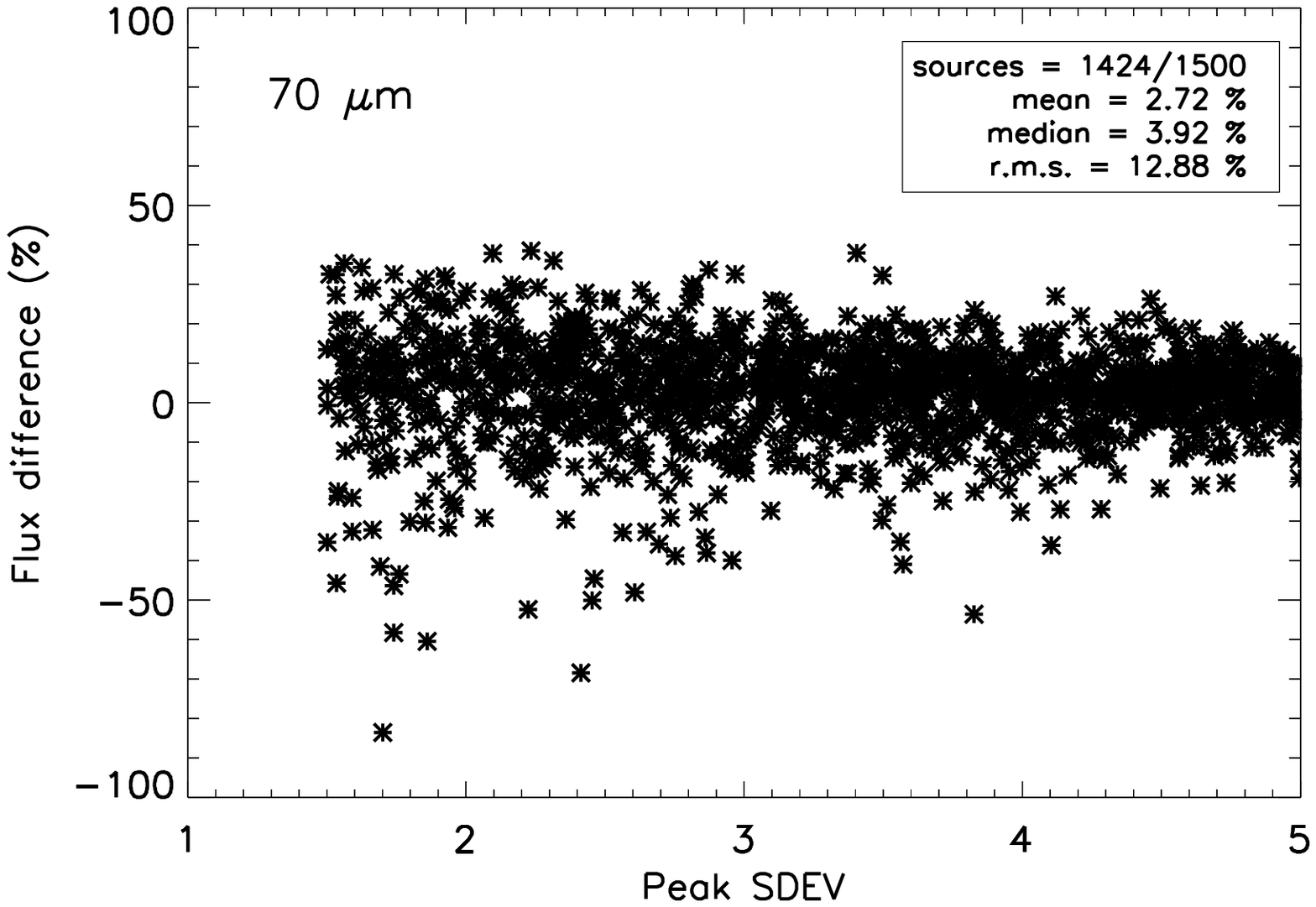}
\includegraphics[width=8cm]{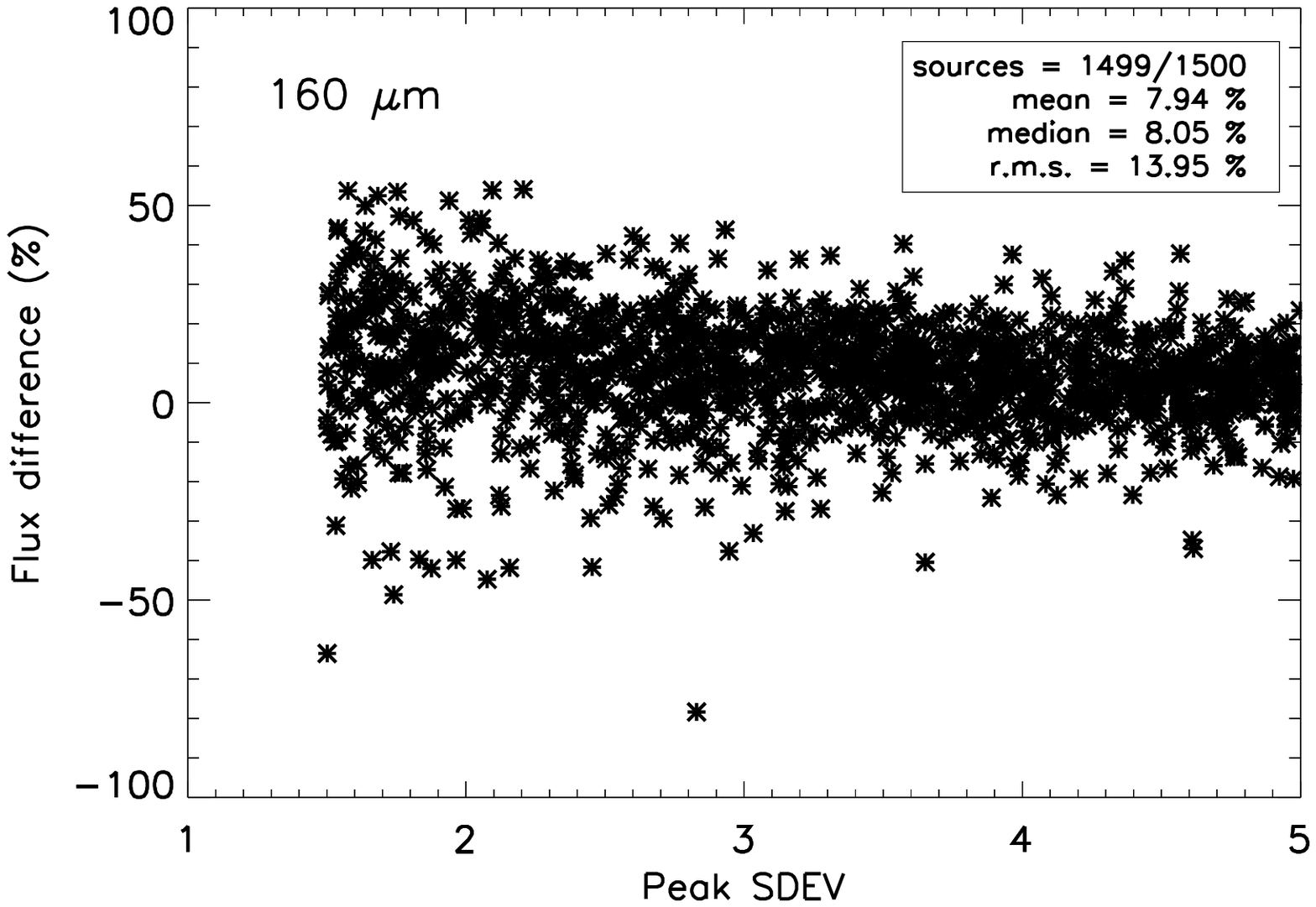}\\
\includegraphics[width=8cm]{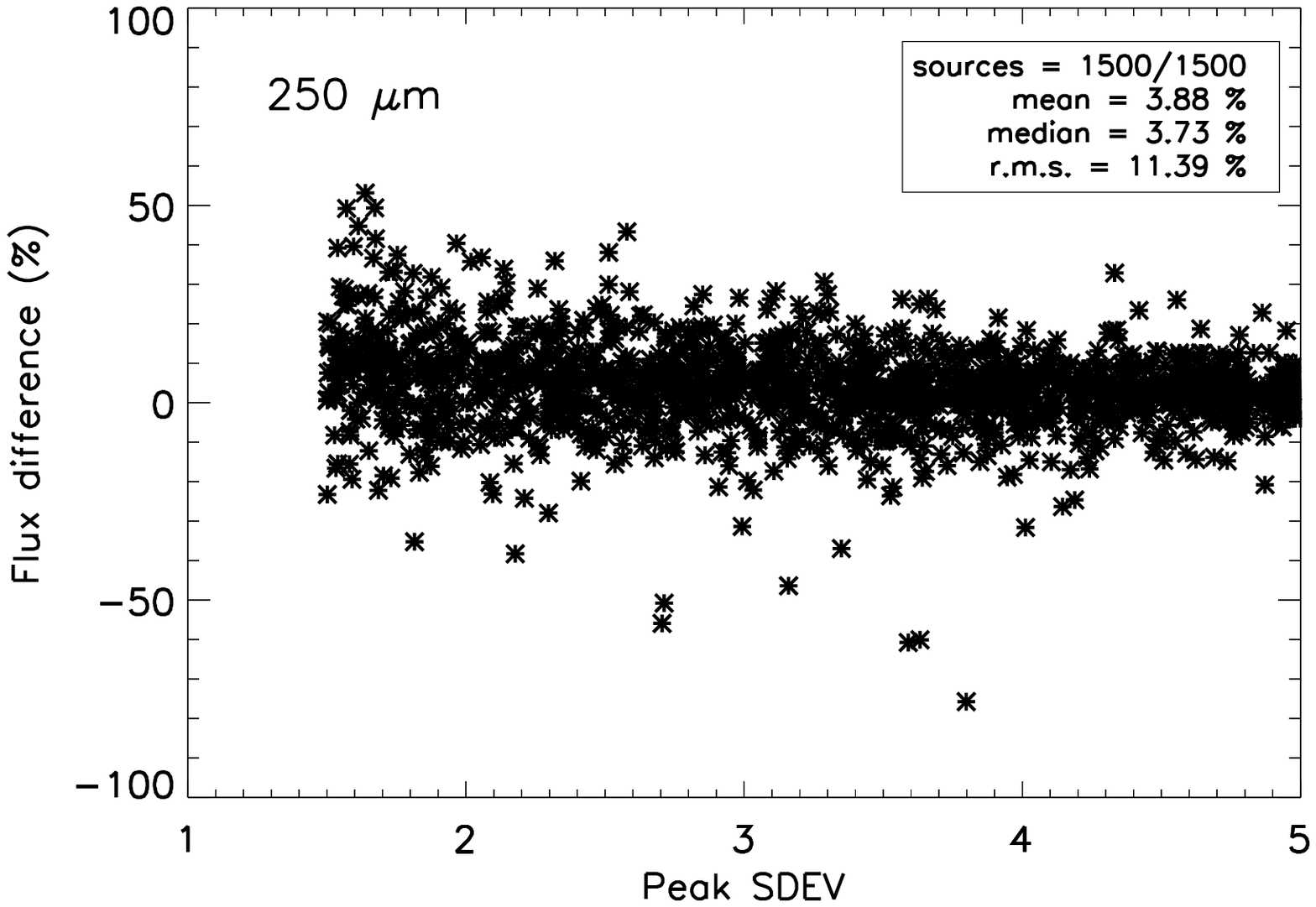}
\includegraphics[width=8cm]{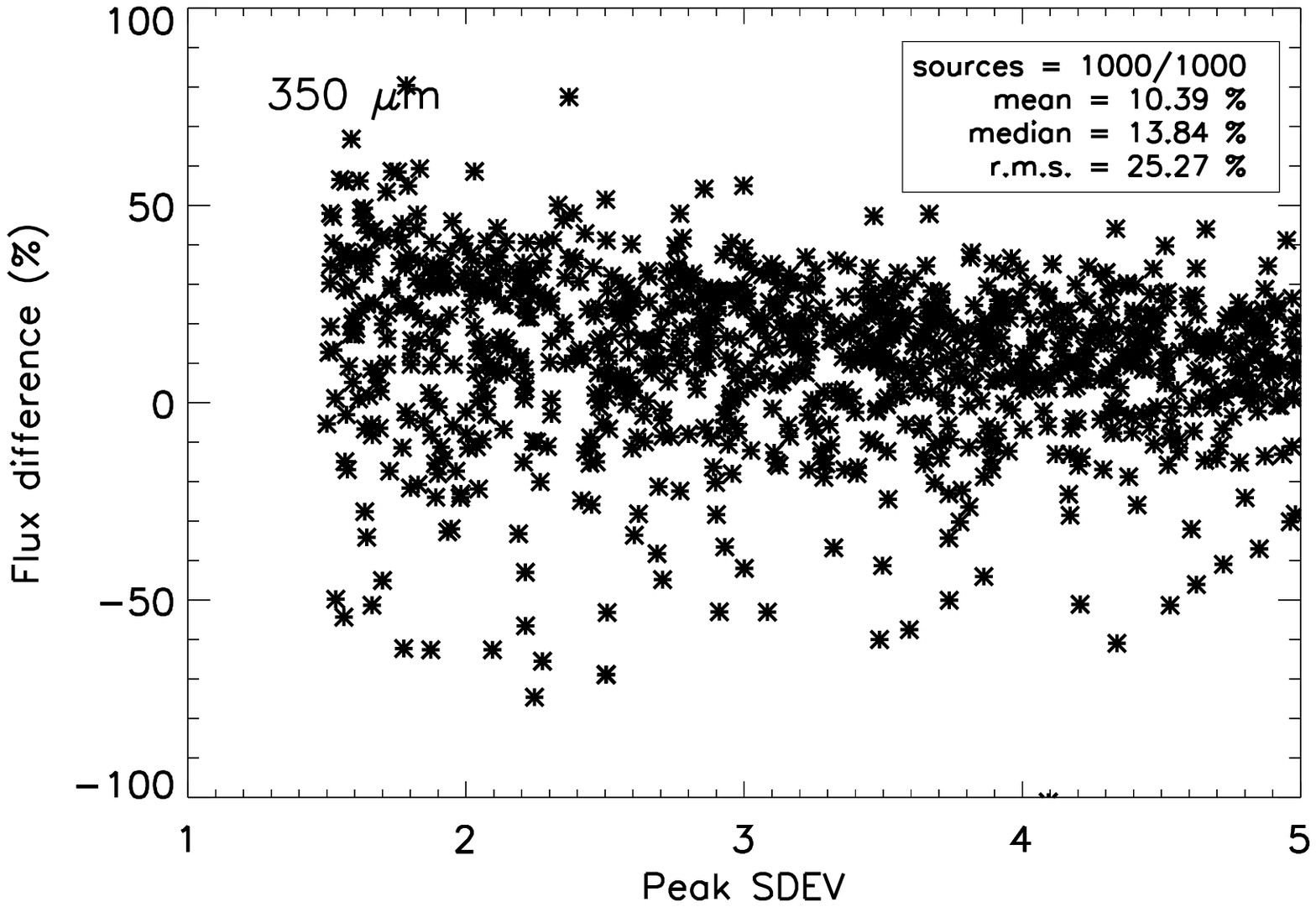}\\
\includegraphics[width=8cm]{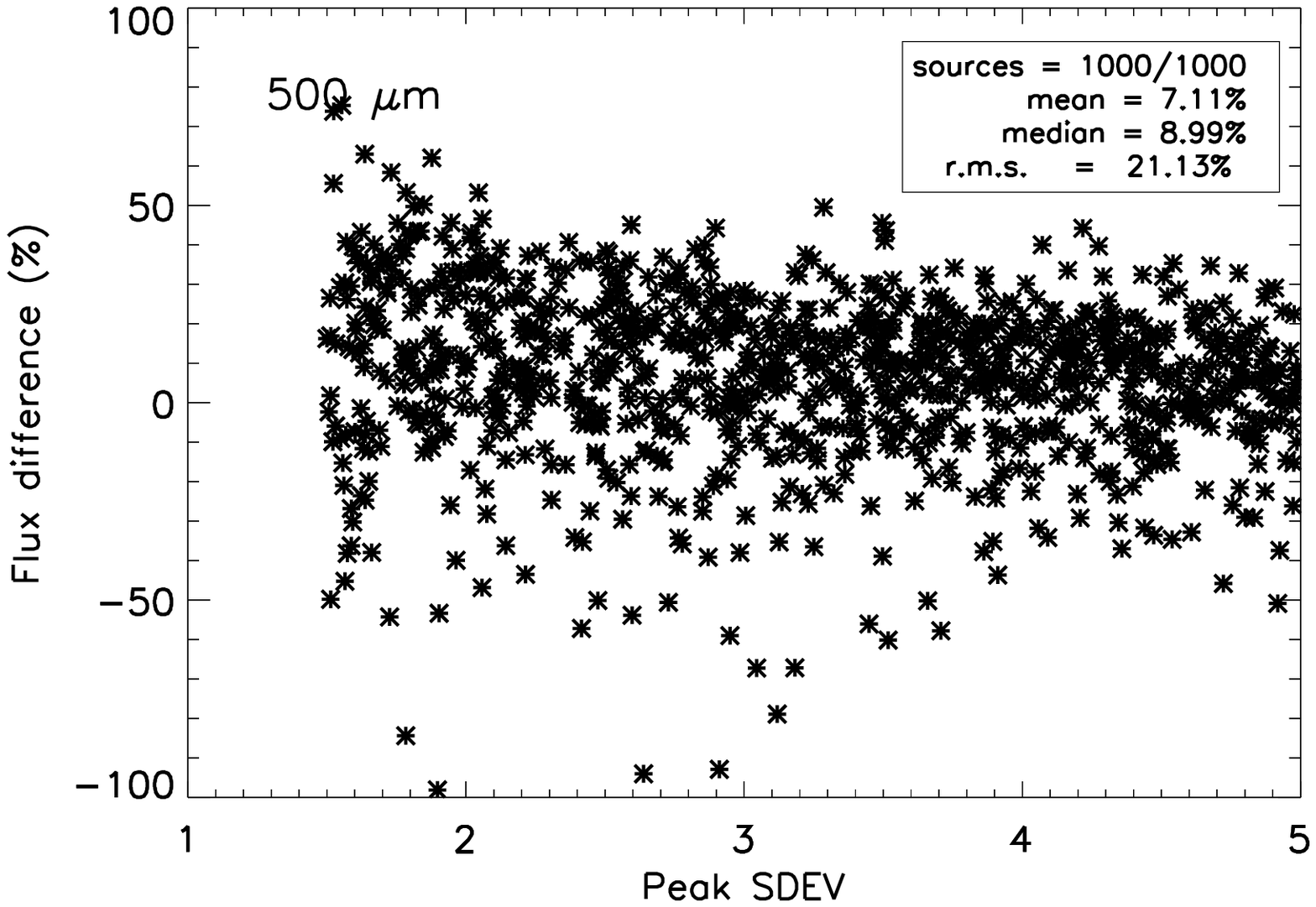}

\caption{Difference between \Hyp\ measured fluxes and the model fluxes as
  function of the source peak fluxes measured as a multiple of the standard
  deviation of the map (peak SDEV), for T1 at all wavelengths. 70 \mum\ flux distribution (upper left panel), 160 \mum\ (upper right), 250 \mum\ (centre left), 350 \mum\ (centre right) and 500 \mum\ (lower panel). The mean and median values are
  only few percent and the r.m.s. of the distribution is higher than 30\% only for the 350 \mum\ map. Mostly for the faint sources the
  flux estimation is worse than 30\%.}
\label{fig:source_sdev_15_5}
\end{figure*}

\begin{figure*}[!ht]
\centering
\includegraphics[width=8cm]{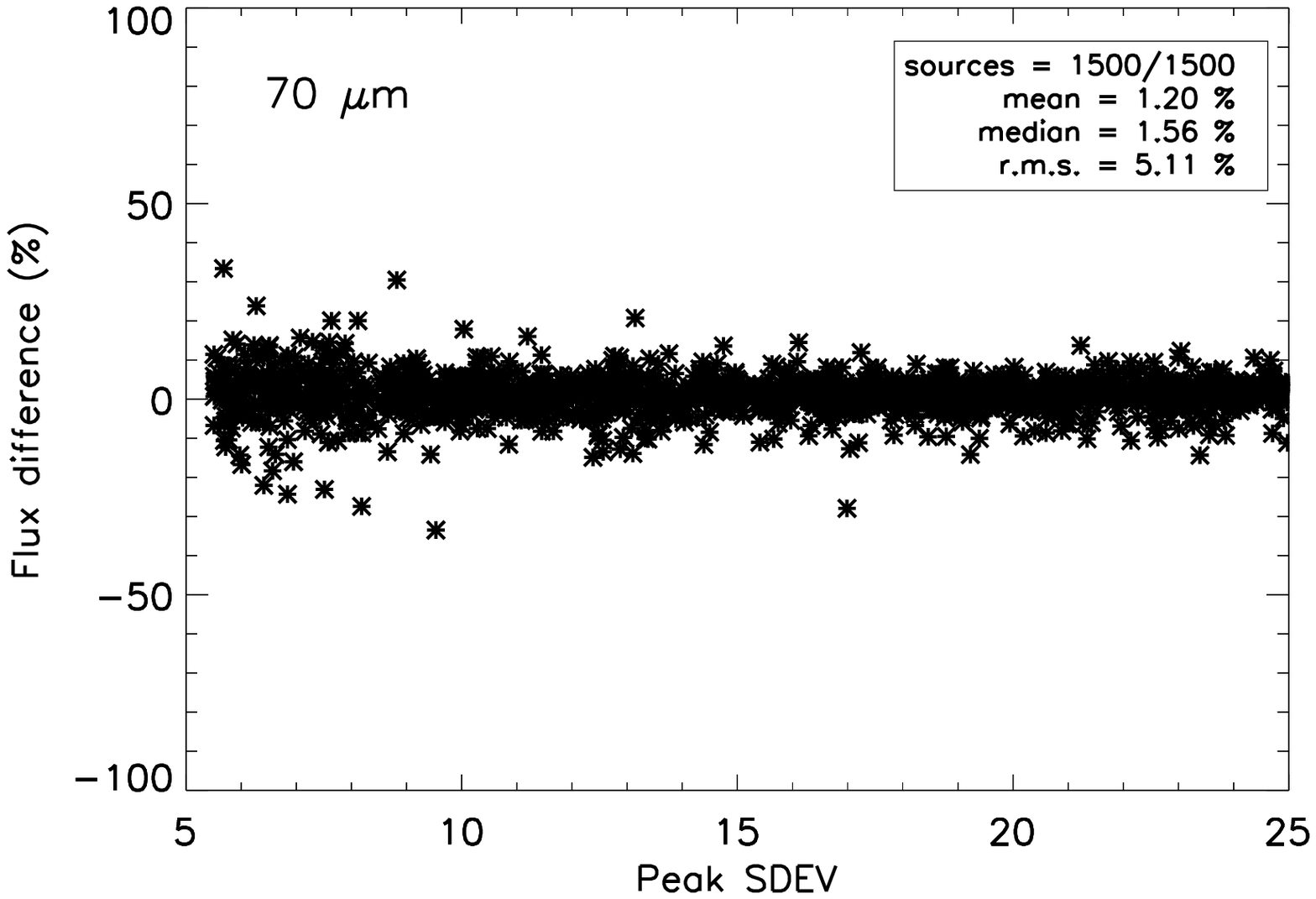}
\includegraphics[width=8cm]{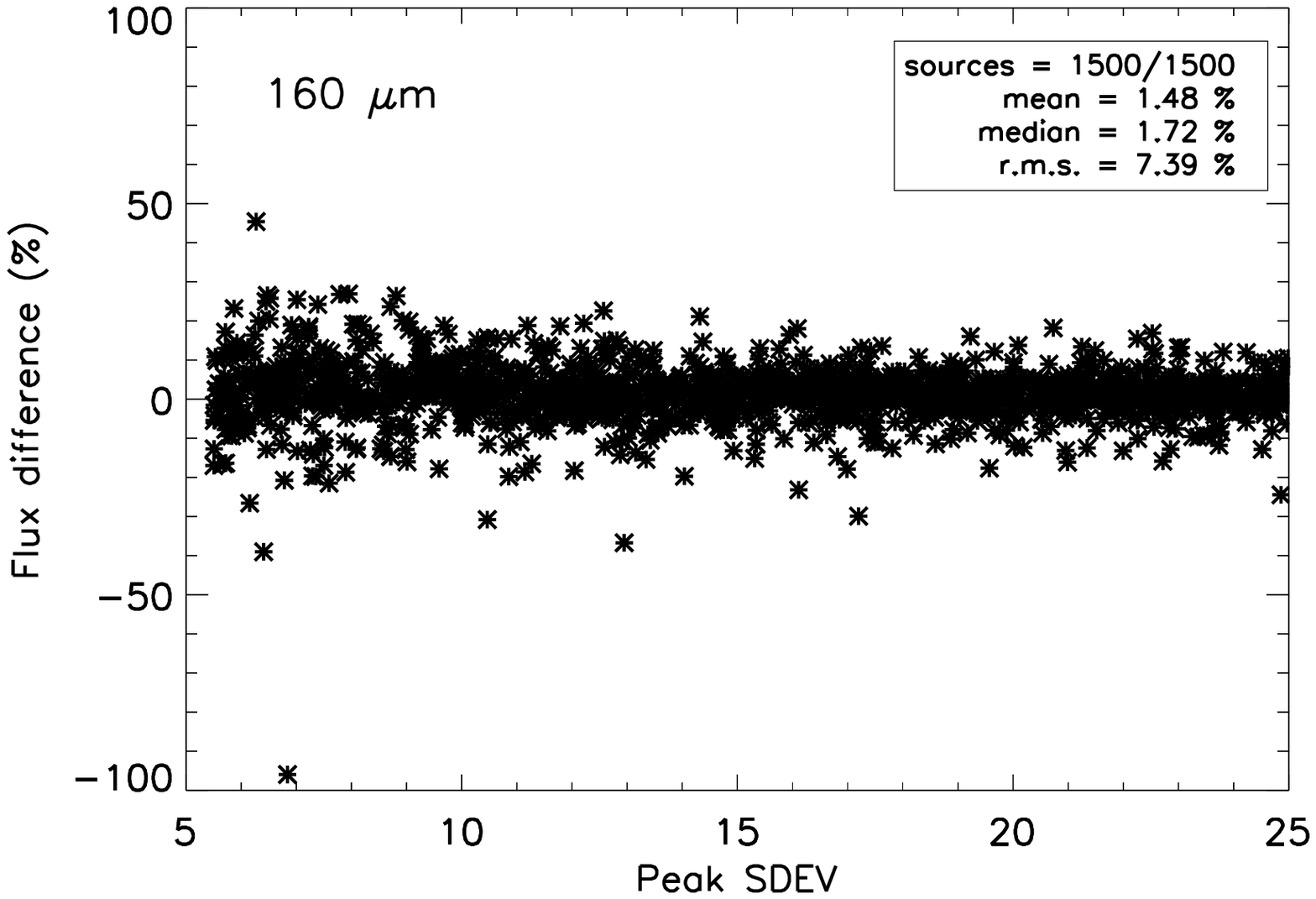}\\
\includegraphics[width=8cm]{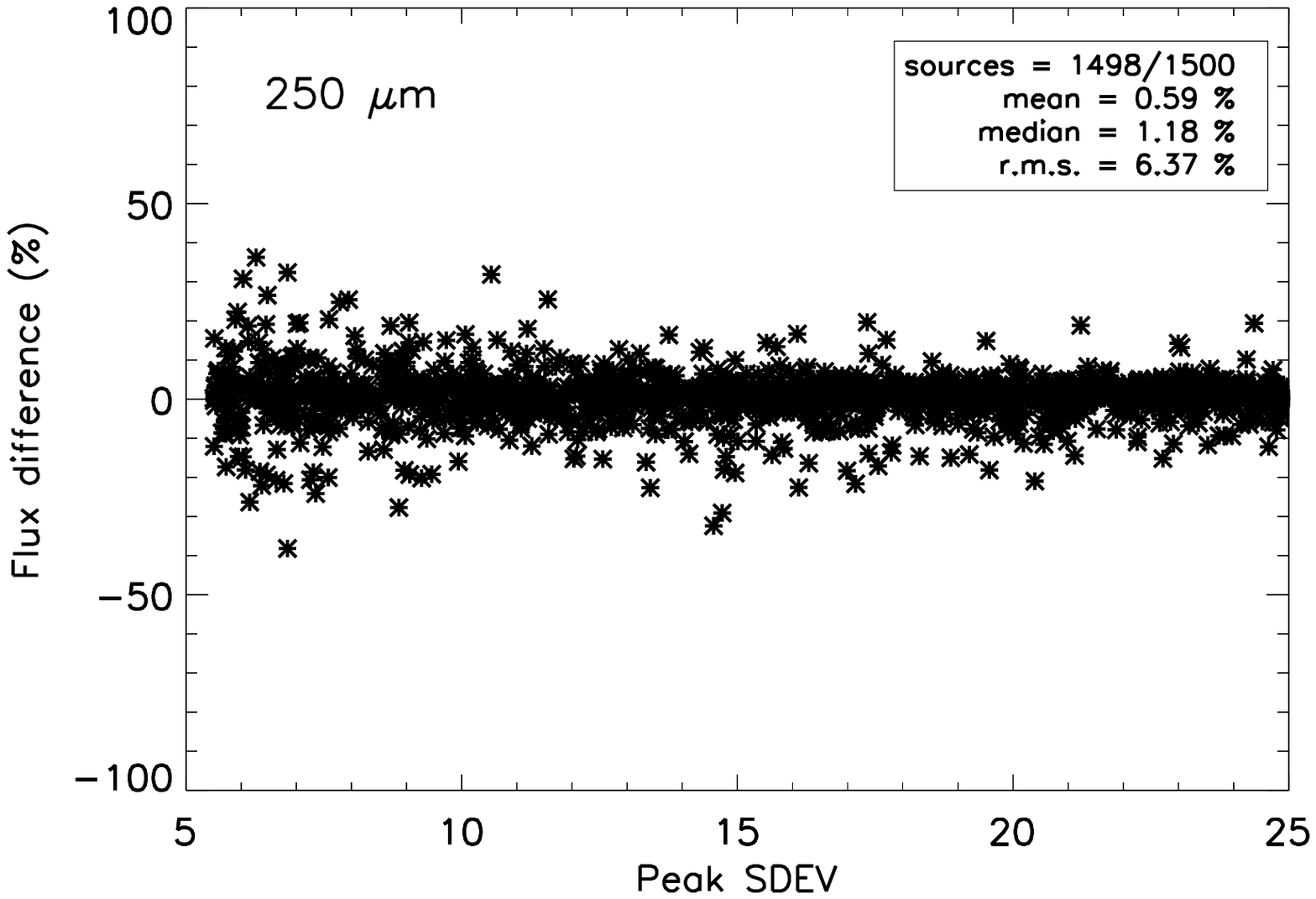}
\includegraphics[width=8cm]{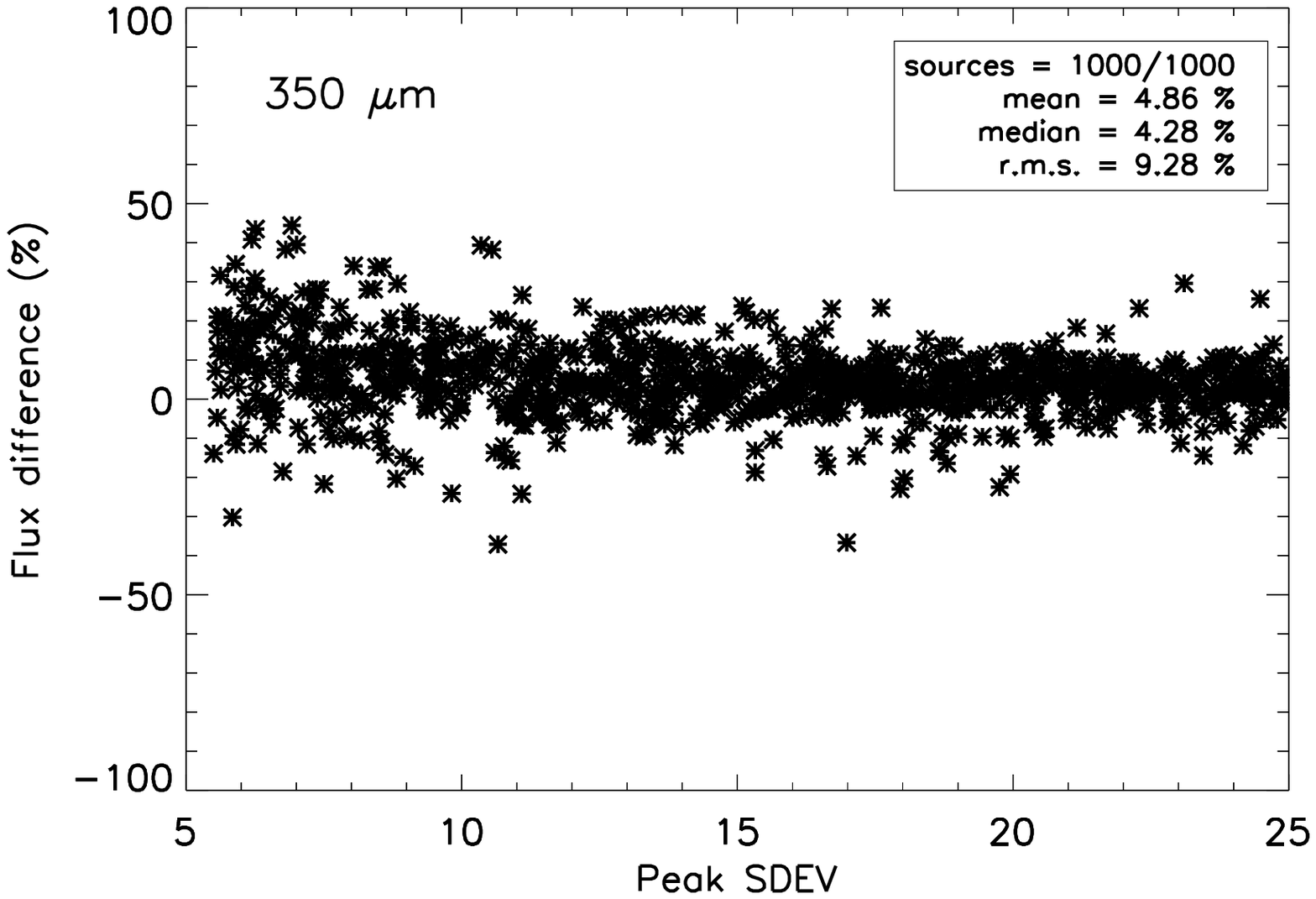}\\
\includegraphics[width=8cm]{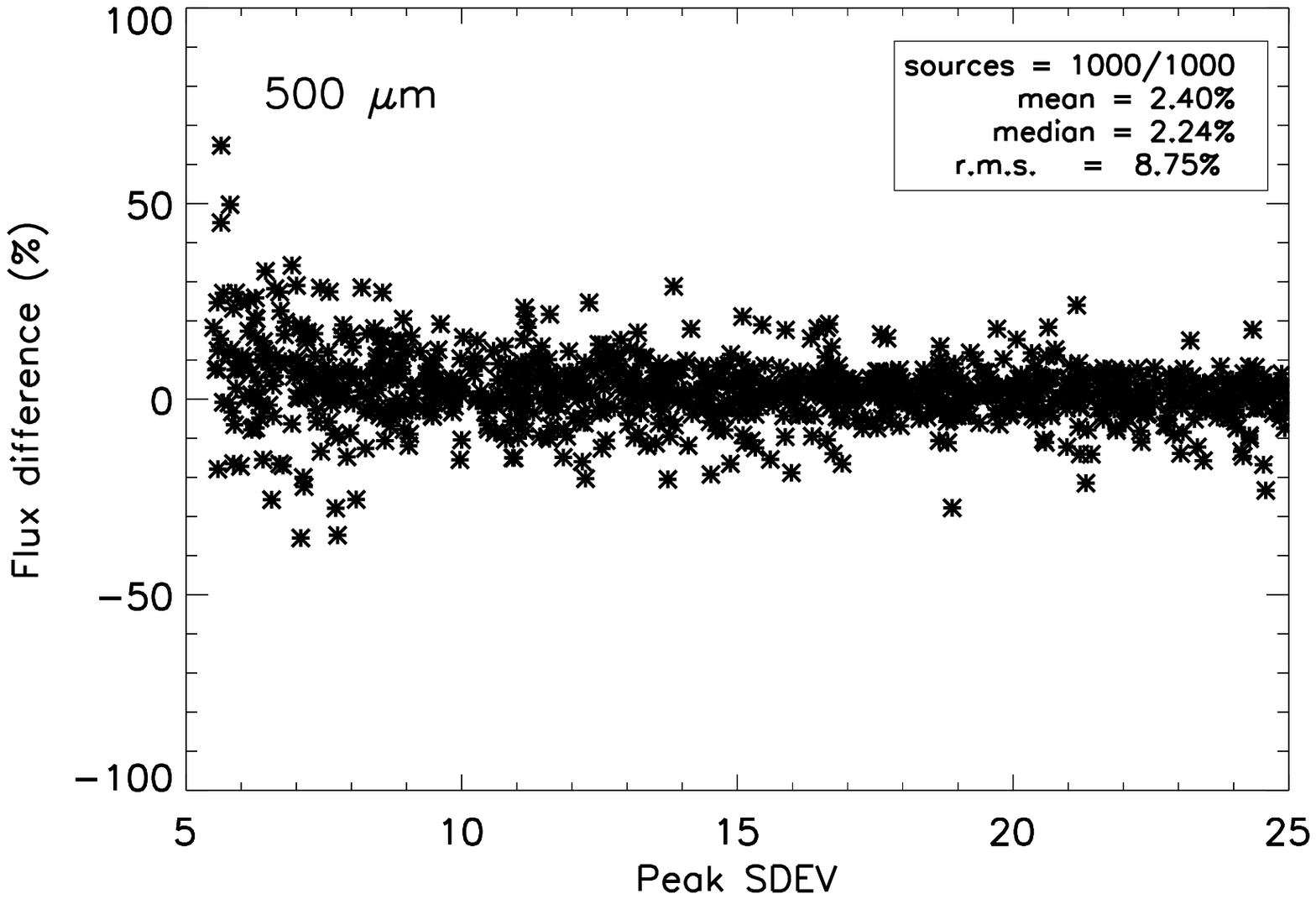}

\caption{Same as Figure \ref{fig:source_sdev_15_5}, but for T2. The distributions are much more regular with a r.m.s. $\leq10$\%. Only few sources have a flux difference higher than 15\%.}
\label{fig:source_sdev_55_25}
\end{figure*}


\newpage
\section{Radius difference between source model and Hyper measurements for T1 and T2}\label{app:radii}

\begin{figure*}[!ht]
\centering
\includegraphics[width=8cm]{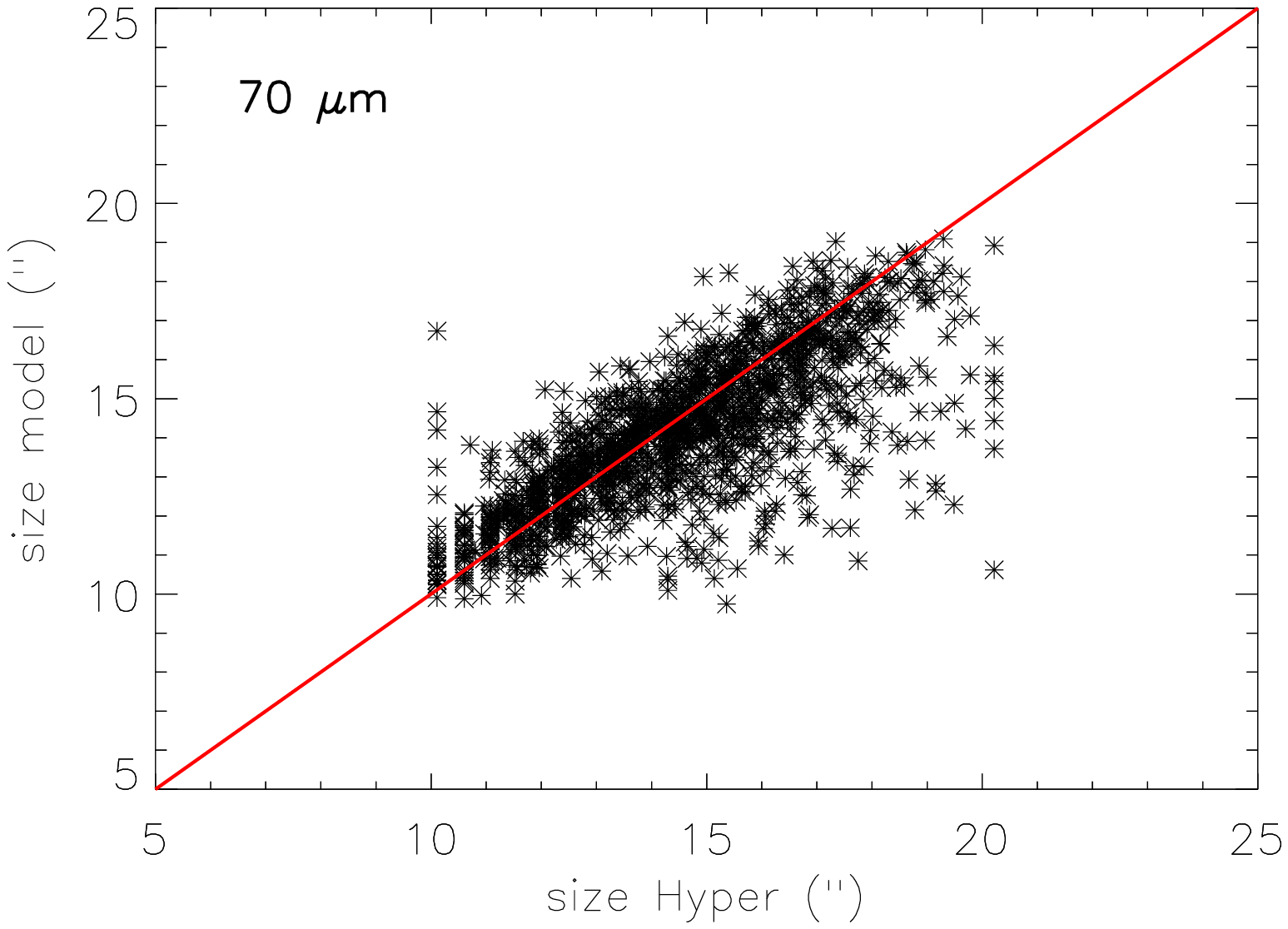}
\includegraphics[width=8cm]{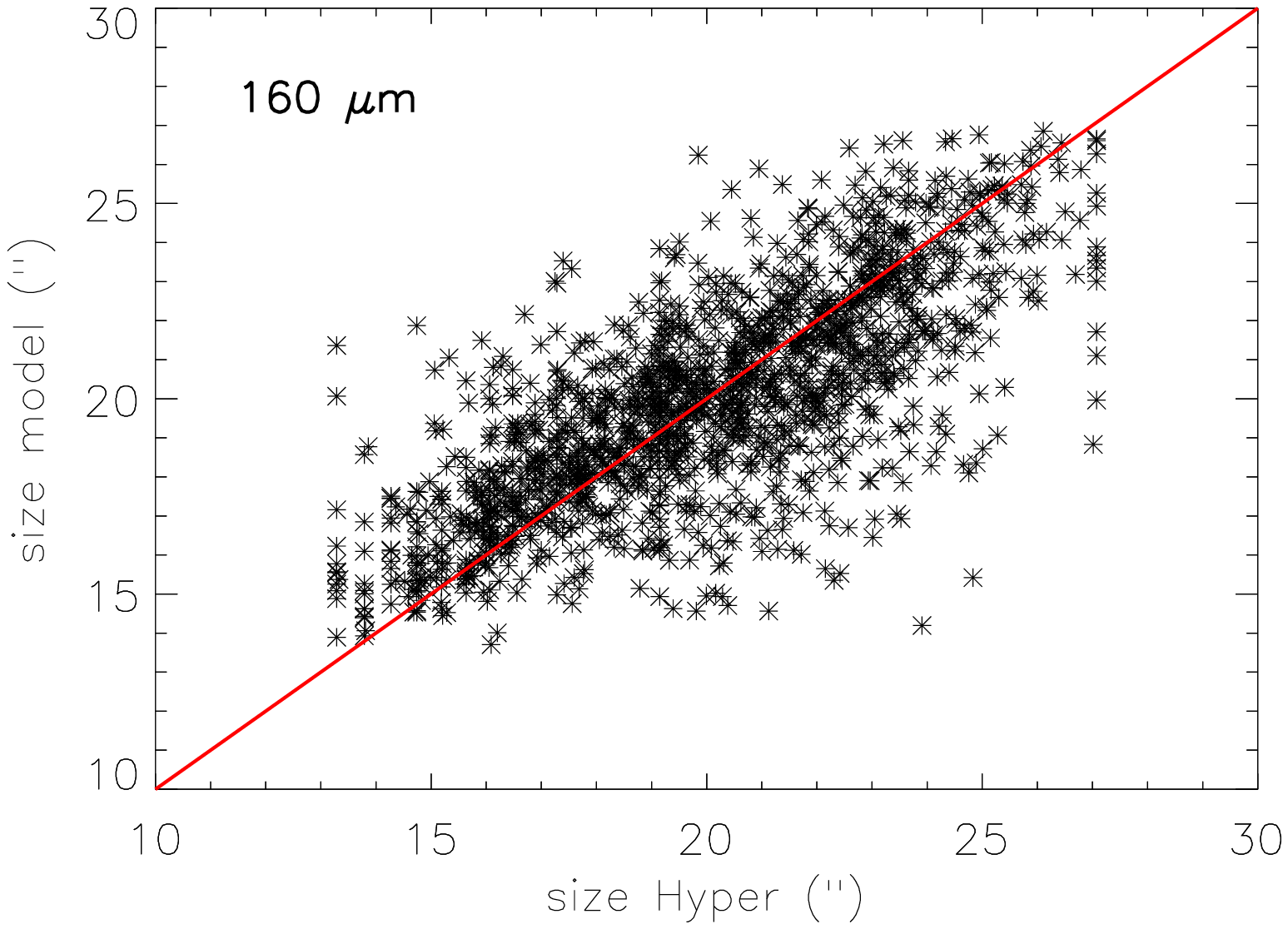}
\includegraphics[width=8cm]{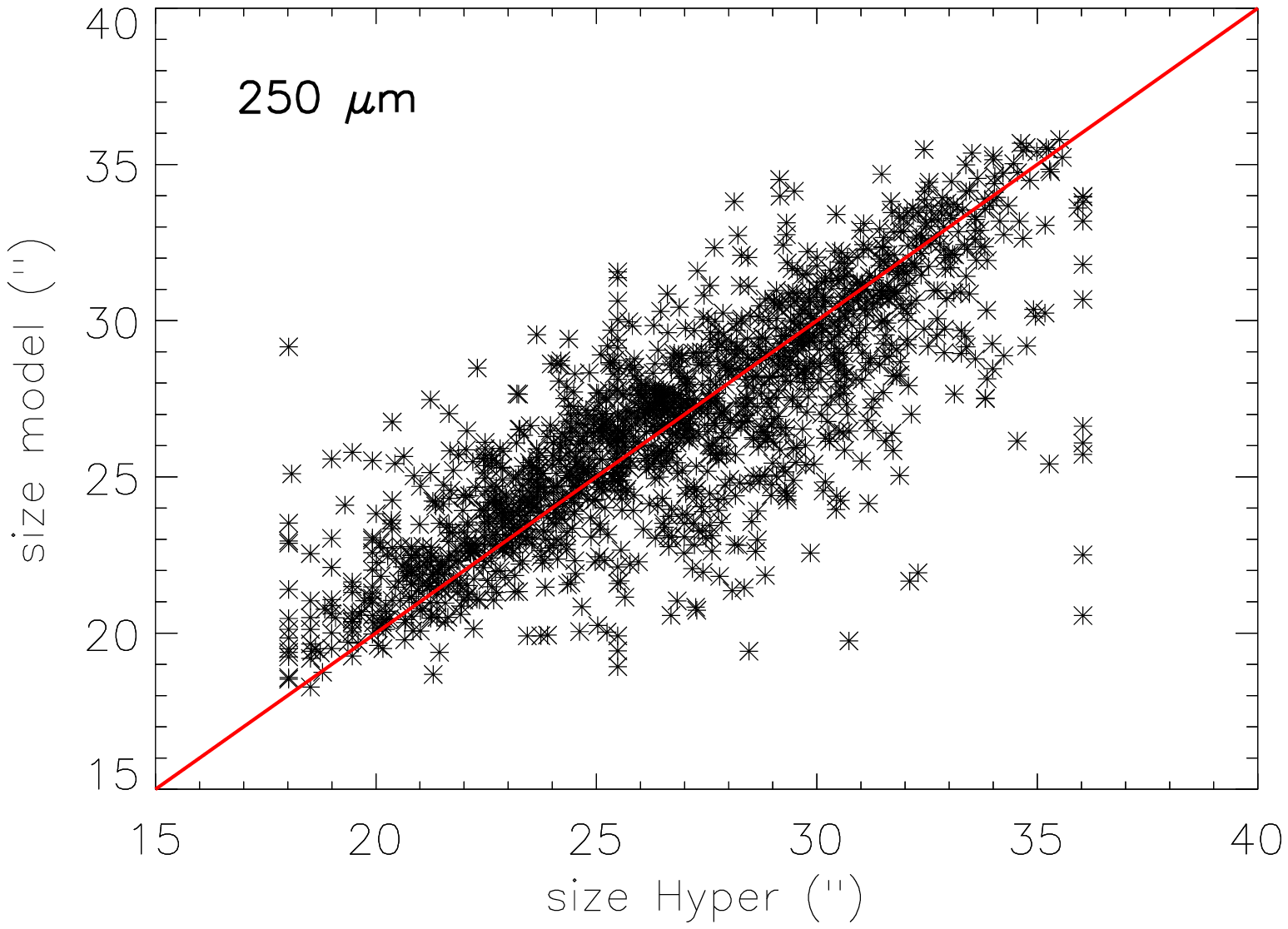}
\includegraphics[width=8cm]{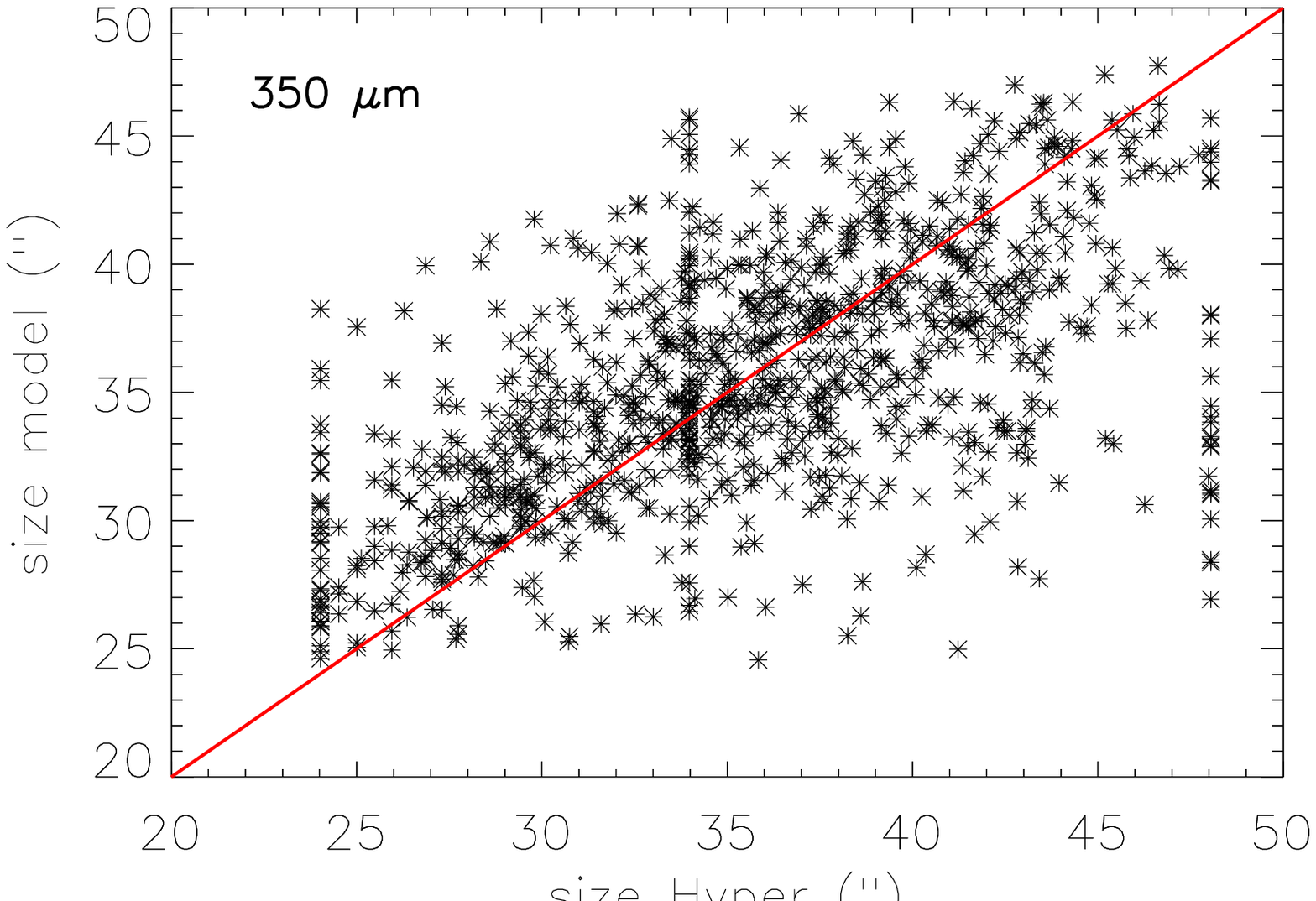}
\includegraphics[width=8cm]{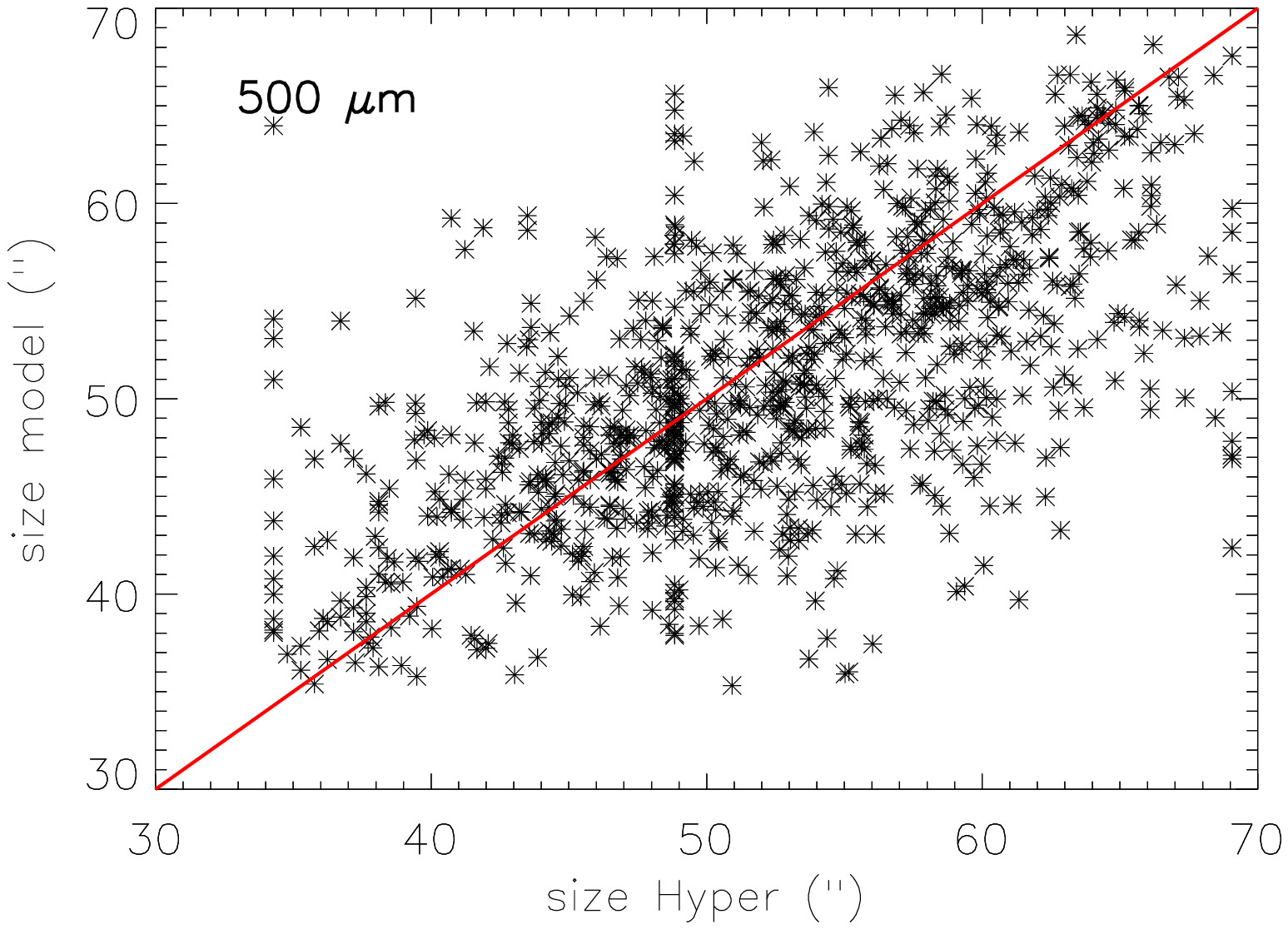}
\caption{Comparison of the injected and measured source size for run T1, evaluated as the geometrical mean between the minimum and the maximum aperture radius. 70 \mum\ size distribution (upper left panel), 160 \mum\ (upper right), 250 \mum\ (centre left), 350 \mum\ (centre right) and 500 \mum\ (lower panel). There are less than 1\% of the points distributed along three vertical lines, corresponding to fixed \Hyp\ source radii. The lowest and the highest radii correspond to the sources that \Hyp\ has forced to be equal to a circular 2d Gaussian with FWHM equal to, respectively, the minimum and the maximum aperture radius chosen in the parameter file. The central points are in correspondence of fits that did not converge. In these cases \Hyp\ has forced the source shape to be circular with a FWHM equal to the geometrical mean of the minimum and maximum FWHM limits set in the parameters file.}
\label{fig:source_radius_15_5}
\end{figure*}

\begin{figure*}[!ht]
\centering
\includegraphics[width=8cm]{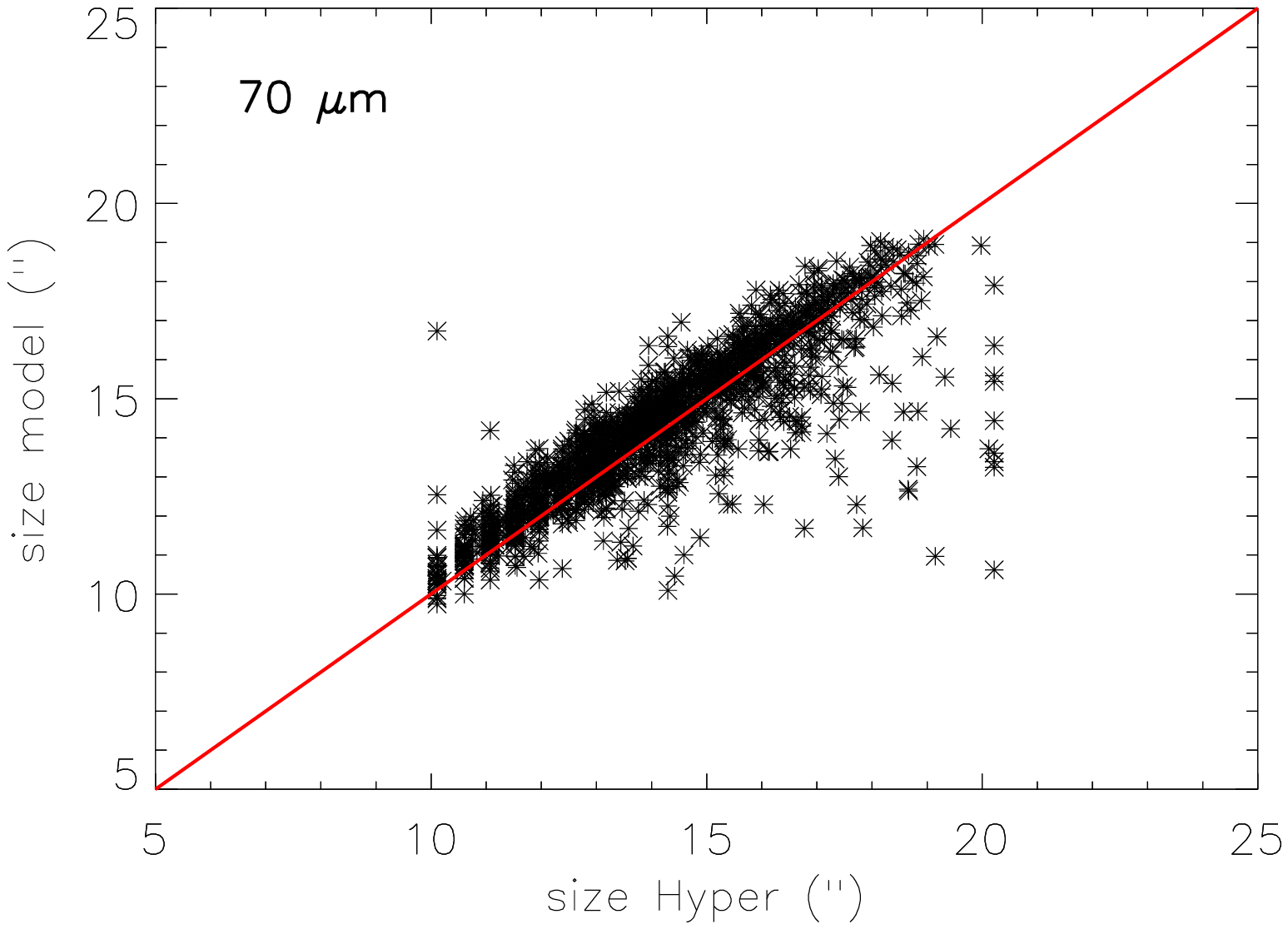}
\includegraphics[width=8cm]{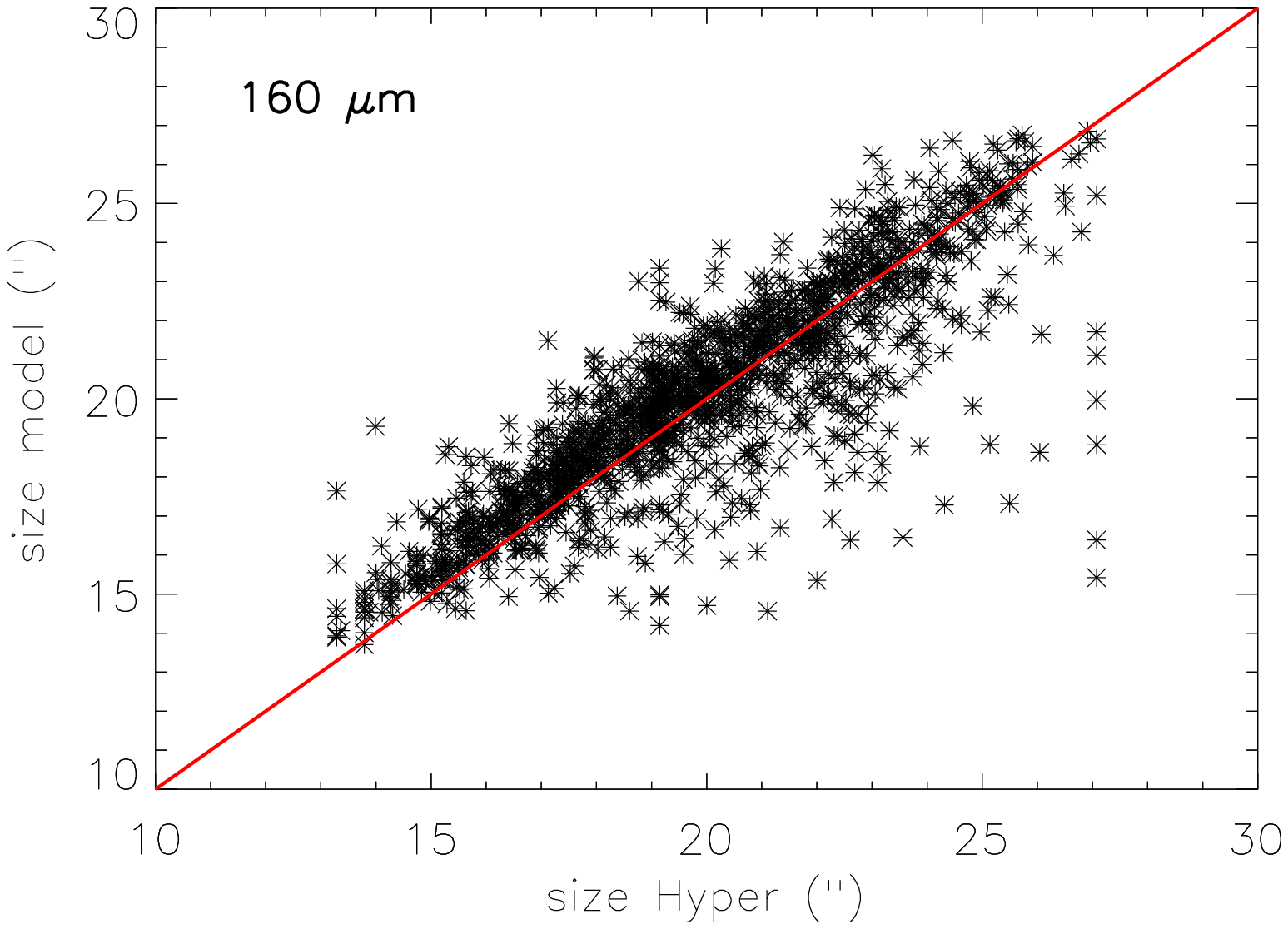}
\includegraphics[width=8cm]{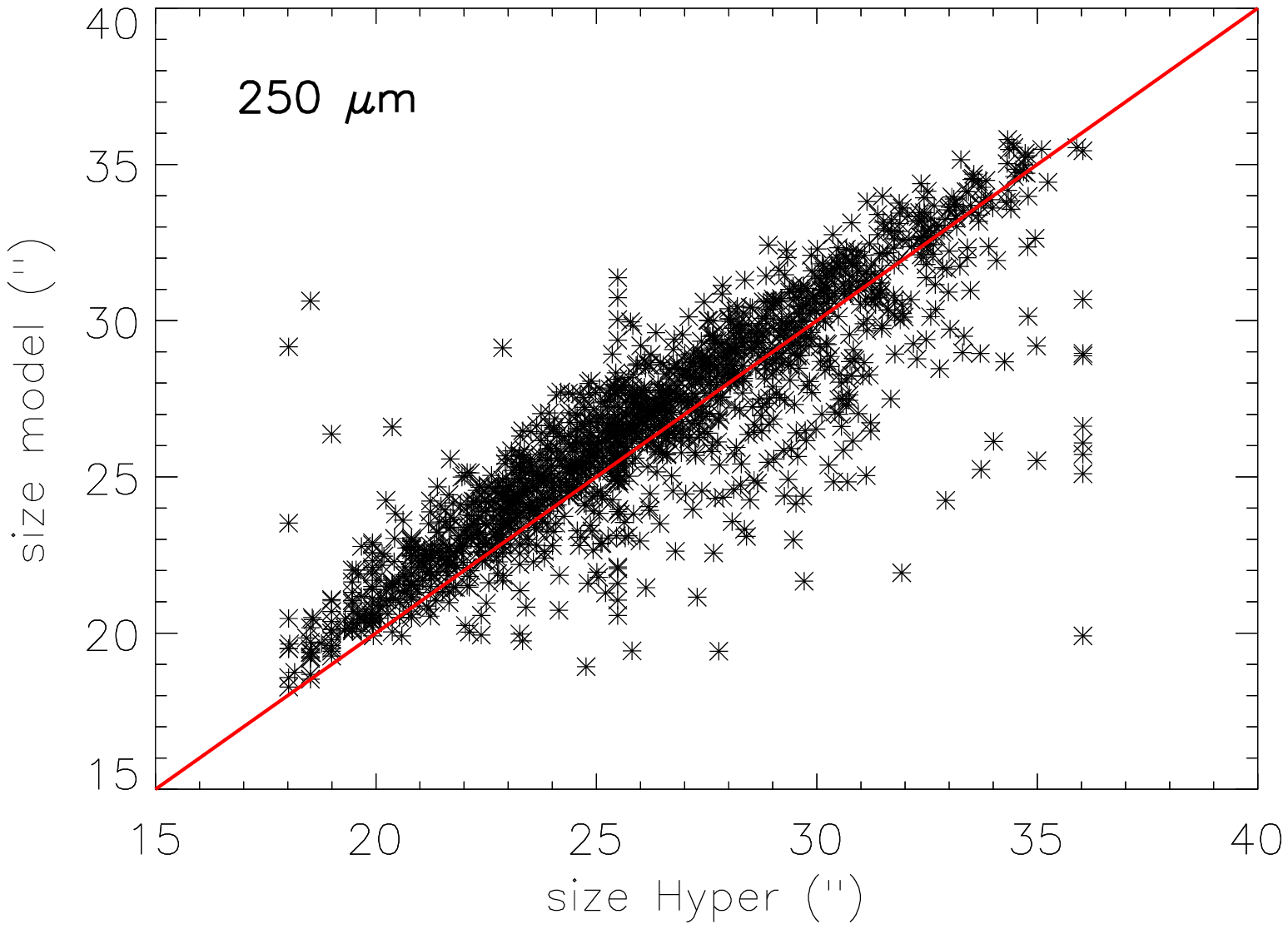}
\includegraphics[width=8cm]{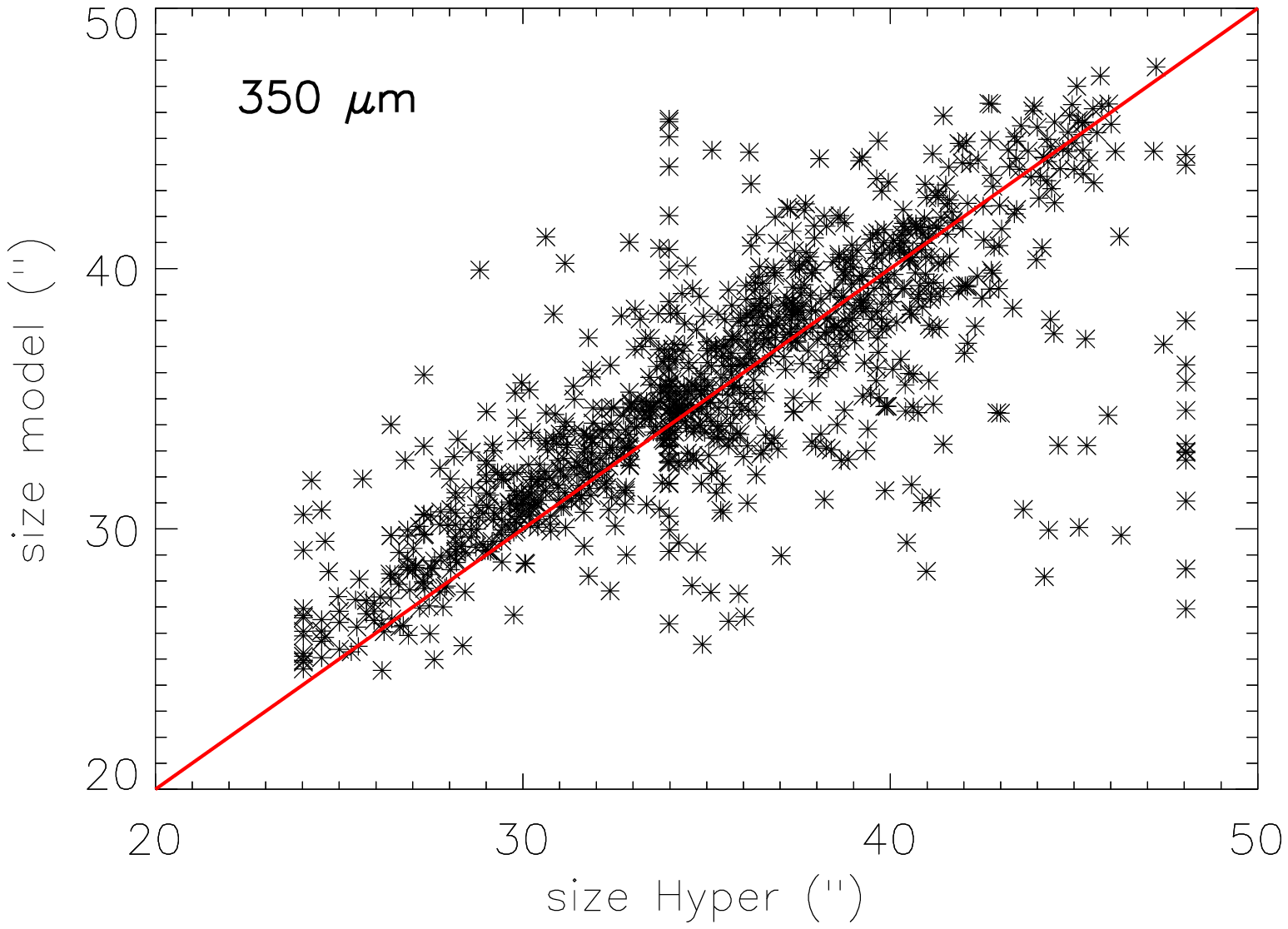}
\includegraphics[width=8cm]{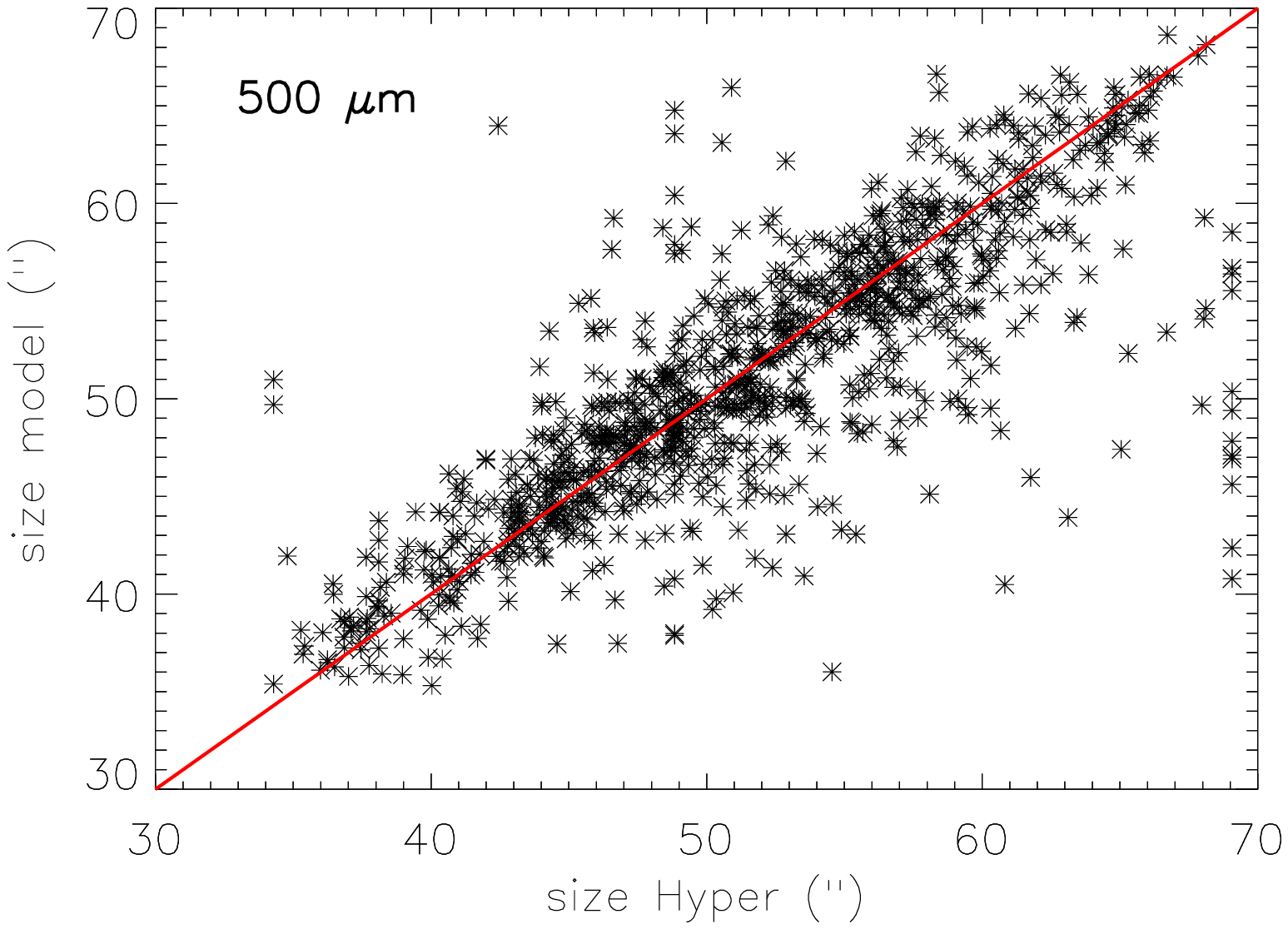}
\caption{Same distributions as Fig. \ref{fig:source_radius_15_5} but for run T2.}
\label{fig:source_radius_55_25}
\end{figure*}  


\newpage

\section{Absolute flux difference distributions between injected sources and \Hyp\ sources in the multi-wavelength approach}\label{app:source_distribution_multiwavelength}

\begin{figure*}[!ht]
\centering
\includegraphics[width=8cm]{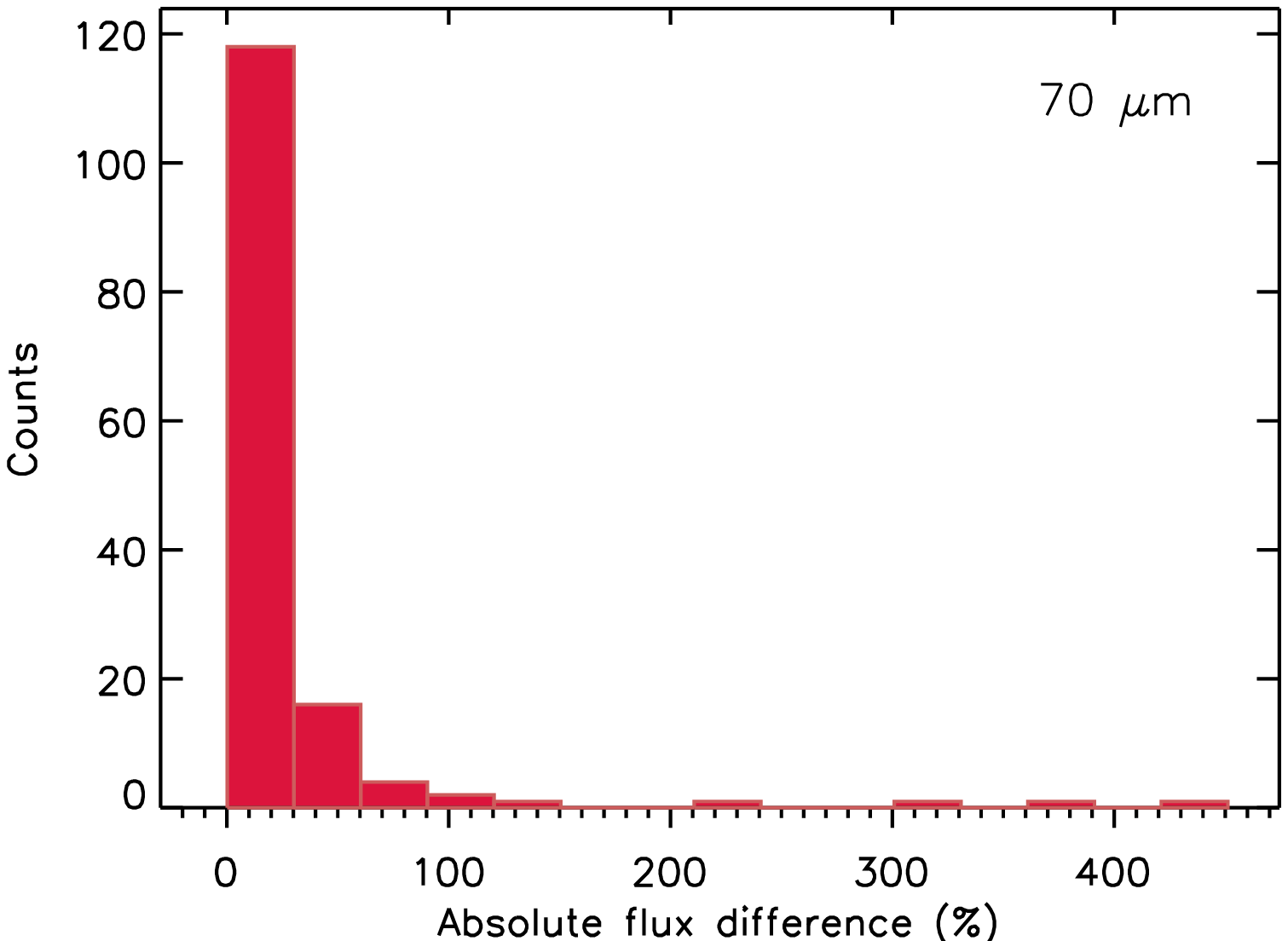}
\includegraphics[width=8cm]{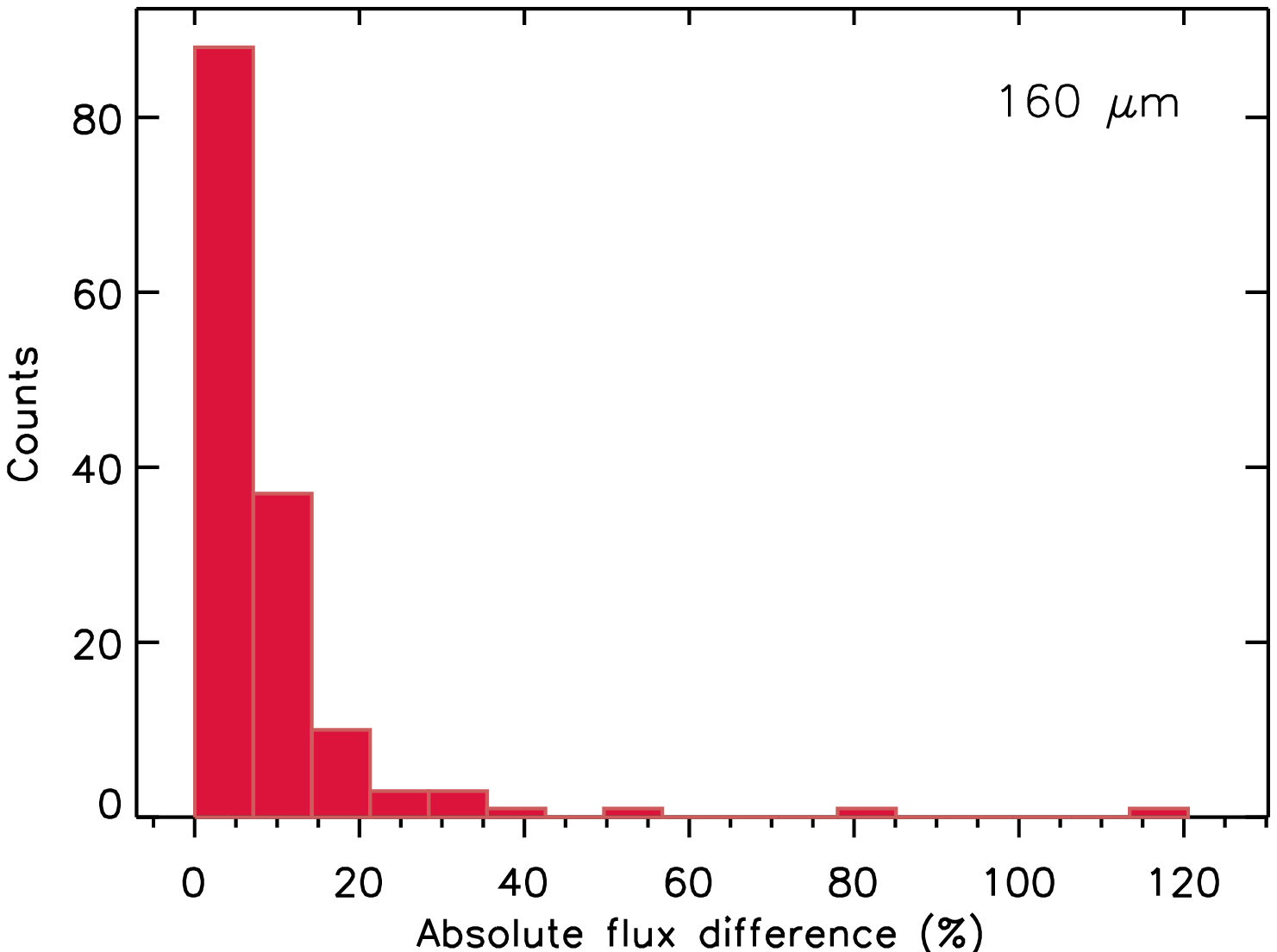}
\includegraphics[width=8cm]{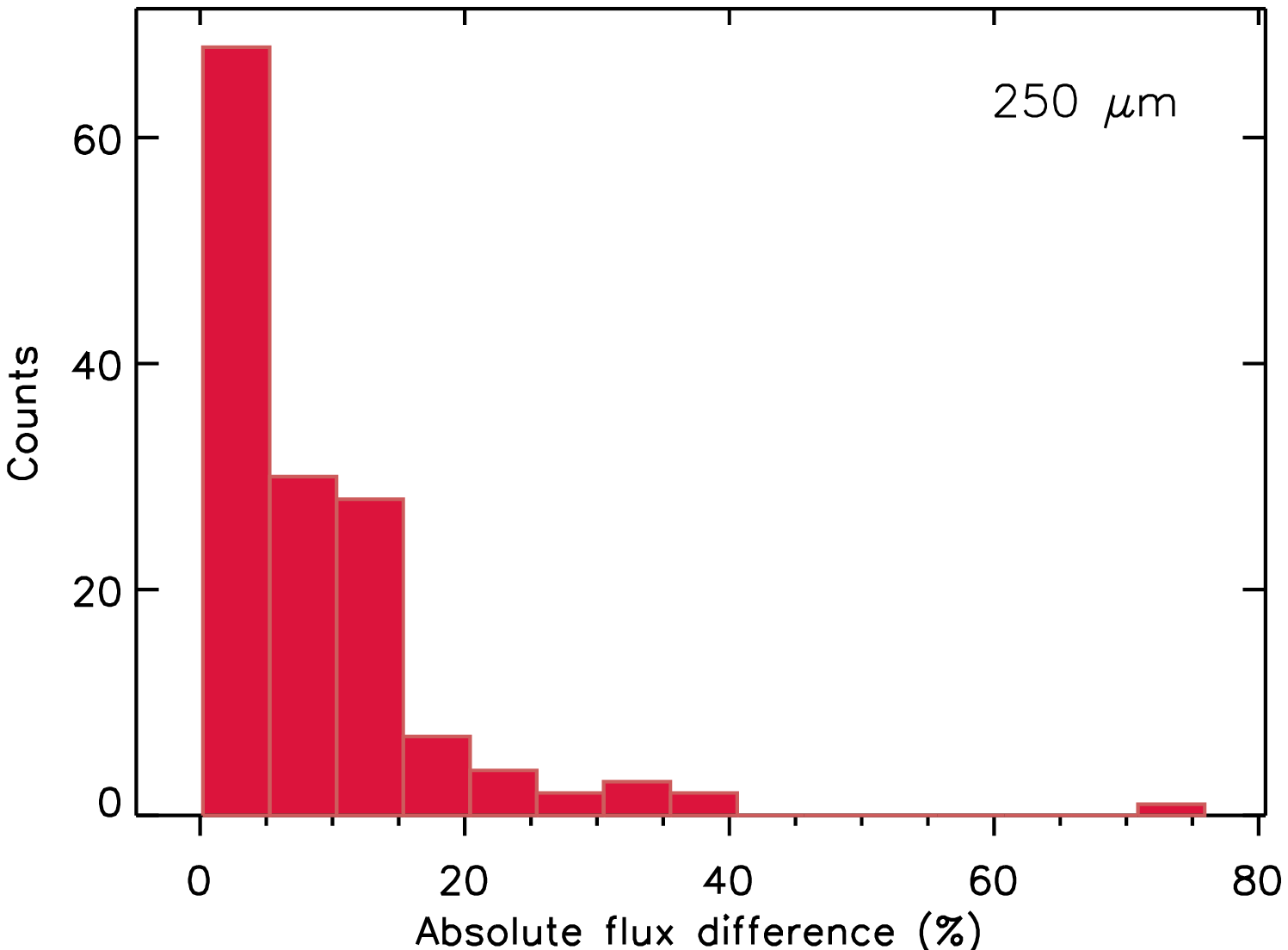}
\includegraphics[width=8cm]{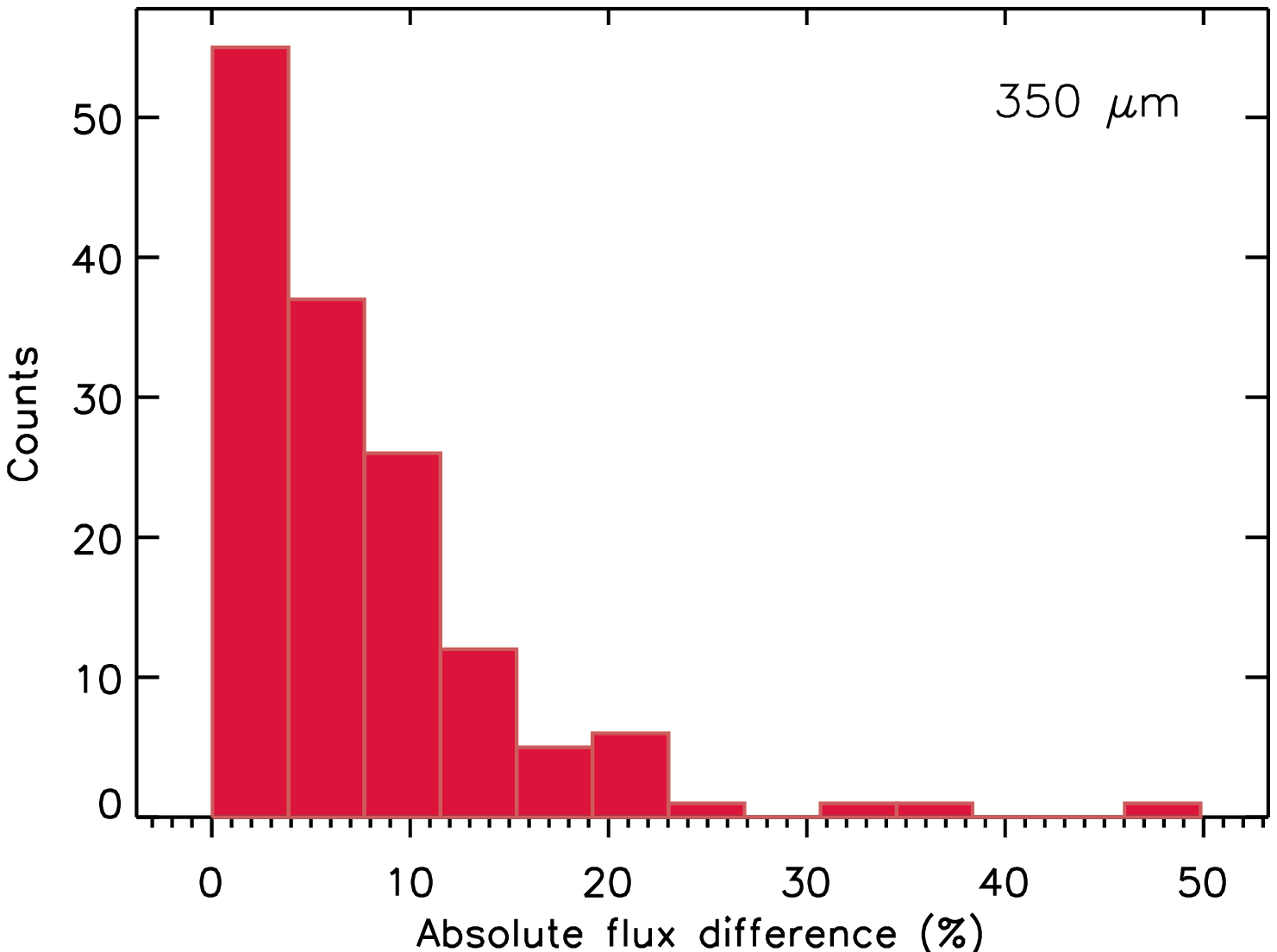}
\includegraphics[width=8cm]{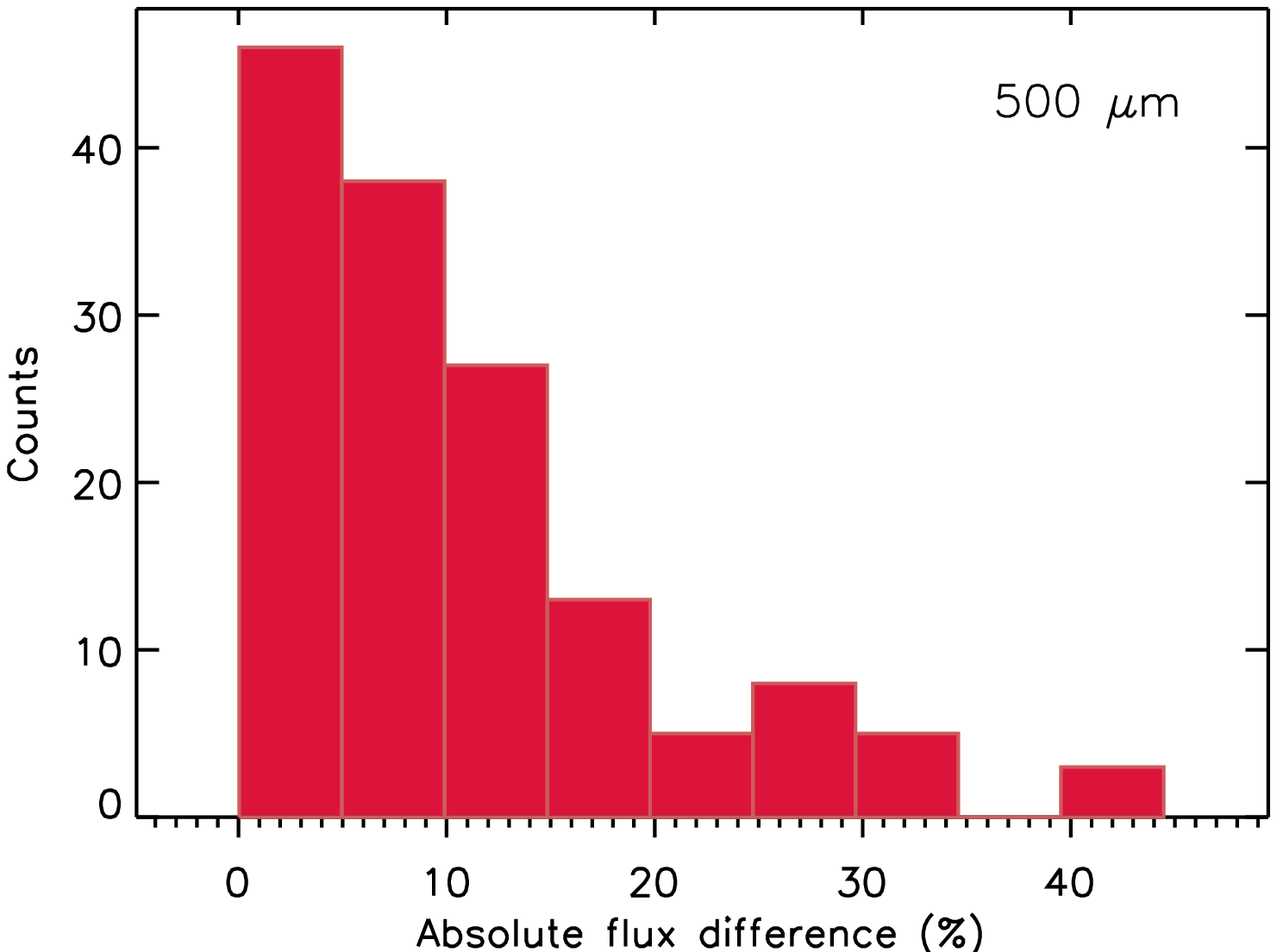}
\caption{Histogram of the absolute flux difference between the flux of the injected sources and the \Hyp\ fluxes at each wavelengths with flux measured in the 500 \mum\ aperture region as described in Sect. \ref{sec:test_multiwave}. The sources are 145 at each wavelength, all the sources with no clustered companions as described in Section \ref{sec:test_multiwave}. The Figures show the 70 \mum\ (upper left panel), 160 \mum\  (upper right), 250 \mum\ (centre left), 350 \mum\ (centre right) and 500 \mum\ (lower panel) distributions.}
\label{fig:source_distribution_multiwavelength}
\end{figure*}  

\end{appendix}

\label{lastpage}
\end{document}